\documentclass[prd,aps,amssymb,amsmath,amsfonts,twocolumn,a4paper,showkeys,nofootinbib]{revtex4-1}

\usepackage{graphicx,psfrag}
\usepackage{mathrsfs}
\usepackage{amsmath,amsfonts,amssymb}
\usepackage{multirow}
\usepackage{comment}
\usepackage[normalem]{ulem}
\usepackage{hyperref}
\usepackage{enumitem}
\usepackage{natbib}
\usepackage{glossaries}
\usepackage{cprotect}
\usepackage{bm}
\usepackage[utf8]{inputenc}
\usepackage[T1]{fontenc}

\makeglossaries

\newcommand{\be}{\begin{equation}}
\newcommand{\ee}{\end{equation}}
\newcommand{\bea}{\begin{eqnarray}}
\newcommand{\eea}{\end{eqnarray}}

\def\lm{{\ell m}}

\def\Msun{{\rm M_{\odot}}}
\def\GPc3yr{{\rm Gpc}^{-3} {\rm yr^{-1}}}
\def\GMc2{{\rm G M_{\odot} c^{-2}}}

\def\kt2{\kappa^\text{T}_2}

\def\Mbh{M_\text{BH}}
\def\abh{a_\text{BH}}

\def\Mns{M_\text{NS}}

\def\Mfbh{M_\bullet}
\def\afbh{a_\bullet}
\def\etal{{\it et al.}}

\usepackage{pifont} 

\newcommand{\teob}{\texttt{TEOBResumS}}
\newcommand{\teobdali}{\texttt{TEOBResumS-Dalí}} \newcommand{\teobgiotto}{\texttt{TEOBResumS-GIOTTO}}
\newcommand{\TEOB}{\teob}
\newcommand{\elliptica}{\texttt{Elliptica}}
\newcommand{\bam}{\texttt{BAM}}

\newacronym{GW}{GW}{gravitational wave}
\newacronym{BHNS}{BHNS}{black-hole--neutron star}
\newacronym{BBH}{BBH}{binary black-hole}
\newacronym{BNS}{BNS}{binary neutron star}
\newacronym{EoS}{EoS}{Equation of State}
\newacronym{EOB}{EOB}{Effective One Body}
\newacronym{EM}{EM}{electromagnetic}
\newacronym{BH}{BH}{black-hole}
\newacronym{NS}{NS}{neutron star}
\newacronym{NR}{NR}{numerical relativity}
\newacronym{AH}{AH}{apparent horizon}
\newacronym{QNM}{QNM}{quasi-normal modes}
\newacronym{ISCO}{ISCO}{Innermost Stable Circular Orbit}
\newacronym{PN}{PN}{post-Newtonian}
\newacronym{NQC}{NQC}{next-to-quasicircular corrections}
\newacronym{SCDD}{SCDD}{Schur complement domain decomposition}
\newacronym{LLF}{LLF}{local Lax-Friedrichs}
\newacronym{WENOZ}{WENOZ}{5th order weighted-essentially-non-oscillatory method}
\newacronym{CFL}{CFL}{Courant-Friedrich-Lewy}
\newacronym{PSD}{PSD}{power spectral density}
\newacronym{ADM}{ADM}{Arnowitt-Deser-Misner}
\newacronym{ET}{ET}{Einstein Telescope}

\usepackage{color}
\usepackage{xcolor}
\definecolor{cyan}{rgb}{0,0.9,0.9}
\definecolor{orange}{rgb}{0.9,0.5,0}
\definecolor{magenta}{rgb}{1,0,1}
\definecolor{purple}{rgb}{0.8,0.4,0.8}
\definecolor{gray}{rgb}{0.8242,0.8242,0.8242}

\usepackage{ulem}

\begin{document}

\title{Black-hole - neutron-star mergers: new numerical-relativity simulations and multipolar effective-one-body model with spin precession and eccentricity}

\author{Alejandra \surname{Gonzalez}$^{1,2}$, 
Sebastiano \surname{Bernuzzi}$^{2}$, Alireza \surname{Rashti}$^{3,4}$, Francesco \surname{Brandoli}$^{2,6}$, Rossella \surname{Gamba}$^{4,5}$}

\affiliation{${}^1$Departament de Física, Universitat de les Illes Balears, IAC3, Carretera Valldemossa km 7.5, E-07122 Palma, Spain}
\affiliation{${}^2$Theoretisch-Physikalisches Institut, Friedrich-Schiller-Universit{\"a}t Jena, Fr{\"o}belstieg 1, 07743 Jena, Germany\hspace{1cm}}
\affiliation{${}^3$Department of Physics, The Pennsylvania State University, University Park, PA 16802}
\affiliation{${}^4$Institute for Gravitation $\&$ the Cosmos, The Pennsylvania State University, University Park, PA 16802}
\affiliation{${}^5$Department of Physics, University of California, Berkeley, CA 94720, USA}
\affiliation{${}^6$Dipartimento di Fisica e Astronomia "Augusto Righi",Università di Bologna, Via Gobetti 93/2, 40129 Bologna, Italy}

\date{\today}
\begin{abstract}
                                    In this paper, we present 52 new numerical-relativity (NR) simulations of
  black-hole-neutron-star merger (BHNS) mergers and employ the data to inform \teobdali:
  a multipolar effective-one-body model also including precession and eccentricity.
    Our simulations target quasicircular mergers and the parameter space
  region characterized by significant tidal disruption of the
  star. Convergent gravitational waveforms are produced with a
  detailed error budget after extensive numerical tests.
  We study in detail the multipolar amplitude hierarchy and identify a
  characteristic tidal signature in the $(\ell,m)=(2,0)$, and $(3,0)$
  modes. 
  We also develop new NR-informed models for the remnant
  black hole and for the recoil velocity. 
    The numerical data is then used to inform next-to-quasicircular
  corrections and the ringdown of \teobdali{} for BHNS.
  We show an overall order of magnitude improvement in the waveform's amplitude
  at merger and more consistent multipoles over our older
  \teobgiotto{} for BHNS.
  \teobdali{} is further validated with a new 12 orbit precessing simulation,
  showing phase and relative amplitude differences below $\sim 0.5$
  (rad) throughout the inspiral. The computed mismatches including all
  the modes lie at the one percent level for low inclinations.
  Finally, we demonstrate for the first time that \teobdali{} can produce
  robust waveforms with both eccentricity and precession, and use the
  model to identify the most urgent BHNS to simulate for
  waveform development. 
  Our new numerical data are publicly released as part of the CoRe database.
\end{abstract}


\maketitle

\section{Introduction} 
Recent years have been marked by the \gls{GW} observations of
\gls{BBH}, \gls{BNS} and most recently \gls{BHNS} mergers by the
LIGO-Virgo interferometers~\cite{LIGOScientific:2014pky,VIRGO:2014yos}.  
With the inclusion of KAGRA in the ongoing fourth \gls{GW} Observing
Run (O4), the estimated merger rate for \gls{BHNS}
is $\mathcal{R}=94^{+109}_{-64}\GPc3yr$~\cite{LIGOScientific:2024elc}.
Nevertheless, detection
of such binaries remains a challenge as current waveform models do not
reach the sophistication that is available for \gls{BBH}.
This is mainly due to tidal effects playing a role on the dynamics
especially towards merger and post-merger. 
Coalescences from \gls{BHNS} where the \gls{NS} is tidally disrupted
are of physical interest as these are expected  
to be the source of \gls{EM} signals such as kilonovae and gamma-ray
bursts. However, (poorly constrained) population studies favour cases
with no detectable \gls{EM} signal, 
corresponding to low spinning and/or highly asymmetric \gls{BHNS}
systems~\cite{Zappa:2019ntl,Kunnumkai:2024qmw,Xing:2024ydg,Colombo:2025sdm}. 
This type of \gls{BHNS} binary produces \gls{GW} signals that 
resemble those from \gls{BBH} with a dimmer \gls{EM} counterpart, thus
making the detection of these BHNS mergers particularly challenging. 

Thus far, the identification of observed \gls{GW} events as \gls{BHNS} has been
possible through the inferred component masses. During the
LIGO-Virgo-Kagra (LVK) O3 run two BHNS events have been observed, 
namely GW200105 and GW200115~\cite{LIGOScientific:2021qlt}. Possible
evidence for eccentricity and precession has been recently
discussed~\cite{Morras:2025xfu}. 
These detections were then followed by the event GW230529 in
O4~\cite{LIGOScientific:2024elc}. The observation was the first of its
kind, providing evidence of the existence of compact objects in the mass range between the heaviest \gls{NS}s and the lightest black-holes. Its low mass ratio has also hinted at the possibility of an \gls{EM}
counterpart~\cite{Pillas:2025pfc,Kunnumkai:2024qmw}, and has sparked
interest in the binary's
origin~\cite{Mahapatra:2025agb,Chandra:2024ila} and implications to
future detections~\cite{Xue:2025ywd,Xing:2024ydg}.

Numerical-relativity simulations have become an essential tool to better
understand the physical properties and dynamics of BHNS mergers~\cite{Faber:2005yg,Grandclement:2006ht,Taniguchi:2007xm,Taniguchi:2007aq,Etienne:2007jg,Duez:2008rb,Etienne:2008re}.
Simulations in full general relativisty have reached a significant
level of sophistication and currently comprise studies with microphysical 
 \gls{EoS}~\cite{Foucart:2016vxd,Brege:2018kii,Matur:2024nwi}, magnetic fields~\cite{Chawla:2010sw,Paschalidis:2014qra,Kiuchi:2015qua,Wan:2016yid,Ruiz:2020elr,East:2021spd,Most:2021ytn,Izquierdo:2024rbb}, 
 neutrino transport~\cite{Foucart:2015vpa,Kyutoku:2017voj,Hayashi:2022cdq} and
 evolutions with extremal \gls{BH} parameters~\cite{Foucart:2010eq,Lovelace:2013vma}. 

Simulations make possible a detailed exploration of the merger
physics; in particular they show how the tidal disruption of the
\gls{NS} influences the merger dynamics and the radiated gravitational
waveforms. \citet{Kyutoku:2011vz} identifies three merger
classes: Type I corresponds to the scenario where the
\gls{NS} is tidally disrupted before merging with 
the \gls{BH}; Type II involves the \gls{NS} directly plunging into the
\gls{BH} without tidal disruption; in Type III the \gls{BH}'s tidal
field  induces unstable mass transfer from the \gls{NS} during mass
shedding. Each of these cases has a distinctive imprint on the remnant
BH and ringdown part of the \gls{GW} spectrum~\cite{Pannarale:2013jua,Zappa:2019ntl}. 
Another effect of astrophysical interest is the kick velocity on the
remnant \gls{BH}, which is due to both the anisotropic \gls{GW}
emission and the mass ejecta. For highly asymmetric binaries (Type
II), the resulting kick velocity has shown to be consistent with
\gls{BBH} fits and simulations~\cite{Foucart:2013psa}. 
Moreover, the ejecta velocity is expected to be dominant over the one
due to \gls{GW}s for Type I BHNS~\cite{Kyutoku:2013wxa}. 
The waveforms extracted from \gls{NR} simulations crucially 
allowed us to fine-tune and validate waveform
models for GW astronomy. Error-controlled simulations with a long inspiral, although
scarce, are necessary for the development of waveform models. 
Significant efforts have been done with the SACRA
code~\cite{Yamamoto:2008js,Kiuchi:2017pte}, that offers a catalogue
containing over 100 configurations. Quasicircular initial data are
simulated for a variety of spins and mass ratios, with the \gls{NS}
modelled employing a piecewise polytropic
\gls{EoS}~\cite{Kyutoku:2010zd,Kyutoku:2011vz,Kyutoku:2013wxa,Hayashi:2020zmn}. The
SACRA waveforms cover about five orbits and do not reach convergence
completely (see Appendix of~\cite{Kyutoku:2010zd}). Only one
resolution with the dominant (2,2) mode is made  
publicly available for each simulation. 
The SXS Collaboration's catalogue offers longer waveforms produced
with the SpEC code~\cite{SpEC} at three different grid resolutions.
Their catalogue comprises six configurations evolved for up to 12 orbits, three simulations with $\sim 15-16$ orbits (including one with a precessing spin \gls{BH}), and a longer \gls{BHNS} simulation with 16.6 
orbits~\cite{SXS:catalog,Foucart:2018lhe,Foucart:2020xkt}. These
binaries have both spinning and nonspinning components and a \gls{NS}
modelled with an ideal gas \gls{EoS}. 
Errors of $1\%$ on the amplitude and 0.01 rad on the phase have been achieved when comparing current extrapolation methods with Cauchy Characteristic Extraction (CCE)
in~\cite{Foucart:2020xkt}.
Nevertheless, detailed waveform accuracy and convergence studies for \gls{BHNS}
waveforms are lacking in the literature. Contrary to \gls{BNS}, see 
e.g.~\cite{Bernuzzi:2011aq,Bernuzzi:2016pie,Kiuchi:2019kzt,Doulis:2022vkx},
achieving convergent waveforms for BHNS still appears a challenging task.

The availability of accurate gravitational waveform templates is essential for identifying the source of the \gls{GW} signal. For \gls{BHNS} coalescences, early development started
with the calibration of a phenomenological \gls{BBH} model with
numerical \gls{BHNS} data~\cite{Lackey:2011vz,Lackey:2013axa}. 
Further progress went into adding amplitude corrections to account for
the different physics at
merger~\cite{Pannarale:2013uoa,Pannarale:2015jka} based on the merger  
classification of~\citet{Pannarale:2015jia}, and a remnant model built
upon estimates of the remnant's disk baryon
mass~\cite{Pannarale:2013jua}. 
Following on this early work, an updated phenomenological model was
developed by \citet{Thompson:2020nei} including a \gls{NR} informed
\gls{GW} phase with tidal contributions. 
\gls{EOB} waveforms in the frequency domain have been computed
by~\citet{Matas:2020wab} by combining a BBH EOB baseline with the NRTidal model~\cite{Dietrich:2019kaq}
and the remnant model of~\citet{Zappa:2019ntl}.
In \cite{Gonzalez:2022prs}, we presented \teob\texttt{-GIOTTO}, an
\gls{EOB} time domain waveform model for \gls{BHNS} consisting of three main
building blocks:
(i)~a \gls{NR}-informed remnant \gls{BH} model,  
(ii)~\gls{BHNS} specific \gls{NQC}, and
(iii)~a "deformable" ringdown model. Compared to other models,
\teob\texttt{-GIOTTO} includes subdominant modes and describes also
precessing binaries.   
A main limitation of all of these models is the availability of
\gls{NR} data to design accurate prescriptions for the merger.

In this paper we present 52 new BHNS simulations together with error-controlled
waveforms and use them to improve the \gls{EOB} model of~\citet{Gonzalez:2022prs}.
In Section~\ref{sec:nr_methods}, we describe the numerical methods
employed for the simulations and the set of simulated
binaries. An extensive study of the grid configuration, convergence
tests and data quality are presented in Appendix~\ref{app:nr}.  
In Section~\ref{sec:nr_res} we discuss the main simulation results.
First, we assess initial data by considering quasiequilibrium sequences
and comparing to EOB predictions.
Second, we give an overview of the dynamics and present an updated model
for the mass and spin of the remnant \gls{BH}.
Third, we investigate in detail the multipolar structure of BHNS
waveforms and identify the hierarchy of modes contributing to the
waveforms.
Fourth, we discuss a model for \gls{GW} recoil of the final \gls{BH}.
Finally, one of our simulations is compatible with GW230529; we thus
discuss some consequences for the interpretation of the event.
In Section~\ref{sec:eob}, we present \teobdali{} for BHNS.
First, we describe the new ringdown model for the $(2,2)$ and
subdominant modes together with \gls{NQC} and other new design choices
for the inspiral-merger-ringdown waveform.
Explicit NR-driven models and the fitting coefficients
developed for all modes are collected in
Appendix~\ref{app:fit_models}.
In Section~\ref{sec:validation} we validate our model by comparing
it with a new precessing \gls{NR} simulation comprising 12 orbits.
In Section~\ref{sec:predict} we showcase the use \teobdali{} for
guiding future NR simulations and for predicting GW merger signals from
arbitrary orbits.
First, we apply \teobdali{} to identify the most urgent regions of the
parameter space where more quasicircular non-precessing simulations
are necessary.
Second, we present the very first BHNS waveforms with eccentricity and
precession.
Finally, we compare the prediction of our model with some of the
recent waveforms best-fitting LVK events.

\paragraph*{Notation.}
Throughout this work, we employ geometric units $c=G=1$ and solar
masses $\Msun$, unless explicitly indicated. 
The binary mass is indicated as 
$M$, $q=m_1/m_2\geq1$ is the mass ratio, and $\nu = q/(1+q)^2$ is the symmetric mass ratio.
We use $\Mbh$ and $\abh=\chi_1$ for the mass and dimensionless spin of the
\gls{BH} in the binary system, with the latter defined as
$\chi_i=S_i/M^2_i$.; $\Mns$ is the \gls{NS} mass (here we
consider irrotational \gls{NS}, $\chi_2=0$).
We denote with $\Mfbh$ and $\afbh=S_{\bullet}/\Mfbh^2$ the remnant \gls{BH}'s mass and dimensionless spin respectively. 
We also employ the effective spin $\tilde{a}_0=\tilde{a}_1 - \tilde{a}_2 = X_1\chi_1 - X_2\chi_2$ and $\tilde{a}_{12}\equiv \tilde{a}_1 + \tilde{a}_2$, where $X_{1,2}$ are the mass fractions: 
$X_i = m_i/M$ and $X_{12}\equiv X_1 - X_2 = \sqrt{1 - 4\nu}$; and $\chi_i$ the dimensionless spins. Aditionally we use the following spins combination,
\be
\hat{S} \equiv \frac{S_1 + S_2}{M^2} = \frac{1}{2}(\tilde{a}_0 + X_{12}\tilde{a}_{12}).
\ee
with the dimensionfull spins $S_i$ along the direction of the orbital momentum. Finally, it is useful to define the dimensionless 
precession spin parameter as in~\cite{Schmidt:2014iyl},
\be
\chi_p = \max \left( |\bm{\chi_{1,\perp}}|, \frac{4+3q}{4q^2+3q} |\bm{\chi_{2,\perp}}| \right).
\ee
The strain is defined as
\be\label{eq:strainh}
h \equiv h_+ - ih_{\times}=\displaystyle\sum_{\ell=2} ^{\infty}\displaystyle\sum_{m=-\ell} ^{\ell}h_{\ell m}\, _{-2}Y_{\ell m},
\ee
where $_{-2}Y_{\ell m}$ are the $s=-2$ spin-weighted spherical harmonics. Multipoles are decomposed in amplitude and phase
\be
h_{\lm} = A_{\lm}e^{-i\phi_{\lm}}\ ,
\ee
the instantaneous \gls{GW} frequency is defined as the time derivative of the phase,  $\omega_{\lm}\equiv \dot{\phi}_{\lm}$. 
We present the \gls{NR} waveforms in terms of the retarded time $u=t-r_*$, where $r_*$ corresponds to the associated tortoise Schwarzschild coordinate at the extraction radius $R$
in the simulation.

\section{Numerical-relativity methods \& Simulations}
\label{sec:nr_methods}

\subsection{Initial Data}

The initial data 
solution is obtained with the publicly available and open-source
pseudospectral code \elliptica~\cite{Rashti:2021ihv,Rashti:2024drr}. 
In contrast to other available initial data solvers, \elliptica~constructs \gls{BHNS} initial data with dimensionless 
spin magnitudes up to $\sim$ 0.8 with arbitrary spin orientations. 

Using the extended conformal thin sandwich method~(XCTS)
formalism~\cite{York:1998hy,Pfeiffer:2002iy}, and 
the velocity potential method~\cite{Tichy:2012rp}, \elliptica{} efficiently solves 
the coupled elliptic partial differential equations of the Einstein-Euler system
in a multi-core parallelism paradigm. This approach leverages employing
 the divide and conquer method of \gls{SCDD}.

\elliptica{} uses excision boundary conditions~\cite{Cook:2004kt}
to solve for the \gls{BH} in the \gls{BHNS} system. 
Consequently, before transferring the initial data to the evolution codes
like the \bam{} code, \elliptica{} fills the excised region by applying
a $C^2$ continuous extrapolation of the metric fields from the surrounding area
of the \gls{BH}.

\subsection{3+1 Evolution}

The initial data in this work is evolved with the \bam~code~\cite{Brugmann:2008zz,Thierfelder:2011yi} using the Z4c formulation of 
Einstein's equations coupled to relativistic hydrodynamics. 

The computational domain in \bam~consists of cell-centered nested Cartesian grids with $n$ points per direction in $L$ refinement levels, labeled as $l=0$,\dots,$L-1$. 
Each refinement level $l$ is composed by one or more overlapping grids with a constant grid spacing $h_l$. These are related by a factor of 2 as $h_l = h_0/2^l$, 
where $h_0$ is the grid spacing at the coarsest level $l=0$. The refinement level grids always stay within the coarser levels. 
Refinement levels above a given user-defined threshold $l^{\rm mv}$, can be dynamically moved, as to follow the orbits of the two objects, adopting a ”moving boxes” 
technique with a number of points per direction $n^{\rm mv}$ arbitrarily selected.
The time evolution of the grid fields relies on the method of lines and Runge-Kutta time integrators with a \gls{CFL} factor of 0.25. The metric variables are approximated 
employing 4th order finite differencing stencils.

For hydrodynamics we employ the \gls{LLF} central scheme~\cite{Kurganov:2000}~\cite{Nessyahu:1990} for the interface fluxes and \gls{WENOZ}
scheme~\cite{Borges:2008a} for the primitive reconstruction.
For the \gls{NS} \gls{EoS} we employ the piecewise polytropic models SLy and MS1b~\cite{Read:2008iy}, and the hybrid model ALF2~\cite{Alford:2004pf} that accounts for deconfined quark matter.

In order to obtain the most accurate
possible waveforms with the least computational cost, several grid configurations employing additional refinement levels on the \gls{BH} 
with respect to the \gls{NS} were tested.
This is described in Appendix~\ref{app:nr_quality}. For the production runs, the configuration M8 (see Table~\ref{tab:grid_configurations}) is chosen as it provides high quality data 
comparable to higher resolutions with the least amount of computational resources. Convergence and comparison of these grid choices are presented in Appendix~\ref{app:nr_quality}.

\subsection{Simulations}

For this work, we simulated 51 quasi-circular and non precessing BHNS and one precessing configuration.  Table~\ref{tab:configurations} summarizes the initial data paramaters of each simulation,
which are chosen so that the configurations are close to our estimated tidal disruption boundary~\cite{Gonzalez:2022prs}.
Since we are interested in extracting information from the merger and post-merger, we evolve the 51 configurations for 3-4 orbits which are used to inform our 
\gls{EOB} model. Given the small length of the waveforms, we do not perform any eccentricity reduction procedure on them, which stays usually below $e\sim 0.02$. 
Similarly for our precessing configuration, evolved for 12 orbits and employed for validation, we do not eccentricity reduce the initial data as it stays at a similar low value. 
Future work will focus on low eccentricity data for precessing \gls{BHNS}.
  
\begin{table}[h!]
  \centering    
  \caption{ \gls{BHNS} configurations simulated in this work:  $M_b$ the baryonic mass of the \gls{NS}, 
  $\Omega_{\rm BHNS}$ the angular velocity of the \gls{BHNS} system, $M_{\rm ADM}$ the total \gls{ADM} mass, and $J_{\rm ADM}$ the total \gls{ADM} angular momentum. 
    For all configurations we consider a nonspinning \gls{NS}, and for the \gls{BH} spin the subscript $^p$ refers instead to $\chi_p$. The last column refers to the merger type
    according to our classification, see text.}
  \scalebox{0.75}{
    \begin{tabular}{c | c | c | c | c | c | c | c | c | c | c}
      \hline\hline
      Name & \gls{EoS} & $q$ & $\Mbh$ & $\abh$ & $M_b$ & $M_{\rm NS}$ & $\Omega_{\rm BHNS}$ & $M_{\rm ADM}$ & $J_{\rm{ADM}}$ & Type\\
      \hline
      \texttt{BAM:0177} & ALF2 & 2.2 & 2.73 & -0.2991 & 1.35 & 1.24 & 0.0069 & 3.9309 & 10.9535 & I\\
      \texttt{BAM:0178} & ALF2 & 2.3 & 2.8 & -0.4971 & 1.35 & 1.24 & 0.007 & 3.9978 & 9.7176 & I\\
      \texttt{BAM:0179} & ALF2 & 2.3 & 2.85 & -0.5944 & 1.35 & 1.24 & 0.0071 & 4.0541 & 9.0849 & I\\
      \texttt{BAM:0180} & ALF2 & 2.4 & 2.93 & -0.6897 & 1.35 & 1.24 & 0.0071 & 4.1381 & 8.4227 & I\\
      \texttt{BAM:0176} & ALF2 & 2.2 & 2.73 & 0.3008 & 1.35 & 1.24 & 0.0069 & 3.9299 & 14.9955 & I\\
      \texttt{BAM:0181} & ALF2 & 2.2 & 3.24 & -0.2991 & 1.6 & 1.44 & 0.0074 & 4.6327 & 14.3109 & III\\
      \texttt{BAM:0182} & ALF2 & 2.3 & 3.38 & -0.597 & 1.6 & 1.44 & 0.0075 & 4.7807 & 11.7118 & III\\
      \texttt{BAM:0185} & ALF2 & 3.4 & 4.86 & -0.299 & 1.6 & 1.44 & 0.0061 & 6.246 & 18.7861 & III\\
      \texttt{BAM:0186} & ALF2 & 3.5 & 5.08 & -0.5974 & 1.6 & 1.44 & 0.0062 & 6.471 & 12.2521 & III\\
      \texttt{BAM:0183} & ALF2 & 3.3 & 4.8 & 0.001 & 1.6 & 1.44 & 0.006 & 6.1868 & 24.8999 & III\\
      \texttt{BAM:0184} & ALF2 & 3.4 & 4.86 & 0.2996 & 1.6 & 1.44 & 0.006 & 6.2423 & 31.5341 & III\\
      \texttt{BAM:0187} & ALF2 & 3.5 & 5.09 & 0.5974 & 1.6 & 1.44 & 0.006 & 6.4753 & 40.1652 & III\\
      \texttt{BAM:0224} & ALF2 & 2.2 & 3.24 & 0.3005 & 1.6 & 1.44 & 0.0073 & 4.631 & 19.8464 & I\\
      \hline
      \texttt{BAM:0188} & MS1b & 2.2 & 2.73 & -0.2991 & 1.35 & 1.25 & 0.0084 & 3.9423 & 10.6347 & I\\
      \texttt{BAM:0189} & MS1b & 2.2 & 2.8 & -0.4971 & 1.35 & 1.25 & 0.007 & 4.0117 & 9.8693 & I\\
      \texttt{BAM:0192} & MS1b & 2.3 & 2.85 & -0.5944 & 1.35 & 1.25 & 0.0071 & 4.068 & 9.2412 & I\\
      \texttt{BAM:0193} & MS1b & 2.3 & 2.93 & -0.6897 & 1.35 & 1.25 & 0.0072 & 4.1521 & 8.5846 & I\\
      \texttt{BAM:0191} & MS1b & 1.9 & 2.83 & -0.2992 & 1.6 & 1.46 & 0.0072 & 4.2504 & 13.3969 & I\\
      \texttt{BAM:0194} & MS1b & 2.0 & 2.96 & -0.5956 & 1.6 & 1.46 & 0.0073 & 4.3785 & 11.5023 & I\\
      \texttt{BAM:0190} & MS1b & 1.9 & 2.8 & 0.0013 & 1.6 & 1.46 & 0.0071 & 4.2174 & 15.3542 & I\\
      \texttt{BAM:0196} & MS1b & 2.2 & 3.24 & -0.2991 & 1.6 & 1.46 & 0.0063 & 4.6546 & 15.0722 & I\\
      \texttt{BAM:0198} & MS1b & 2.3 & 3.38 & -0.5951 & 1.6 & 1.46 & 0.0064 & 4.8024 & 12.4563 & I\\
      \texttt{BAM:0195} & MS1b & 2.2 & 3.2 & 0.0012 & 1.6 & 1.46 & 0.0073 & 4.6134 & 17.1394 & I\\
      \texttt{BAM:0197} & MS1b & 2.2 & 3.24 & 0.3004 & 1.6 & 1.46 & 0.0073 & 4.6499 & 20.0667 & I\\
      \texttt{BAM:0200} & MS1b & 3.3 & 4.86 & -0.2989 & 1.6 & 1.46 & 0.0061 & 6.2649 & 19.123 & III\\
      \texttt{BAM:0203} & MS1b & 3.5 & 5.08 & -0.5974 & 1.6 & 1.46 & 0.0062 & 6.49 & 12.6149 & III\\
      \texttt{BAM:0199} & MS1b & 3.3 & 4.8 & 0.0011 & 1.6 & 1.46 & 0.006 & 6.2058 & 25.2409 & III\\
      \texttt{BAM:0201} & MS1b & 3.3 & 4.86 & 0.3007 & 1.6 & 1.46 & 0.006 & 6.2624 & 31.9143 & III\\
      \texttt{BAM:0202} & MS1b & 3.5 & 5.08 & 0.5988 & 1.6 & 1.46 & 0.006 & 6.4818 & 40.433 & I\\
      \texttt{BAM:0225} & MS1b & 2.3 & 2.94 & 0.6936 & 1.35 & 1.25 & 0.007 & 4.1533 & 19.3349 & I\\
      \texttt{BAM:0226} & MS1b & 2.0 & 2.96 & 0.5975 & 1.6 & 1.46 & 0.0071 & 4.3767 & 20.7158 & I\\
      \hline
      \texttt{BAM:0204} & SLy & 2.0 & 2.8 & 0.0013 & 1.6 & 1.43 & 0.0071 & 4.1913 & 15.0943 & III\\
      \texttt{BAM:0205} & SLy & 2.0 & 2.83 & 0.3009 & 1.6 & 1.43 & 0.0071 & 4.2236 & 17.3788 & III\\
      \texttt{BAM:0208} & SLy & 2.5 & 3.24 & -0.2991 & 1.4 & 1.27 & 0.0062 & 4.4713 & 12.9242 & III\\
      \texttt{BAM:0210} & SLy & 2.3 & 3.24 & -0.2991 & 1.6 & 1.43 & 0.0074 & 4.6256 & 14.1989 & III\\
      \texttt{BAM:0211} & SLy & 2.7 & 3.38 & -0.5952 & 1.4 & 1.27 & 0.0063 & 4.6184 & 10.1689 & III\\
      \texttt{BAM:0206} & SLy & 2.2 & 3.2 & 0.0012 & 1.6 & 1.43 & 0.0073 & 4.5873 & 16.8223 & III\\
      \texttt{BAM:0207} & SLy & 2.5 & 3.24 & 0.3007 & 1.4 & 1.27 & 0.0061 & 4.4702 & 18.6489 & III\\
      \texttt{BAM:0209} & SLy & 2.3 & 3.24 & 0.3004 & 1.6 & 1.43 & 0.0073 & 4.624 & 19.7719 & III\\
      \texttt{BAM:0212} & SLy & 2.7 & 3.38 & 0.5976 & 1.4 & 1.27 & 0.0062 & 4.6164 & 22.5826 & I\\
      \texttt{BAM:0213} & SLy & 2.4 & 3.38 & 0.5989 & 1.6 & 1.43 & 0.0073 & 4.7694 & 23.7182 & III\\
      \texttt{BAM:0216} & SLy & 3.4 & 4.86 & -0.299 & 1.6 & 1.43 & 0.0053 & 6.241 & 19.1908 & II\\
      \texttt{BAM:0219} & SLy & 4.0 & 5.08 & -0.5973 & 1.4 & 1.27 & 0.0062 & 6.3087 & 9.1763 & III\\
      \texttt{BAM:0220} & SLy & 3.5 & 5.08 & -0.5964 & 1.6 & 1.43 & 0.0054 & 6.469 & 12.5726 & II\\
      \texttt{BAM:0221} & SLy & 4.1 & 5.23 & -0.6946 & 1.4 & 1.27 & 0.0062 & 6.4669 & 6.516 & III\\
      \texttt{BAM:0214} & SLy & 3.3 & 4.8 & 0.0011 & 1.6 & 1.43 & 0.006 & 6.1799 & 24.7838 & II\\
      \texttt{BAM:0215} & SLy & 3.4 & 4.86 & 0.2996 & 1.6 & 1.43 & 0.006 & 6.235 & 31.4135 & III\\
      \texttt{BAM:0217} & SLy & 4.0 & 5.08 & 0.5987 & 1.4 & 1.27 & 0.0059 & 6.3012 & 37.3741 & III\\
      \texttt{BAM:0218} & SLy & 3.5 & 5.08 & 0.5988 & 1.6 & 1.43 & 0.006 & 6.4557 & 39.9675 & III\\
      \texttt{BAM:0222} & SLy & 4.1 & 5.23 & 0.6971 & 1.4 & 1.27 & 0.006 & 6.4622 & 41.3462 & III\\
      \hline
      \texttt{BAM:0223} & ALF2 & 2.5 & 3.57 & 0.613$^p$ & 1.6 & 1.44 & 0.0042 & 4.9824 & 27.0892 & I\\
      \hline\hline
    \end{tabular}}
 \label{tab:configurations}
\end{table}

\begin{figure}[t]
  \centering 
    \includegraphics[width=0.49\textwidth]{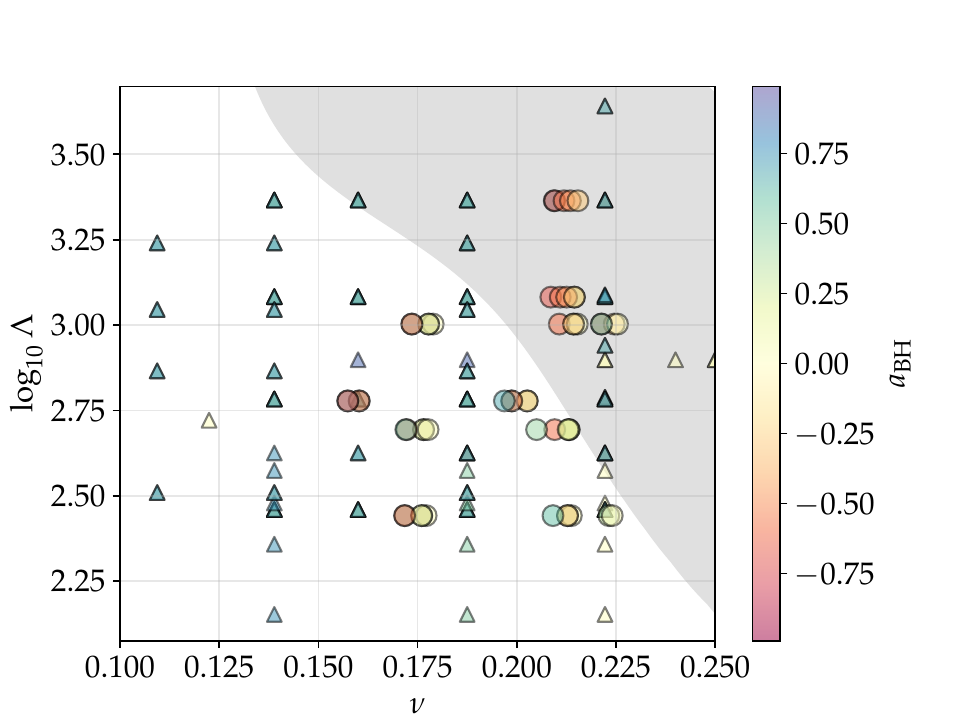}
    \caption{ Available \gls{NR} simulations for different \gls{BHNS} configurations. Circle markers show the simulations done for this paper.
    The grey shaded area covers the binaries where we estimate tidal disruption to occur, see Sec.~\ref{sec:eob_imr}.}
 \label{fig:parameterspace}
\end{figure}

Figure~\ref{fig:parameterspace} shows the parameter space that current publicly available simulations 
(including the ones produced in this work as circular markers) cover in terms of the tidal parameter $\Lambda$, symmetric mass ratio $\nu$ and \gls{BH} dimensionless spin $\abh$. 
A key challenge in waveform modelling of mixed binaries is classifying the different merger types according to the initial binary paramaters. Hence, we focus our attention 
on the region of the parameter space between the boundary of tidal disruption, Type I binaries, (grey shaded area on Fig.~\ref{fig:parameterspace}) and Type III cases where the \gls{QNM} are still
excited and present in the ringdown, but are slightly damped due to mass shedding close to merger.
Contrary to earlier 
simulations, we also consider a variety of anti-aligned spins down to $\abh \geq -0.7$. Low values of retrograde spin are consistent with 
the expected $|\chi_{\rm eff}|\approx 0$ from astrophysical studies~\cite{Broekgaarden:2021iew}, and higher values help us inform and extend our analytical models.

 \section{Simulation results}
 \label{sec:nr_res}

\subsection{Quasiequilibrium sequences}
\label{sec:nr_quseq}

\begin{figure*}
   \centering 
    \includegraphics[width=0.99\textwidth]{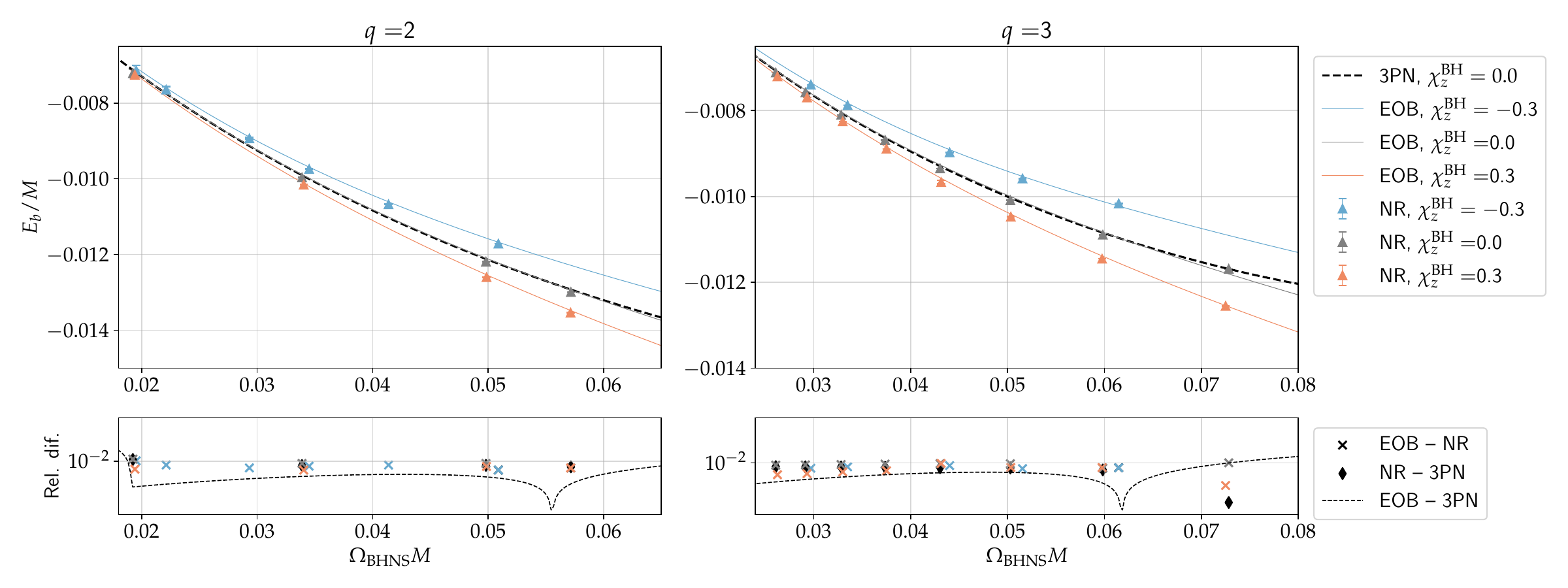}
     \caption{ Quasiequilibrium sequences of \gls{BHNS} configurations with the SLy EoS. These are computed for two mass ratios $q=2$ and $q=3$ with 
     aligned (blue) and anti-aligned (orange) spins. The results from our simulations are compared directly with those of \gls{EOB} and 3\gls{PN} (see text).
     We see an agreement between the initial data and the prediction from \gls{EOB}, thus asserting the correctness of the numerically produced data.}
  \label{fig:sly_sequences}
\end{figure*}

We start discussing quasiequilibrium configurations of \gls{BHNS} initial data with different mass ratios and spins. The study of these sequences serves as a tool to diagnose the consistency of the initial data constructed for our evolutions to predictions from analytical approximations (e.g. \gls{PN} and \gls{EOB}), e.g.~\cite{Taniguchi:2007aq,Damour:2011fu,Bernuzzi:2013rza,Topolski:2024beu}.
For \gls{BHNS} binaries, sequences of this type have served especially to study the onset of mass shedding and tidal disruption~\cite{Taniguchi:2007xm,Topolski:2024beu}.

We produce quasiequilibrium data for \gls{BHNS} configurations
employing the SLy \gls{EoS}~\cite{Read:2008iy}. The baryonic mass of
the \gls{NS} is set to be $M_b = 1.6$ in all cases and we show the
comparison with $q=2$ and $q=3$. We consider a nonrotating \gls{NS},
whereas for the \gls{BH} we choose both aligned and anti-aligned spins
as well as nonspinning punctures with $\chi^{\rm BH}_z=\abh=-0.3$,
$0.0$, and $0.3$. In the following, we focus on the relation between the
binding energy $E_b/M=M_{\rm ADM}/M - 1$ and orbital angular velocity
$M\Omega_{\rm BHNS}$. 

Figure~\ref{fig:sly_sequences} shows the energy curves for the
adiabatic configuration and compare them to the \gls{BHNS} model in \teobgiotto{}~\cite{Gonzalez:2022mgo} and the
post-Newtonian (3PN) point-mass nonspinning prediction.
Larger values of orbital angular velocity indicate smaller separations
between the \gls{BH} and \gls{NS} (closer to merger). The \gls{EOB} prediction
agrees with the produced initial data with errors around $\sim 0.01\%$. 
As expected, tidal effects take over the dynamics with increasing $\Omega_{\rm BHNS}M$
(towards the late inspiral) as the \gls{EOB} and 3PN curves start
deviating from each other. As the two objects approach each other, the effect of
the spin-orbit coupling starts increasing, thus making the system more or less bound according to the spin magnitudes
and orientation as seen in the Figure~\ref{fig:sly_sequences}. 
This effect is also captured by the prediction from \gls{EOB} in agreement with the numerical sequences.

\subsection{Dynamics}
 
Here we give an overview of the results from the NR evolutions produced in this work.
As described in Sec.~\ref{sec:nr_methods}, these simulations comprise for the most part the parameter space close to the boundary between 
tidal disruption and mass shedding. Earlier systematic BHNS studies have performed 
similar evolutions with different $\abh$ and \gls{EoS} and have focused on studying the ejecta properties of these systems~\cite{Kyutoku:2011vz,Kyutoku:2015gda,Hayashi:2022cdq,Kawaguchi:2024hdk,Chen:2024ogz}.

\begin{figure*}
   \centering 
   \includegraphics[width=\textwidth]{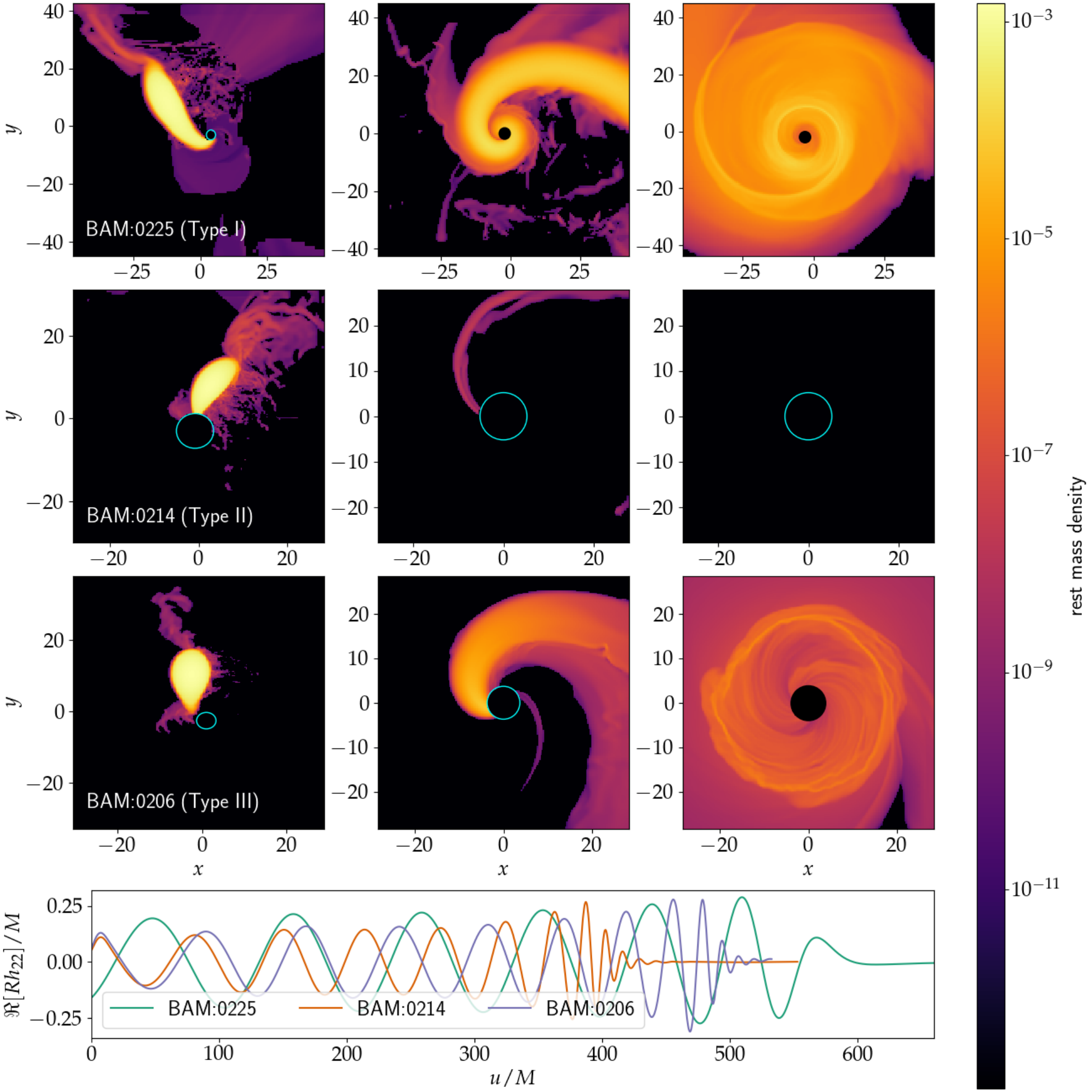}
     \cprotect\caption{Density profile at three different moments of the merger and postmerger of \verb|BAM:0225| (top), \verb|BAM:0214| (middle) and \verb|BAM:0206| (bottom),
     corresponding to Type I, II and III respectively. 
     The cyan contour indicates the location of the \gls{AH}. The bottom panel shows the resulting gravitational radiation from each coalescence. 
     Note that the retarded time axis is shifted for better visualization of the waveforms.}
  \label{fig:q2den_2d}
 \end{figure*}

Figure~\ref{fig:q2den_2d} illustrates the merger dynamics of fiducial simulations. The plot shows the rest mass density profile during the merger 
and postmerger of three different merger scenarios, including the corresponding \gls{GW} signal for the (2,2) mode for each case. According to our 
previously defined classification in~\cite{Gonzalez:2022prs} (we present here an update on its boundaries in Sec.~\ref{sec:eob_imr}), the top panels correspond to \texttt{BAM:0225}, a 
Type I merger. This type of binaries experience tidal disruption from the interplay of the effects due to masses, spins and tides. Parallel to our discussion on the quasiequilibrium
sequences in Sec.~\ref{sec:nr_quseq}, attractive tidal effects decrease the orbital separation more rapidly for stiffer \gls{EoS}. 
Higher aligned spins on the other hand have a repulsive effect. Consequently, the encounter of the two objects is delayed, prompting
the \gls{NS} to reach the onset of tidal disruption before the \gls{ISCO}. The star is thus deformed while 
leaving most of its material outside the \gls{BH}, seen as a bright disk of mass.
Consequently, the gravitational radiation does not show a ringdown as the material outside the \gls{BH} dampens it. 
The disk formed for this simulation reached a mass of $M_{\rm disk}\approx1\times10^{-6}\Msun$ (right most top panel) and 
ejected mass of $M_{\rm ej}\approx 0.02\Msun$ with velocities around $v_{\rm ej}\approx 0.1c - 0.2c$. 

The middle panels show the merger of \texttt{BAM:0214} (Type II), with the \gls{NS} directly plunging into the \gls{BH}. Here, the mass ratio's repulsive effect and the low magnitude (or zero) spins contribute to the 
\gls{NS} reaching the \gls{ISCO} first, leaving no ejected material outside the \gls{BH} and increasing the mass of the remnant \gls{BH} (notice the size increase of the \gls{AH}
from the left panel where the two objects are merging to the middle one after merger).
The perturbed remnant is thus responsible for the clear ringdown signal of the gravitational waveform.

The last panels correspond to \texttt{BAM:0206}. This is an intermediate case, close to (but not reaching) the onset of tidal disruption: a Type III merger. As seen in the figure, after the \gls{NS} is swallowed by the 
\gls{BH}, leftover (low density) material from the shed outer layers surrounds the remnant. The \gls{GW} thus shows a partially dampened ringdown, where excited \gls{QNM} are still present
but are supressed compared to Type I.

\subsection{Remnant \gls{BH}}
\label{sec:nr_bh}

The properties of the remnant \gls{BH} from a \gls{BHNS} merger are of great interest for waveform modelling and inference. We therefore extract this information 
from the \gls{AH} data of \bam~\cite{Brugmann:2008zz}. The remnant mass is obtained from the Christodoulou formula~\cite{Christodoulou:1971pcn} that involves the
irreducible mass from the \gls{AH} area, $M_{\rm irr}=\sqrt{A/16\pi}$, and a contribution from the BH spin,
\be
\label{eq:mchr}
\Mfbh^2 = (M_{\rm irr})^2 + \frac{S_{\bullet}^2}{4(M_{\rm irr})^2},
\ee
where $S_{\bullet}$ is the spin of the puncture. The results from our simulations are summarized in Tab.~\ref{tab:rem_bh}.
Note that the \gls{AH} finder was not successful in finding the \gls{AH} in all cases, hence we present measured values of fewer configurations than the ones 
shown in Tab.~\ref{tab:configurations}.

\begin{table}[h!]
   \centering    
   \caption{Remnant \gls{BH}'s properties obtained for the different \gls{BHNS} configurations.}
   \begin{tabular}{c | c | c}
    \hline\hline
    Name & $\Mfbh$ & $\afbh$ \\
    \hline\hline
    \verb|BAM:0176| & 3.7167 & 0.684 \\
    \verb|BAM:0177| & 3.8163 & 0.5313 \\
    \verb|BAM:0178| & 3.9095 & 0.464 \\
    \verb|BAM:0179| & 3.9745 & 0.4249 \\
    \verb|BAM:0184| & 6.132 & 0.6639 \\
    \verb|BAM:0189| & 3.8021 & 0.4359 \\
    \verb|BAM:0190| & 4.0024 & 0.6259 \\
    \verb|BAM:0192| & 3.9061 & 0.416 \\
    \verb|BAM:0193| & 4.025 & 0.3791 \\
    \verb|BAM:0194| & 4.3095 & 0.4852 \\
    \verb|BAM:0195| & 4.4366 & 0.6195 \\
    \verb|BAM:0196| & 4.5437 & 0.5318 \\
    \verb|BAM:0197| & 4.392 & 0.6951 \\
    \verb|BAM:0199| & 6.091 & 0.5304 \\
    \verb|BAM:0200| & 6.218 & 0.3919 \\
    \verb|BAM:0201| & 6.0721 & 0.6503 \\
    \verb|BAM:0202| & 6.1971 & 0.7759 \\
    \verb|BAM:0204| & 4.1177 & 0.6676 \\
    \verb|BAM:0205| & 4.1218 & 0.7553 \\
    \verb|BAM:0206| & 4.5299 & 0.6254 \\
    \verb|BAM:0210| & 4.5515 & 0.5117 \\
    \verb|BAM:0214| & 6.0871 & 0.5199 \\
    \verb|BAM:0215| & 6.1254 & 0.6591 \\
    \verb|BAM:0216| & 6.1551 & 0.3716 \\
    \hline\hline
    \end{tabular} 
  \label{tab:rem_bh}
 \end{table}

Models to predict the mass and spin of the remnant have been developed as more \gls{NR} data of \gls{BHNS} evolutions have been proposed in various works~\cite{Kyutoku:2011vz,Pannarale:2013jua,Zappa:2019ntl,Gonzalez:2022prs}. Here, we present yet another updated version of our previous model from~\cite{Zappa:2019ntl,Gonzalez:2022prs} by including the newly produced data. Similarly to these previous works, we factorize the \gls{BBH} contribution in order to obtain a model that smoothly connects to the BBH case. The remnant mass can be well represented by the parameters $\{\Lambda, \nu, \abh\}$ as 
\be
\frac{M^{\rm BHNS}_\bullet}{M^{\rm BBH}_\bullet} = \frac{1 + \Lambda p^{(2)}_{1}(\nu,\abh) + \Lambda^2 p^{(2)}_{2}(\nu,\abh) }{1 + \Lambda^2 c_{312}\nu^2  }
\ee
with the low-order polynomials 
\begin{subequations}
  \label{eq:mfbh_p}
  \begin{align}
  p^{(2)}_{k}(\nu,\abh)&= p^{(2)}_{k1}(\abh)\nu + p^{(2)}_{k2}(\abh)\nu^2 ,\\
  p^{(2)}_{kj}(\abh)&= c_{kj2}\abh^2 + c_{kj1}\abh + c_{kj0}.
  \end{align}
\end{subequations}
The model clearly captures the \gls{BBH} remnant mass for $\Lambda\rightarrow 0$.
The polynomial coefficients fitting the NR data can be found in Table~\ref{tab:remnant_coef} in Appendix~\ref{app:final_bh}.

\begin{figure*}
  \centering 
  \includegraphics[width=0.8\textwidth]{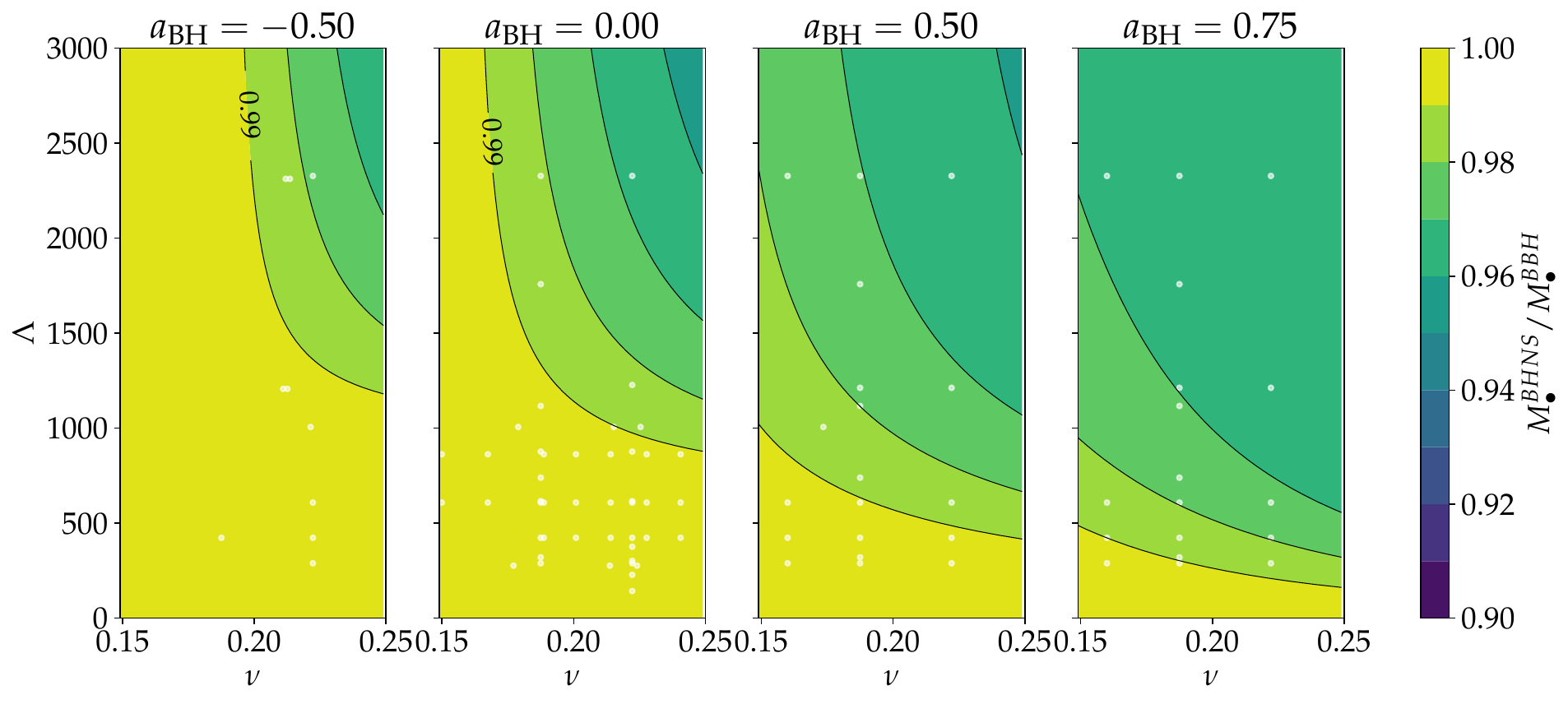}
  \includegraphics[width=0.8\textwidth]{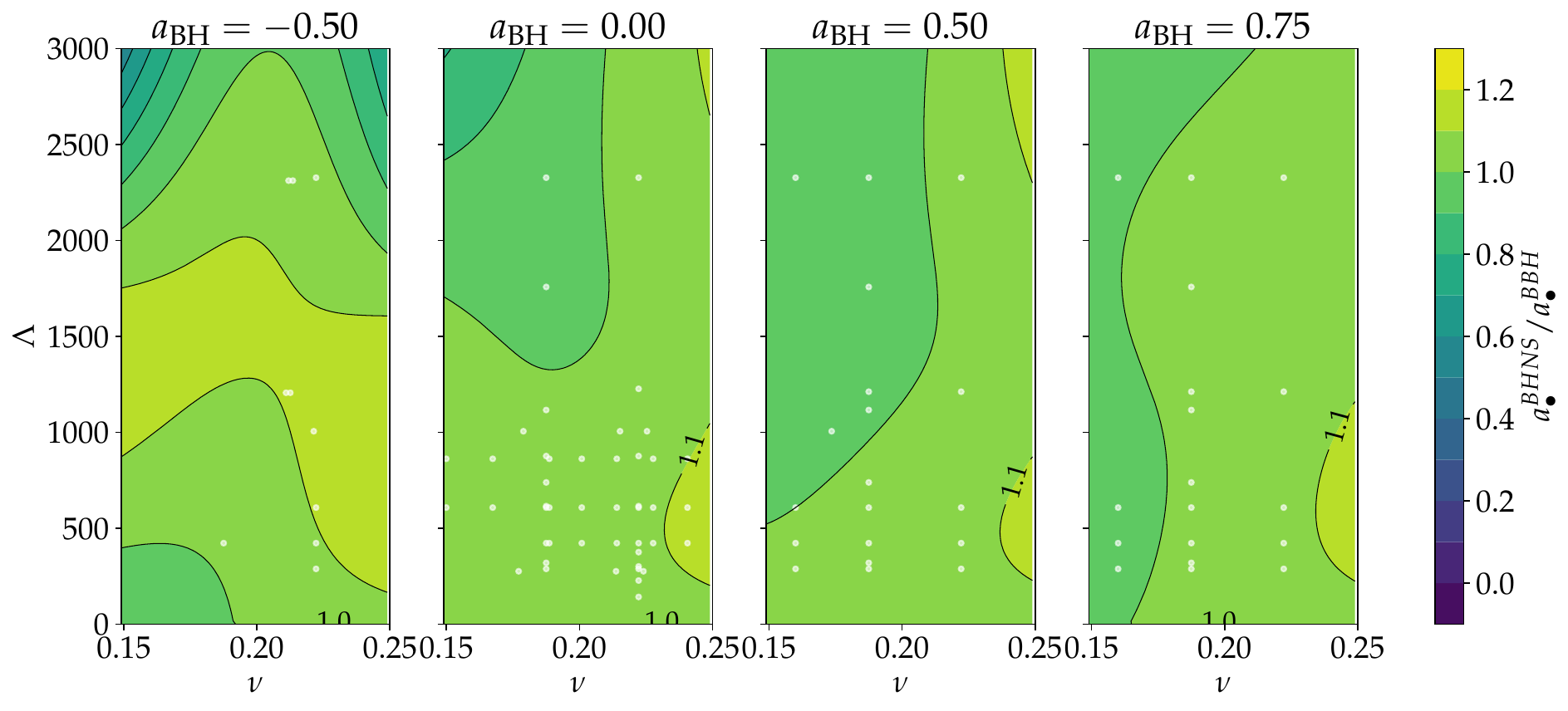}
    \caption{Remnant \gls{BH} mass (top) and spin (bottom) model as a function of the tidal polarizability $\Lambda$ and the symmetric mass ratio $\nu$ for different \gls{BH} initial
      spin values $\abh$. White markers are the NR data extracted from the AH.}
 \label{fig:mfbh}
\end{figure*}

Figure~\ref{fig:mfbh} shows the remnant \gls{BH} mass model together with the data employed for the fitting.
For nonspinning \gls{BH}, \gls{NS}s with $\Lambda\lesssim1000$ directly plunge with no significant tidal disruption (Type II). \gls{NS}s with larger $\Lambda$ also directly plunge as 
far as the mass ratio remains larger then $q\sim 3$. 
Aligned \gls{BH} spins trigger the tidal disruption of the \gls{NS} at a mass ratio as high as $q\sim 5$. This is because the \gls{ISCO} radius is smaller for larger positive $\abh$ and the \gls{NS} is disrupted well before merger (Type II). 
Small deviations from $\Mfbh^{\rm BBH}$, corresponding to Type III where some material from the \gls{NS} is shedded before merger, are highly sensitive to the spin's alignment.
Namely, for a nonspinning $q\sim2$ and $\Lambda\sim1000$ binary the remnant mass would be identical to that of a \gls{BBH}. A spin of $\abh=0.5$ would imply 
a smaller remnant mass than in the \gls{BBH} case for $\Lambda\gtrsim 500$, whereas if it's antialigned $\abh=-0.5$ binaries with $\Lambda$ as high as $\Lambda\sim1200$ would have instead 
a remnant mass as in Type I.
For $\abh\sim+0.75$, only NS with very soft EoS ($\Lambda\lesssim500$) directly plunge into the companion BH.

For the remnant's final spin $\afbh \equiv S_{\bullet}/\Mfbh^2$, we developed a model in a similar fashion as the final mass,
\begin{widetext}
\be
  \frac{a^{\rm BHNS}_\bullet}{a^{\rm BBH}_\bullet} = \frac{1 + \Lambda p^{(3)}_{1}(\nu,\abh)+ \Lambda^2 p^{(3)}_{2}(\nu,\abh) + \Lambda^3 p^{(3)}_{2}(\nu,\abh)}{\left(1 + \Lambda^2 c_{412}\nu\right)^2} 
\ee
\end{widetext}
where the polynomials are defined as 
\begin{subequations}
  \label{eq:mfbh_p}
  \begin{align}
  p^{(3)}_{k}(\nu,\abh)&= p^{(2)}_{k1}(\abh)\nu + p^{(2)}_{k2}(\abh)\nu^2 \nonumber \\
  &\quad + p^{(2)}_{k3}(\abh)\nu^3,\\
  p^{(2)}_{kj}(\abh)&= c_{kj2}\abh^2 + c_{kj1}\abh + c_{kj0}.
  \end{align}
\end{subequations}

The resulting fits for the remnant's spin are shown in Fig.~\ref{fig:mfbh}.
Due to the remaining
material outside of the \gls{AH} influencing the angular momentum of the system, the final spin decreases or increases  depending on the alignment of the
initial spin. When tidal disruption occurs the formed hot disk surrounding the remnant can increase 
the magnitude of the final spin with respect to $\afbh^{\rm BBH}$. These are the regions in the plot where $\afbh^{\rm BHNS}/\afbh^{\rm BBH}>1$, especially for mass ratios close to
equal mass and high $\Lambda$ values. Lower remnant spins are expected above $\Lambda\gtrsim2000$ and $q>2$ for nonspinning cases, whereas higher aligned spins and lower values of 
$\Lambda$ are enough to deviate from \gls{BBH}. The same can be said for $\abh=0.75$ for any $\Lambda$ value and $q>4$.
\gls{BHNS} with antialigned spins will remain close to the \gls{BBH}'s remnant spin for the most part, as long as $\Lambda\lesssim3000$.

\subsection{Gravitational waves}
\label{sec:nr_gw}
  
Waveforms are extracted from the Weyl's scalar $\Psi_4$ curvature modes $\psi_{lm}$ which are then integrated to obtain the strain, $\ddot{h}_{\lm}=\psi_{lm}$,
using the fix-frequency integration method~\cite{Reisswig:2010di}. We consider modes up to $\ell=4$. Convergence and error budget of our waveforms are discussed in detail in Appendix~\ref{app:nr_quality}.
 
Since it is the first time so many (publicly available) multipolar waveforms are extracted from \gls{BHNS} simulations, we study the contribution each 
mode has to the waveform. In particular, earlier works have suggested that the \gls{BBH} waveform amplitude peaks of each mode, $A^{\rm peak}_{\lm} = \max(A_{\lm})$, show a 
structured behaviour of the form~\cite{Bernuzzi:2010xj,Nagar:2022icd}
\be 
\frac{A^{\rm peak}_{\lm}}{\nu |c_{\ell + \epsilon}(\nu)|} \approx e^{c_1(\ell)m + c_2(\ell)\ell}
\ee
where the leading $\nu$ dependence is factorized in the denominator $\nu|c_{\ell + \epsilon}(\nu)|$.
The functions $c_1(\ell), c_2(\ell)$ are quasiuniversal and can be computed in the test-mass limit, see Tab.VI of~\cite{Bernuzzi:2010xj}.
Identifying this kind of pattern 
proves useful when modelling the ringdown part of the waveform.

\begin{figure}
  \centering
  \includegraphics[width=0.5\textwidth]{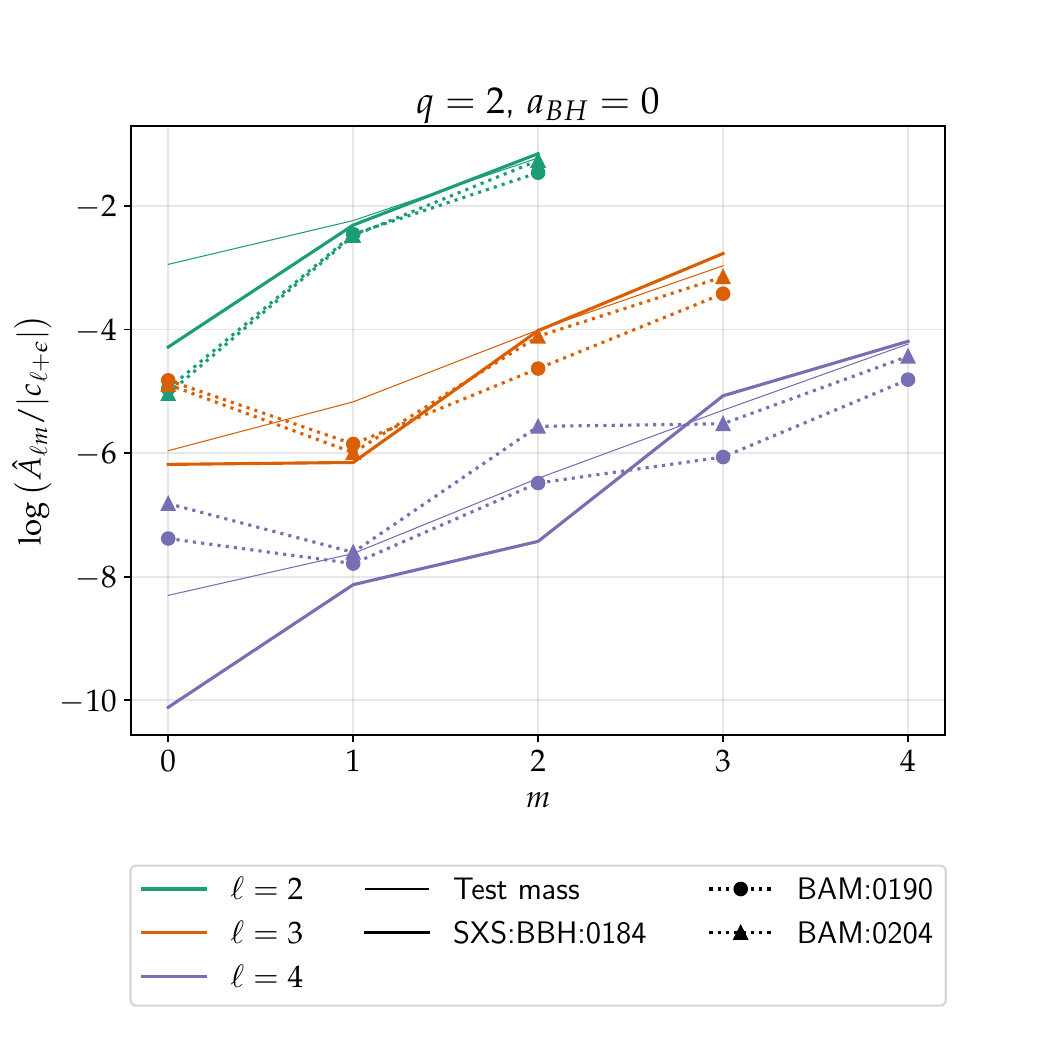}
  \caption{Multipolar amplitude hierarchy comparing $q=2$ nonspinning configurations: \texttt{SXS:BBH:0184}, \texttt{BAM:0190}, \texttt{BAM:0204} and the test mass case.
  Note that the \gls{BHNS} simulations are not evolved from the same initial data (different \gls{EoS}),
  as such their mass ratios are not exactly the same, $q=1.91$ for \texttt{BAM:0190} and $q=1.95$ for \texttt{BAM:0204}. Although small differences, the amplitude is extremely 
  sensitive to these changes, thus showing a low degree of scatter on the plot. These differences do not affect the multipolar pattern described in the main text.}
 \label{fig:Alm_q2a0}
\end{figure}

Figure~\ref{fig:Alm_q2a0} compares the multipolar hierarchy 
structure between a \gls{BBH} and \gls{BHNS} of the same parameters ($q=2$, nonspinning), namely \texttt{BAM:0190} and \texttt{BAM:0204}.
As a reference, we also add the corresponding test mass case as a thin solid line on the plot. For the \gls{BBH} data we employ the simulation \texttt{SXS:BBH:0184} 
from the SXS catalogue~\cite{Blackman:2015pia,Boyle:2019kee,SXS:catalog}. 
The \gls{BHNS} amplitude peaks are smaller than BBH for all the $\ell=m$ modes, 
and decrease according to the stiffness of the \gls{EoS}. As we go to the $m<2$ and $m<3$ cases for $\ell=3$ and $4$ respectively, the amplitude peaks 
tend to surpass those of the \gls{BBH}. Particularly, the ($\ell=3,4$,$m=0$) peaks tend to approach those of ($\ell+1$,0) and correspond to a significant 
contribution to the waveform comparable to that of the (3,2) and (4,4) modes.

  \begin{figure*}
    \centering
    \includegraphics[width=\textwidth]{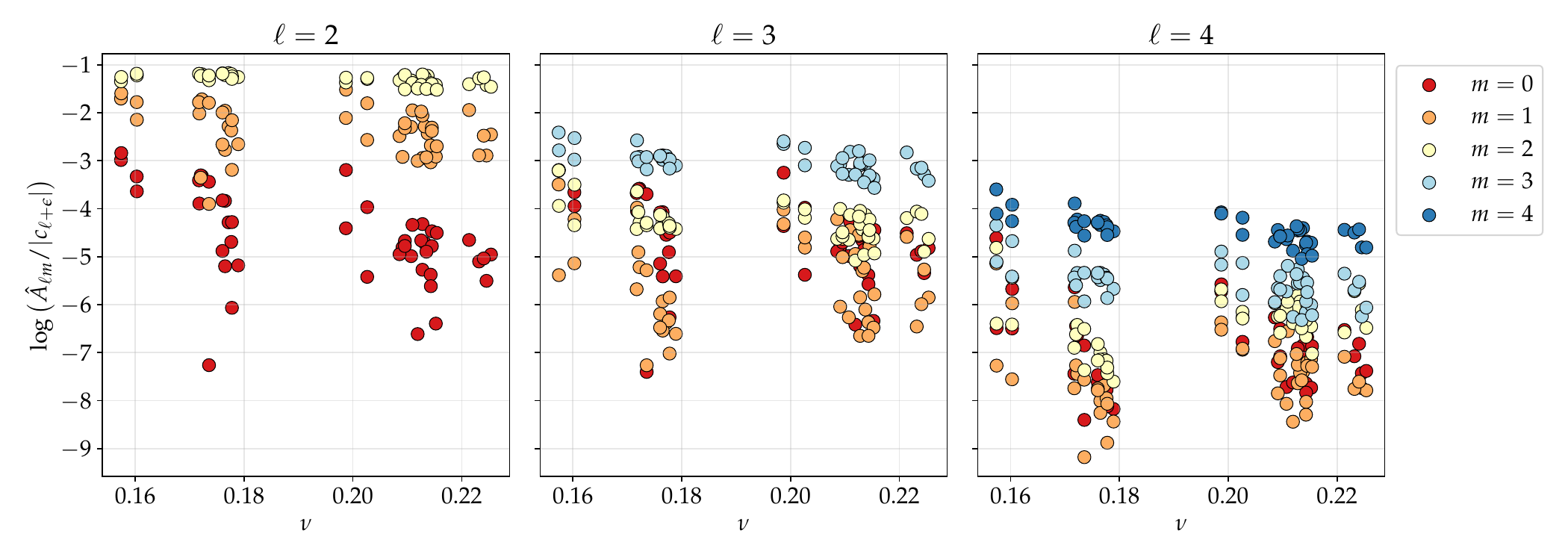}
          \caption{Overview of the multipolar amplitudes of the data produced for this work as a function of $\nu$. Each color represents a different value of $m$ 
        for each $\ell$.}
   \label{fig:Alm_summary}
  \end{figure*}
 
 In order to quantify the multipolar amplitude peak structure's dependence on mass ratio, spin and tidal effects, we 
 employ the following rescaling of the peak amplitude \cite{Nagar:2020pcj}: 
 \begin{subequations}
  \label{eq:amplm}
  \begin{align}
  \hat{A}_{22}&\equiv A^{\rm peak}_{22}/[\nu(1-\hat{S}\omega_{22})],\\
  \hat{A}_{21}&\equiv A^{\rm peak}_{21}/\nu,\\
  \hat{A}_{33}&\equiv A^{\rm peak}_{33}/\nu,\\
  \hat{A}_{32}&\equiv A^{\rm peak}_{32}/[\nu(1-\tilde{a}_0(\omega_{32}/2)^{1/3})],\\
  \hat{A}_{44}&\equiv A^{\rm peak}_{44}/\left[\nu\left( 1 - \frac{1}{2}\hat{S}\omega_{44} \right)\right].
  \end{align}
  \end{subequations}

Figure~\ref{fig:Alm_summary} summarizes the contribution of each mode as a function of the symmetric mass ratio $\nu$. From this plot and Fig.~\ref{fig:Alm_q2a0}
we identify the most significant subdominant modes for a \gls{GW} coming from a \gls{BHNS} merger: (2,1), (3,3), (4,4), and (3,2) (in order of higher contribution). 
In contrast, the modes (4,1) and (4,0) contribute the least to the full waveform (dark blue squares and circles respectively on the figure). Noteworthy is the mode
(3,0) (red circles on middle panel) as mentioned above, with magnitudes comparable to those of the (4,4) and (3,2) modes which decrease towards equal mass cases ($\nu\rightarrow 1/4$).
This is a significant difference from what one would expect in the \gls{BBH} case, where the (3,0) has a much lower contribution. 
In Figure~\ref{fig:Alm_summary} for instance, the (3,0) amplitude is of the same order as the (3,2) mode for $q\gtrsim2$. Similarly, the contribution of the (2,0) approaches that of the
(2,1) for the same mass ratio range.
The fact that the dominance of the (2,0) and (3,0) modes in \gls{BHNS} waveforms is larger for increasing mass ratio, could potentially help us in distinguishing the 
source of binaries where no electromagnetic counterparts are observed. 
Therefore, the modelling of these modes in analytical waveform templates would prove useful for future potential observations. The $m=0$ modes are characterized by the nonlinear 
memory effects arising from general relativity. However, further study on this effect is only possible with additional \gls{NR} simulations with waveforms 
extracted at null infinity employing methods such as Cauchy-characteristic evolution (CCE)~\cite{Bishop:1996gt,Moxon:2020gha,Mitman:2020pbt}.

 \subsection{Kick velocity from \gls{GW}s}
 \label{sec:nr_vkick}
 
 Gravitational waves carry energy, angular and linear momentum away from a system. The kick velocity $v^{\rm GW}_{\rm kick}$ imparted to the final \gls{BH}
 is a response to the loss of the latter. Estimating these velocities started with the work of Fitchett~\cite{Fitchett:1984qn} and has been studied 
 for \gls{BHNS} systems~\cite{Kyutoku:2011vz,Kyutoku:2015gda}. 
 In the following, we discuss the effects that the initial spin and tidal polarizability has on $v^{\rm GW}_{\rm kick}$
 and present a fitting model to estimate kick velocity values for \gls{BHNS} mergers. 
 Results focus solely on the recoil due to radiation of \gls{GW}s and we leave the discussion of the ejecta velocity for future studies.
 Effects due to resolution on the measurement of $v^{\rm GW}_{\rm kick}$ are discussed in Appendix~\ref{app:vkick_res}.
 
 We obtain $v^{\rm GW}_{\rm kick}$ directly from the extracted waveforms to compute the linear momentum fluxes $\dot{P}_x$ and $\dot{P}_y$~\cite{Favata:2004wz,Pollney:2007ss},
 \be
 \dot{P}_i = \frac{R^2}{16\pi}\int d\Omega  \left( \dot{h}^2_+ + \dot{h}^2_{\times} \right)n_i\, 
 \ee
 and then integrate them to obtain the kick velocity vector $v$ (see~\cite{Reisswig:2009rx} for the full expression of the integrand)
 \be
 v \equiv v_x + iv_y = -\frac{1}{M}\int^t_{-\infty} \left( \dot{P}_x + i\dot{P}_y \right)dt\,
 \ee
where $v$ is a complex quantity with modulus $v^{\rm GW}_{\rm kick}=|v|=\sqrt{v^2_x + v^2_y}$ and $n_i=x_i/r$ is the unit radial vector pointing from the source to the
observer. By fixing an initial time $t_0$ for integration
(when the two objects are at a large separation), we are not taking into account the net linear momentum of the binary from $t\rightarrow\infty$ to 
$t=t_0$. As earlier works have shown~\cite{Pollney:2007ss,Bernuzzi:2010ty}, finding an appropiate vectorial integration constant $v_0$ can reduce the 
error of the kick velocity measurement. This fix is of special relevance for our data since we are evolving binaries for just a few orbits.
The procedure is described in detail in Appendix~\ref{app:vkick_err}.

\begin{figure}
  \centering
                \includegraphics[width=0.45\textwidth]{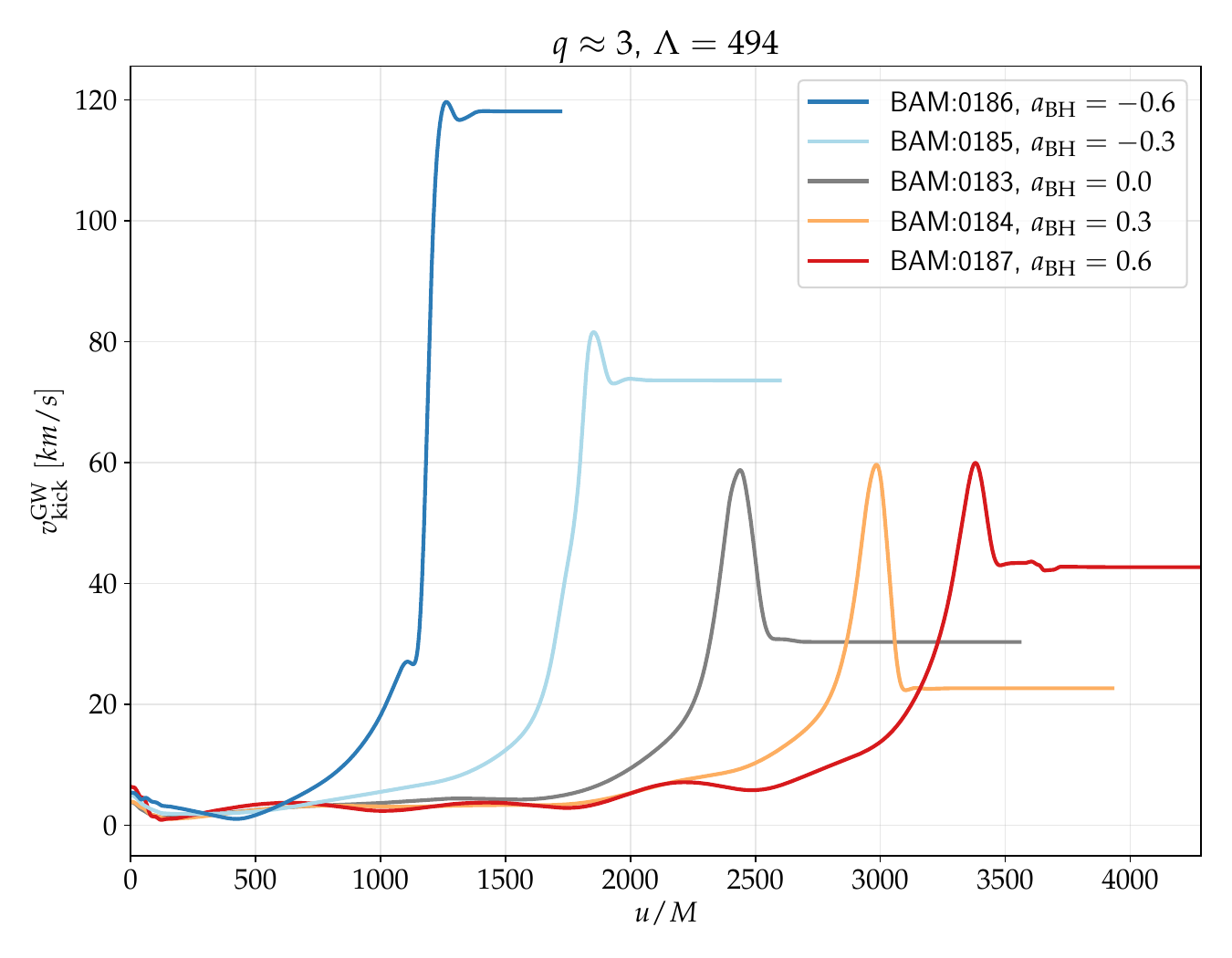}
    \cprotect\caption{Kick velocity obtained for different configurations with $q\approx 3$, $\Lambda=494$, and a variety of spins. The antialigned spins have a higher impact on the resulting
      recoil velocity of the remnant.}
 \label{fig:vk_spineff}
\end{figure}

 Our simulations allow us to illustrate the behaviour of $v^{\rm
   GW}_{\rm kick}$ with respect to the initial spins of the \gls{BH} showing consistency with our estimates for \gls{BBH}.
 Figure~\ref{fig:vk_spineff} shows the \gls{GW} 
 kick velocities for different configurations with $q\approx 3$ and $\Lambda=494$: \verb|BAM:0186|, \verb|BAM:0185|, \verb|BAM:0183|, \verb|BAM:0184|, and \verb|BAM:0187| with 
 spins $\abh=-0.6$, $-0.3$, $0.0$, $0.3$, and $0.6$ respectively. 
 The binaries with initial antialigned spin, and increasing spin magnitude, induce a significantly larger kick on the remnant, reaching velocities of almost 
 $v^{\rm GW}_{\rm kick}\approx120~km/s$. This result goes on par with earlier \gls{BBH} where they measure superkicks for antialigned 
 configurations~\cite{Gonzalez:2007hi,Campanelli:2007ew,Campanelli:2007cga}.
 However, contrary to our estimates for \gls{BBH} where the recoil on the remnant increases significantly with the initial spin, 
 for a \gls{BHNS} with this mass ratio and tidal polarizability the kick velocity stays below $v^{\rm GW}_{\rm kick}\approx60~km/s$ for both nonspinning and aligned spins 
 up to $\abh=0.6$. This suggests that tides can have a "suppressing" effect on the remnant's \gls{GW} kick.

For future estimates of the remnant's kick velocities, we develop a fitting model based on the one developed in Varma \etal~for \gls{BBH} mergers. 
We again factorize the BBH kick, $v^{\rm BBH}_{kick}$, and represent the data as
\be
\frac{v^{\rm BHNS}_{kick}}{v^{\rm BBH}_{kick}} = \frac{\Lambda p^{(2)}_{1}(\abh,\nu) }{ ( 1 + \Lambda p^{(1)}_2(\abh)\nu^2 )^2   },\,
\ee
where we define the polynomials as
\begin{subequations}
  \begin{align}
  p^{(2)}_{1}(\nu,\abh)&= p^{(3)}_{11}(\abh)\nu + p^{(3)}_{12}(\abh)\nu^2 ,\\
  p^{(3)}_{1j}(\abh)&= c_{1j3}\abh^3 + c_{1j2}\abh^2 + c_{1j1}\abh + c_{k0}, \\
  p^{(1)}_{2}(\abh)&= c_{221}\abh + c_{220},
  \end{align}
\end{subequations}

\begin{figure*}
  \centering 
    \includegraphics[width=0.8\textwidth]{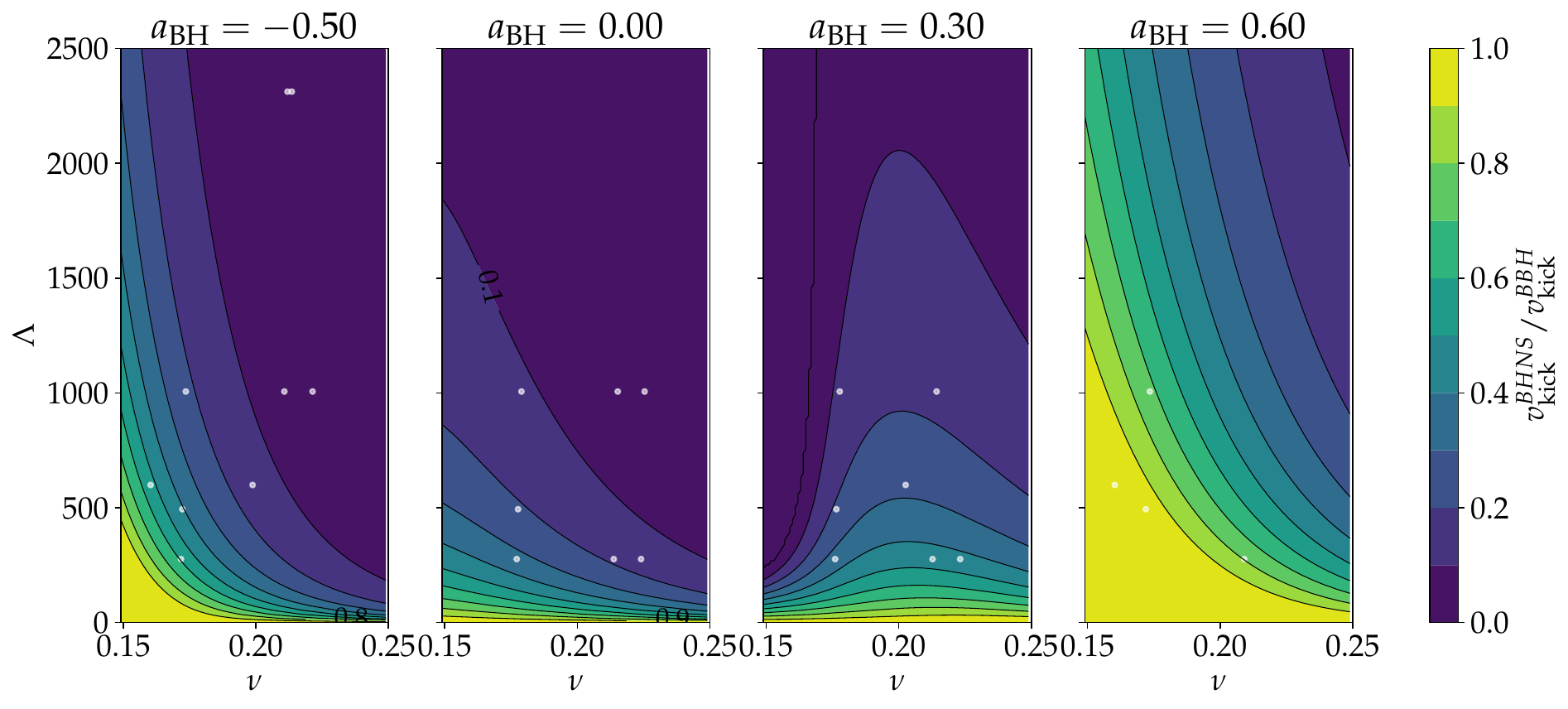}
    \caption{Kick velocity model as a function of symmetric mass ratio $\nu$ and tidal polarizability $\Lambda$ for different initial spins $\abh$.}
 \label{fig:vk_fits}
\end{figure*}

The coefficients $c_{kji}$ are listed in Table~\ref{tab:vkick_coef} in Appendix~\ref{app:vkick}. 
Figure~\ref{fig:vk_fits} summarizes the behaviour of the kick velocity values on a ($\Lambda$,$\nu$) parameter space for different initial spins,
with the white dots representing the \gls{NR} data employed for fitting. For nonspinning and low spin cases, the remnant's kick velocity from \gls{BHNS} significantly differs from 
the estimated for \gls{BBH}, dramatically decreasing in value with increasing values of $\Lambda$ regardless of the mass ratio. 
However, with higher aligned spins the $v^{\rm BHNS}_{\rm kick}$
starts approaching that of \gls{BBH} for $q\gtrsim 2$ and
$\Lambda\lesssim1000$.

For $q\lesssim2$ and $\Lambda\gtrsim500$
the \gls{BHNS} kick velocity is remarkably lower than for the nontidal case. We could then argue on the impact of the ejecta in these scenarios, it has been shown 
in earlier works that tidal disruption suppresses the kick coming from \gls{GW}s and consequently the backreaction of the mass ejection will result in a higher 
magnitude for the kick velocity due to the ejecta's momentum~\cite{Kyutoku:2013wxa,Kyutoku:2015gda}. The \gls{GW} recoil will thus be more suppressed for binaries presenting earlier 
tidal disruption than those when it occurs later closer to merger.
This correlates with the fact that highly spinning \gls{BH}s and more compact \gls{NS}s experience tidal disruption early in the evolution, thus inducing a higher ejecta velocity.
Nonetheless, since we compare directly with the \gls{BBH} case, we only take into account the recoil due to \gls{GW}s.

\subsection{Simulation compatible with GW230529}
\label{sec:nr_gw230529}

During the completion of the present work, a \gls{BHNS} \gls{GW} event, labeled GW230529, was detected in the O4a observation run~\cite{LIGOScientific:2024elc}. Among the simulations produced for this campaign, the 12 orbit evolution, namely \verb|BAM:0223|, has a chirp mass of $\mathcal{M}_c=1.94$, a total mass of $M=5~\Msun$,
$q=2.47$, $\chi_1=\abh=0.74$ and precessing spin of $\chi_p=0.6$. 
These parameters coincide exactly within the best values to 90$\%$ CI
found in the LVK analysis. We compute the mismatch between the \gls{NR} waveform and the best waveform obtained in~\cite{LIGOScientific:2024elc}
with the \texttt{SEOBNRv5PHM} model~\cite{Mihaylov:2023bkc,Ramos-Buades:2023ehm} (see Sec.~\ref{sec:mismatch} for a description of how we obtain this quantity).
The mismatch is calculated from $f_{\rm low}=276$~Hz, corresponding to the initial frequency of the simulation, up to $f_{\rm high}=2048$~Hz; and employing the noise curve
of \texttt{aLIGOZeroDetHighPower}~\cite{aLIGODesign_PSD}. 
The resulting mismatches lie around $\sim0.3$ including subdominant modes for different inclinations. In the same way, we obtain similar mismatches of $\sim0.2$ with \teobdali~and no tides
(See~\cite{Albanesi:2025txj} for a description of the model). Given the symmetric nature of this event, we obtain again the mismatch against our \gls{BHNS} model within 
\teobdali~(see Sec.~\ref{sec:eob}). These results are presented in Sec.~\ref{sec:mismatch}, which include tides ($\Lambda\neq 0$). The mismatches stay around the order of $\sim 0.01$.
These results highlight the need for accurate modelling of the high frequency regime for upcoming next-generation detectors, which will be able to uncover the physics at higher frequencies. 

Finally, we obtain the kick velocity directly from our numerical simulation data and obtain a recoil of $v^{\rm GW}_{\rm kick}=211$~km/s.
This is a much higher value than our results from our spin-aligned configurations, as one expects precessing systems to produce large kicks~\cite{Kesden:2010ji,Borchers:2021vyw,Borchers:2024tdi}.

\section{\teob~model}
\label{sec:eob}

In this section we describe a new EOB model for BHNS.
Our earlier work \cite{Gonzalez:2022prs} was based on \teob-\texttt{GIOTTO}, a version of \teob~for spin-aligned quasicircular compact binary 
coalescences~\cite{Damour:2014sva,Nagar:2015xqa,Nagar:2018zoe,Nagar:2019wds,Nagar:2020pcj,Riemenschneider:2021ppj,Nagar:2023zxh}.
The present model has been implemented within the framework of \teobdali~\cite{Chiaramello:2020ehz,Nagar:2021gss,Nagar:2021xnh,Nagar:2023zxh,Albanesi:2025txj}, an \gls{EOB} model 
for generic compact binaries and arbitrary orbits, including eccentricity, precession and scattering.
Similarly to our previous work, we focus on developing a ringdown 
model for the merger and postmerger part of the \gls{BHNS} waveform. This is achieved by employing the same strategy: Extract information from the new \gls{NR} 
data presented in Sec.~\ref{sec:nr_methods} together with the old simulations (used in~\cite{Gonzalez:2022prs}), and represent the deviation of relevant quantities to the \gls{BBH} 
case. From our discussion in Sec.~\ref{sec:nr_gw}, we model explicitely the modes (2,1), (2,2), (3,2), (3,3) and (4,4), which contribute the most to the overall waveform.
At the time of this work's development, \teob~did not provide a description for the $m=0$ modes making it impossible to build a \gls{BHNS} extension. However, the implementation of these modes
into \teobdali~is currently ongoing~\cite{Grilli:2024lfh,Albanesi:2024fts}.
We will leave its extension for \gls{BHNS} for future improvement of the model. 

\subsection{Ringdown model}
\label{sec:eob_ringdown}

Following the usual recipe to model the ringdown based on the procedure from~\cite{Damour:2014yha} and used in \teobdali, 
we employ the \gls{QNM} rescaled waveform seen in Eq. (4) of~\cite{Gonzalez:2022prs}. Furthermore, for the modes (2,1), (3,3) and (4,4), we use 
instead the strategy implemented in the \texttt{SEOBNR} waveform family~\cite{Cotesta:2018fcv,Pompili:2023tna}, where the coefficients $c^{A,\phi}_i$ are constrained in the following manner

\begin{subequations}
  \label{eq:const_pompili}
  \begin{align}
  c^A_1 &= (\dot{A}_{t^{\rm match}_{\lm}} + \alpha_1 A_{t^{\rm match}_{\lm}})\cosh^2 c^A_3/c^A_2,\\
  c^A_4 &= A_{t^{\rm match}_{\lm}} - (\dot{A}_{t^{\rm match}_{\lm}} + \alpha_1 A_{t^{\rm match}_{\lm}})\cosh c^A_3 \sinh c^A_3/c^A_2,\\
  c^{\phi}_1 &= (\omega_1 - \omega_{t^{\rm match}_{\lm}})\frac{1 + c^{\phi}_3}{c^{\phi}_2c^{\phi}_3},\\
  c^{\phi}_4 &= 0,
  \end{align}
\end{subequations}
and using ($c^A_2$,$c^A_3$,$c^{\phi}_2$,$c^{\phi}_3$) as free coefficients. Note that for these modes, the matching time corresponds to the time of the amplitude peak of
the (2,2) mode, $t^{\rm match}_{\lm}\equiv t^{\rm peak}_{22}$. Additionally, we account for the time-delay, between the time of merger $t_{\rm mrg}$ and the time where the 
amplitude of each peak occurs, by defining it as~\cite{Nagar:2020pcj}
\be 
\Delta t_{\lm} \equiv t^{\rm peak}_{\rm \lm} - t_{\rm mrg} .
\ee

For all the modes, we fit the quantities ($\alpha_{\lm 1}$, $\omega_{\lm 1}$, $A^{\rm peak}_{\lm}$, $\omega^{\rm peak}_{\lm}$, $\Delta t_{\lm}$) as a function of 
($\nu$,$\Lambda$,$\abh$) as described in Appendix~\ref{app:fit_models}. The fits for the peak amplitudes employ the same rescaling as in
Eq.~\ref{eq:amplm}.

\begin{figure*}
  \centering 
   \includegraphics[width=0.45\textwidth]{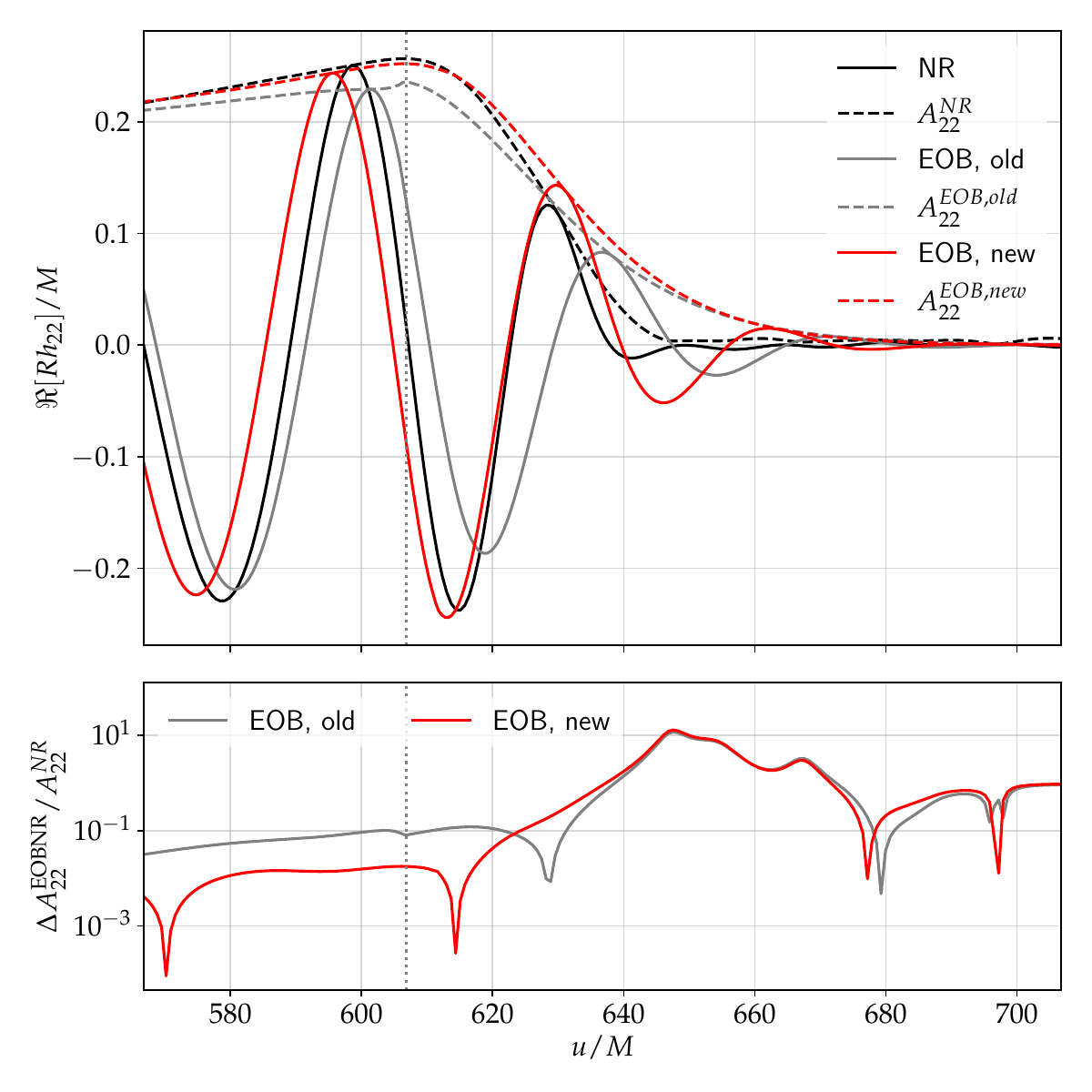}
        \includegraphics[width=0.45\textwidth]{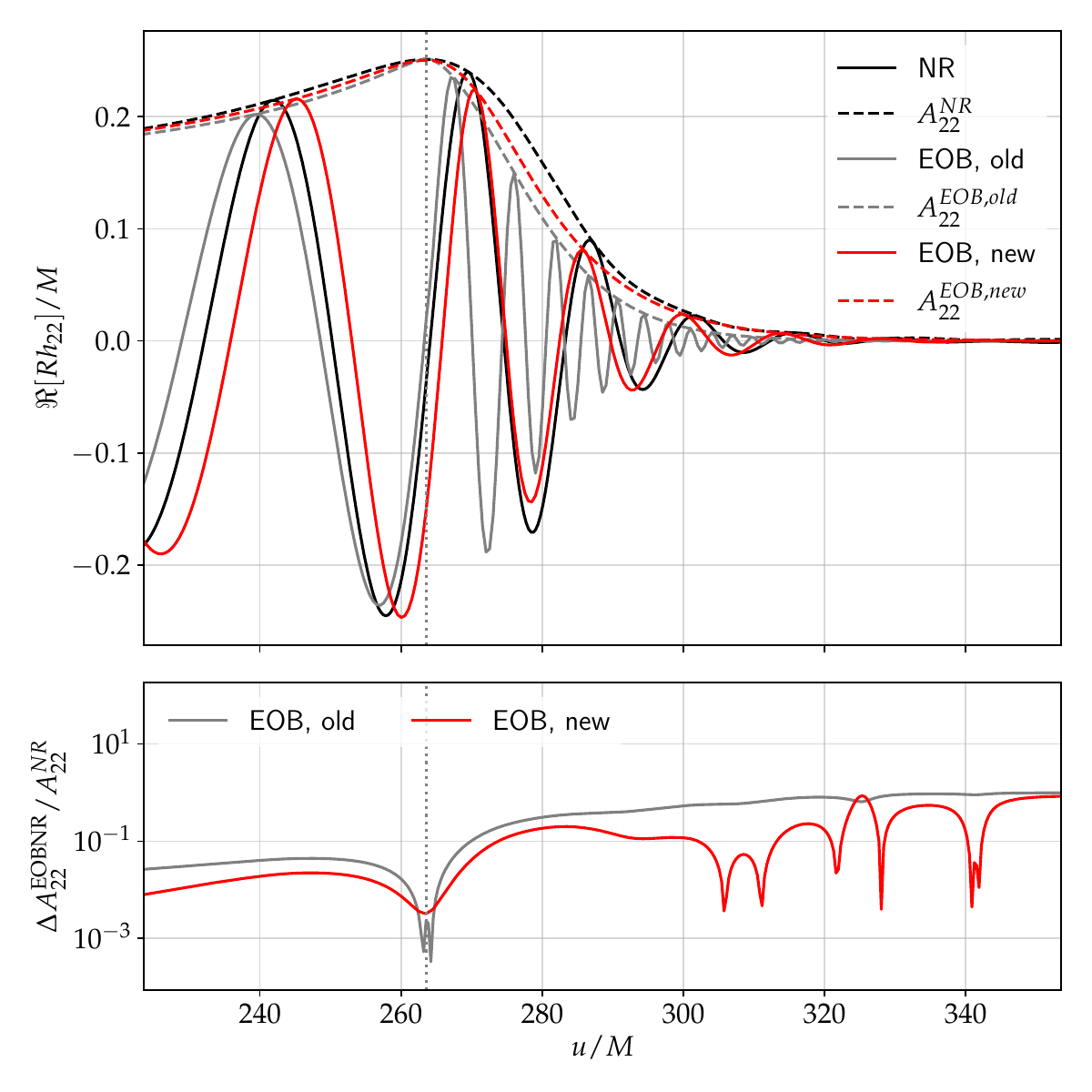}
                \cprotect\caption{Ringdown waveform of the (2,2) mode for different \gls{BHNS} binary configurations from the fitting set: \verb|BAM:0190| (left) and \verb|BAM:0200| (right). The solid lines show the real part of the strain while the dashed lines indicate their respective amplitudes.
    Black lines represent the \gls{NR} waveform, the grey ones come from the old model of~\cite{Gonzalez:2022prs}, and the red lines are produced with the new model presented in this paper. The smaller bottom panels show the 
    amplitude differences with respect to the \gls{NR} simulation for both models. As reference, we add a dotted verticle line to indicate the moment of merger.}
    \label{fig:ringdown_old_new}
\end{figure*}

The new ringdown waveform is shown in Fig.~\ref{fig:ringdown_old_new}, where we compare the (2,2) mode with  the new \gls{NR} simulations and our 
previous model from~\cite{Gonzalez:2022prs}. We show fiducial simulations \verb|BAM:0190| (left) and \verb|BAM:0200| (right).
Our new model shows amplitude differences at merger well below $10\%$ and a phase that qualitatively agrees to that of the \gls{NR} waveform for both configurations. 
The ringdown is accurately modelled with the new fits employed to classify the different \gls{BHNS} merger types within the code (see next subsection).
This is particularly the case for \verb|BAM:0200|, where the fits adequately deform the \gls{BBH} waveform to model the suppressed \gls{BHNS} ringdown. 
In contrast, the old model (shown in grey) misclassifies this configuration and produces a ringdown with very excited \gls{QNM} as opposed to the damped signal from \gls{NR}. 
Furthermore, for \verb|BAM:0190| one can notice a small attachment artifact at merger which is no longer present for this case in \teobdali.

\begin{figure*}
  \centering 
    \includegraphics[width=0.45\textwidth]{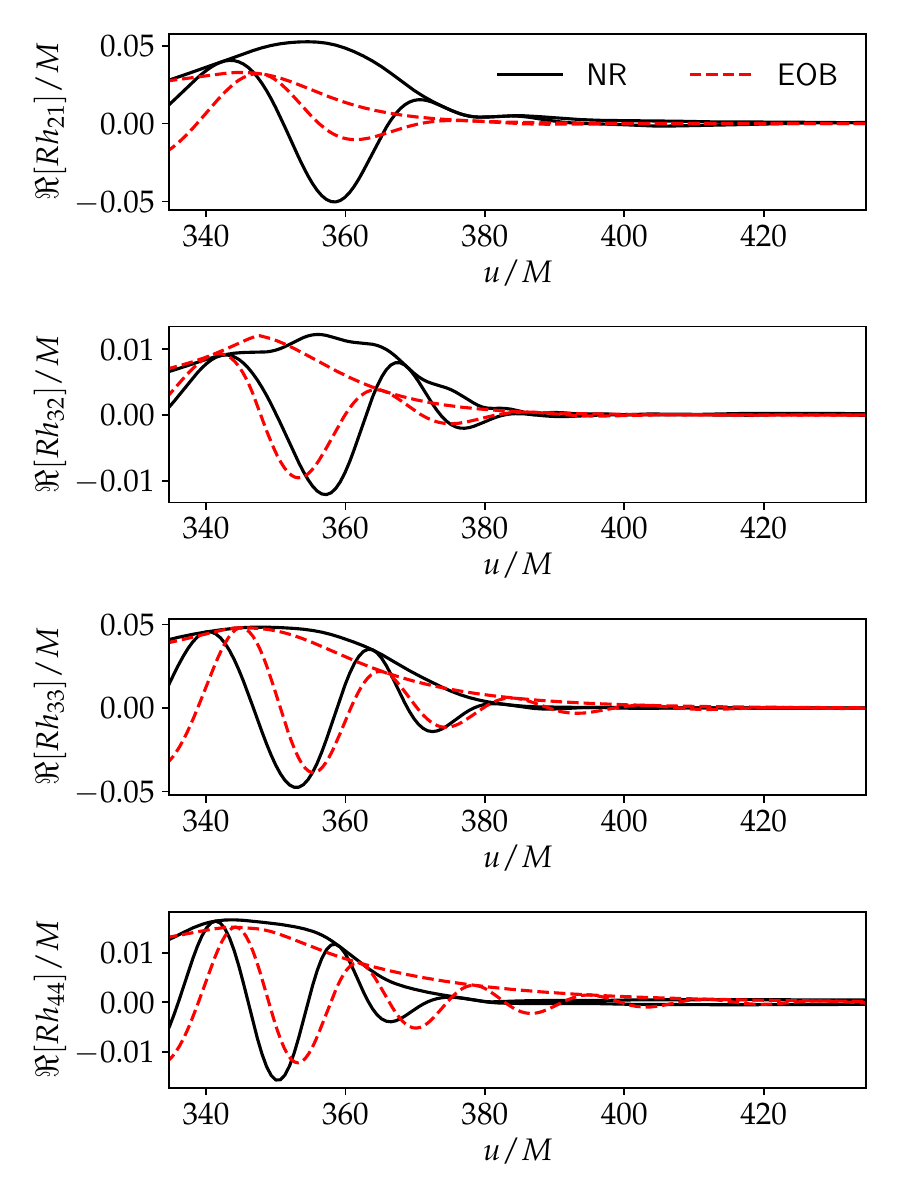}
    \includegraphics[width=0.45\textwidth]{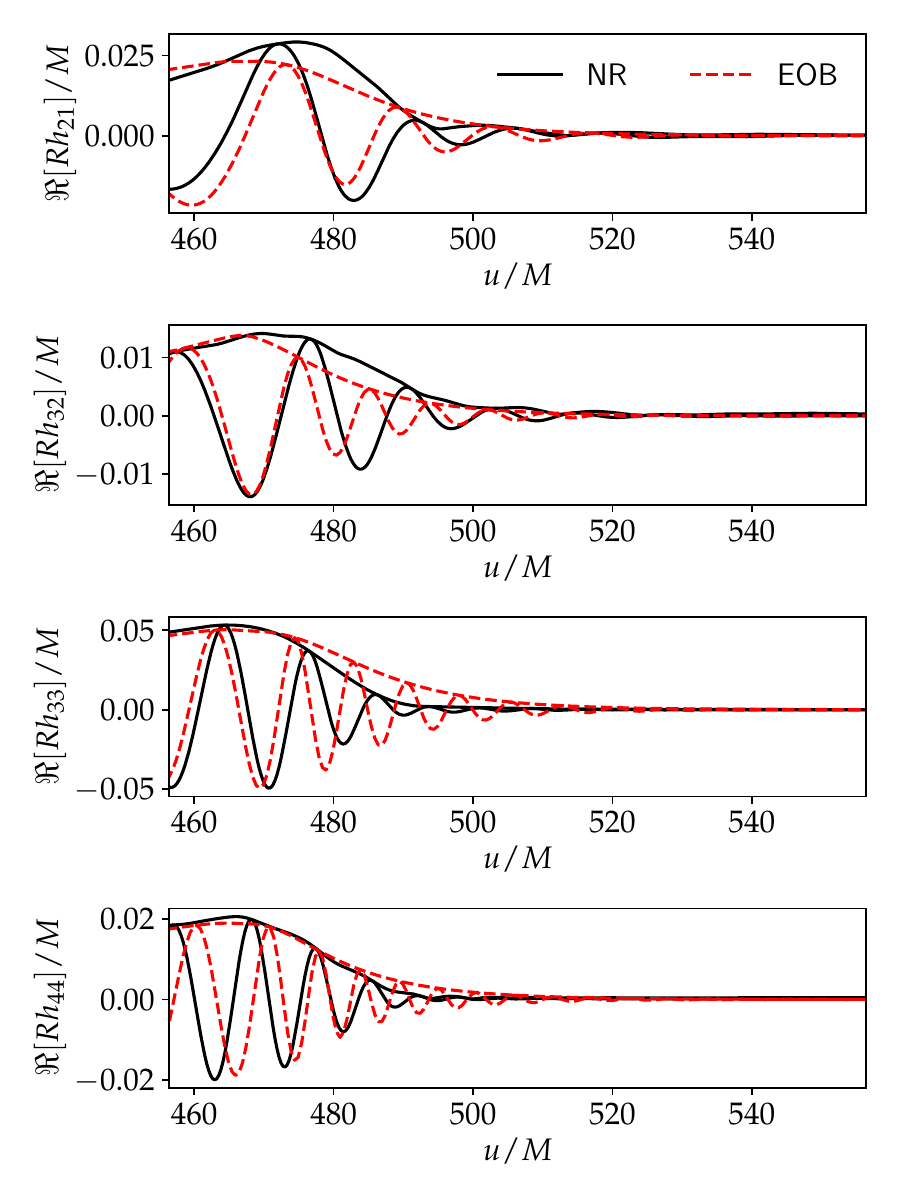}
    \caption{Similar to Fig.~\ref{fig:ringdown_old_new} but showing the correponding subdominant modes' ringdown waveforms and amplitudes for \gls{NR} (black solid lines) 
    and the new model (red dashed lines).}
 \label{fig:ringdown_lm}
\end{figure*}

We show the ringdown waveform for the subdominant modes in Fig.~\ref{fig:ringdown_lm} for two different binary configurations. 
We find qualitatively good agreement between the \gls{NR} and
\gls{EOB} amplitudes towards merger and a ringdown signal approaching
the numerical one.

\subsection{Inspiral-Merger-Ringdown Waveform}
\label{sec:eob_imr}

The main new elements entering \teobdali{} inspiral-merger-ringdown for BHNS are:
(i) the remnant model presented in Sec.~\ref{sec:nr_bh}, 
(ii) an updated \gls{NQC} to the (2,2)  waveform and new, specific NQC for the other multipoles, and (iii) the ringdown template described in Sec.~\ref{sec:eob_ringdown} and (iv) an update classification of the binary in Type I, II and III.
Tidal effects are incorporated in the same fashion as for BNS, namely using 2PN and GSF3 models for gravitomagnetic and gravitoelectric tidal effects respectively~\cite{Bernuzzi:2012ci,Bernuzzi:2014owa,Akcay:2018yyh}.
Spin, spin-precession and eccentricity effects are incorporated in the same way as for BBH (and BNS). We discuss in the following items (ii) and (iv).

NQC are fixed by extracting from NR multipolar waveform the quantities 
($A^{\rm NQC}_{\lm}$, $\dot{A}^{\rm NQC}_{\lm}$, $\omega^{\rm NQC}_{\lm}$, $\dot{\omega}^{\rm NQC}_{\lm}$) at times 
\be
t^{\rm NQC}_{\lm} \equiv t^{\rm peak}_{\lm} + 2,
\ee
with exception of the (2,1), (3,3) and (4,4) modes that are extracted instead at $t^{\rm peak}_{\lm}\equiv t^{\rm peak}_{22}$. These quantities are fitted as described in Appendix~\ref{app:fit_models}, i.e. by factorizing the \gls{BBH} values and representing them as a functions of $\{\nu$,$\Lambda$,$\abh\}$. 
In \teob~the time-shift between $t^{\rm NQC}_{22}$ (time at which the \gls{NQC} parameters are computed on the \gls{EOB} time axis)
and the peak of the orbital frequency $t^{\rm peak}_{\Omega_{\rm orb}}$ is defined as~\cite{Nagar:2018zoe}
\be
t^{\rm EOB}_{\rm NQC} = t^{\rm peak}_{\Omega_{\rm orb}} -  \Delta t_{\rm NQC}
\ee
where $\Delta t_{\rm NQC} = 1$ inspired by test-particle results. We find however, that setting this quantity to $\Delta t_{\rm NQC} = 4$ for \gls{BHNS} yields better
results. We note that this choice is purely technical and has no physical meaning in the dynamics.
 
\begin{figure*}
  \centering 
    \includegraphics[width=0.8\textwidth]{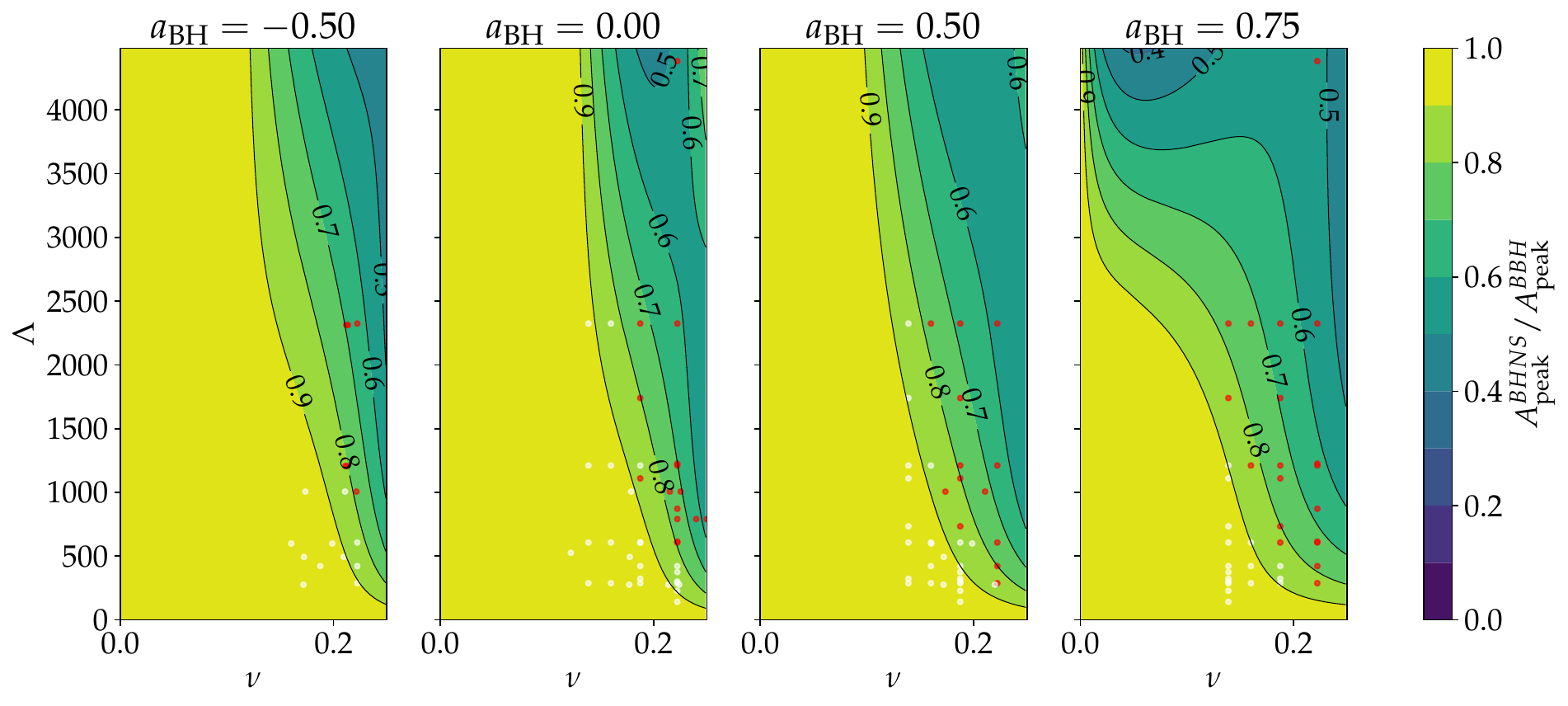}
    \caption{Fits for the peak of the (2,2) amplitude $A^{\rm BHNS}/A^{\rm BBH}$ as a function of ($\nu$,$\Lambda$,$\abh$). The dots represent the \gls{NR} data used
    to inform the fits, the red ones indicate the presence of tidal disruption (see text).}
 \label{fig:apeak22}
\end{figure*}

The classification in Type I, II and III BHNS is used in \teob{} to select the ringdown waveform and obtain the best possible description with minimal NR-information.
Contrary to our earlier work where we employ the \gls{QNM} inverse damping time $\alpha_{221}$ fit to approximately identify a binary among the three types of mergers, 
we now make use of $A^{\rm peak}_{22}$. This fit is shown in Fig.~\ref{fig:apeak22}, red dots indicate the binaries going 
through tidal disruption, Type I. We therefore consider a contour around $A^{\rm BHNS}/A^{\rm BBH}<0.85$ to identify a binary as Type I and use all fits
developed in this work. All points falling above the contour $A^{\rm BHNS}/A^{\rm BBH}>0.99$ (Type II) go through the \gls{BBH} pipeline of 
\teob~but employ the remnant model developed for \gls{BHNS}. For the cases falling in the middle of these limits, Type III, we use all fits in this work except for
$\alpha_{\lm 1}$ and $\omega_{\lm 1}$ as we saw that the ones designed for \gls{BBH} do a better job at simulating the ringdown produced in \gls{NR} data.

\section{Waveform model validation}
\label{sec:validation}

To validate our new model, we use \verb|BAM:0223|: a 12 orbit inspiral-merger-ringdown waveform with a precessing spin \gls{BH}.
The binary has a mass ratio of $q=2.3$, a spin parameter of $\chi_p=0.61$ (indicating a highly precessing
spin) and $\Lambda=494$. Waveform convergence and error budget for this specific waveform is discussed in Appendix~\ref{app:errorbud_prec}. 
We discuss below time-domain phasing and faithfulness with \teobdali{}.

\subsection{Phasing}

We follow the same phasing procedure as performed in our previous work~\cite{Gonzalez:2022prs}: minimize the functional $\Xi^2 (\delta t, \delta \phi)$ of the 
\gls{EOB} and \gls{NR} waveform phases,
\be 
\Xi^2 (\delta t, \delta \phi)=\int ^{t_f}_{t_i}[\phi_{\rm{NR}}(t)-\phi_{\rm{EOB}}(t+\delta t)+\delta\phi]^2dt,
\ee
in an alignment window [$t_i$,$t_f$] and extracting the optimal values for the time $\delta t$ and phase $\delta \phi$ shift. These are then used to shift the \gls{EOB} waveform which is
then compared to the waveform from \gls{NR}.

\begin{figure*}
  \centering 
    \includegraphics[width=0.9\textwidth]{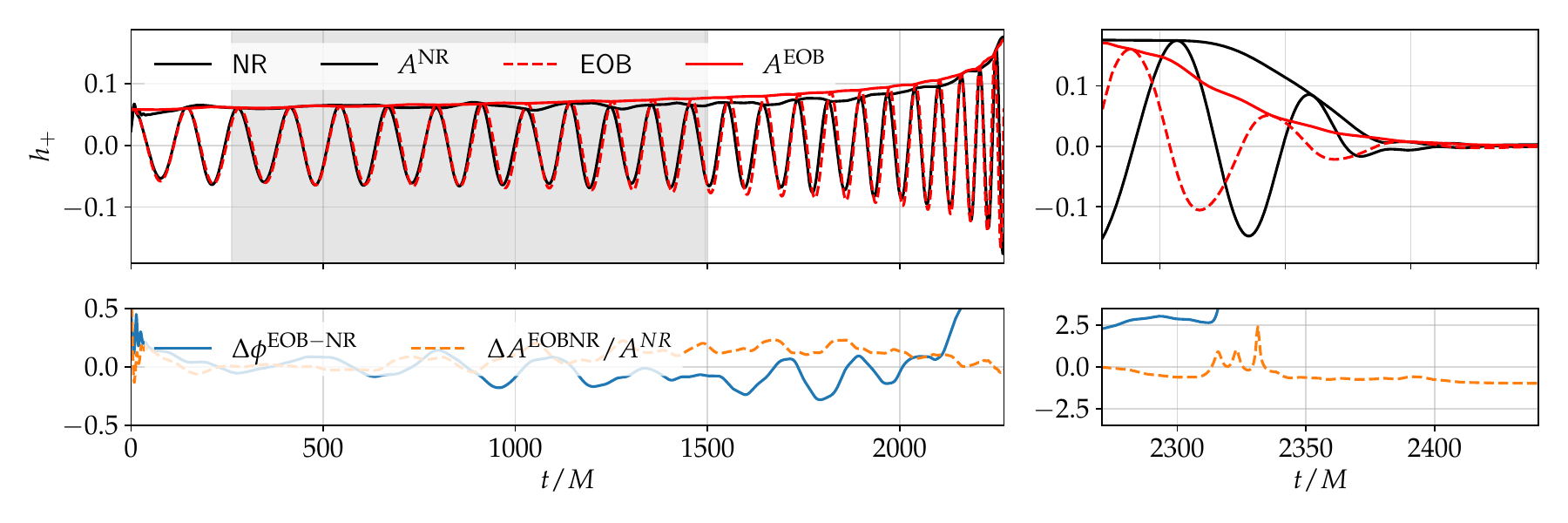}
    \cprotect\caption{Alignment against \gls{EOB} of the real part of the waveform $h_{+}$ of \verb|BAM:0223| computed with
    the modelled modes. 
    Amplitude (orange dashed line) and phase differences (blue solid line) are shown in the bottom panels. The grey area indicates the 
    alignment window.}
    \label{fig:wvf2_prec22}
\end{figure*}

The phasing using this alignement is shown in Fig.~\ref{fig:wvf2_prec22}. One can see the consistency between the \gls{EOB} and the \gls{NR} waveforms 
especially during the inspiral with minimal phase and amplitude deviations towards merger and ringdown. 
Despite this, the model accurately deforms the \gls{BBH} ringdown to accomodate the morphology resulting from the numerical
simulation. The phase difference stays below 0.5 rad throughout the inspiral and increases to $\sim 2.3$ rad at merger. 

\subsection{Mismatch}
\label{sec:mismatch}

As a check for validity of the model, we employ the unfaithfulness (or mismatch) defined as
\be 
\bar{\mathcal{F}}\equiv 1-\mathcal{F}=1-\mathop{\rm{max}}_{t_0,\phi_0}\frac{\langle h^{\rm{EOB}},h^{\rm{NR}}\rangle}{\|h^{\rm{EOB}}\| \|h^{\rm{NR}}\|},
\ee
where $t_0$ and $\phi_0$ are the initial time and phase, and $\|h\|\equiv \sqrt{\langle h,h\rangle}$. We define the inner product with the 
the \gls{PSD} of the detector $S_n(f)$ and the Fourier transformed waveform $\tilde{h}(f)$
\be 
\langle h_1,h_2\rangle \equiv 4\Re{ \int \frac{\tilde{h}_1(f)\tilde{h}^*_2(f)}{S_n(f)} }df.
\ee
Since the binary experiences precession, the unfaithfullness will also depend on the extrinsic parameters of the binary. 
Hence, we report the sky-maximized faithfulness including all modes at different inclinations, and optimizing  
over the coalescence angle $\phi$ and the rotations on the in-plane component of the spin, as described in~\cite{Gamba:2021ydi,Gamba:2024cvy}.
The mismatch is obtained from the initial frequency of the simulation $f_{\rm low}=276$~Hz up to $f_{\rm{high}}=2048$~Hz,
and employing the \gls{PSD} from both the \gls{ET}~\cite{Hild:2011np} and the zero-detuned high-power Advanced LIGO~\cite{aLIGODesign_PSD} noise curves. 

We present the resulting mismatches in Table~\ref{tab:match_HM} for each noise curve employed.
The lowest mismatches are obtained with $\iota=\pi/8$, with values as low as $\sim 0.01$. For $\iota=0$ and  $\iota=\pi/4$ the mismatch stays well below $\sim 0.1$, 
whereas for $\iota=\pi/2$ it reaches $\sim 0.1$ for both detectors. These values showcase the performance of the model for precessing binaries with 
spins as high as $\chi_p=0.6$ especially at low inclinations.

\begin{table}[h!]
  \centering    
  \cprotect\caption{Unfaithfulness of \TEOB{} including HMs with NR waveform from \verb|BAM:0223| for different inclinations $\iota$ and noise curves. }
    \scalebox{0.75}{
  \begin{tabular}{c|c|c|c|c}  
    \hline\hline
    \gls{PSD} & $\iota=0$ & $\iota=\pi/8$ & $\iota=\pi/4$ & $\iota=\pi/2$\\
    \hline
    \texttt{EinsteinTelescopeP1600143} & 4.7$\times$10$^{-2}$ & 1.3$\times$10$^{-2}$ & 3.4$\times$10$^{-2}$ & 9.9$\times$10$^{-2}$ \\
    \texttt{aLIGOZeroDetHighPower} & 4.9$\times$10$^{-2}$ & 1.4$\times$10$^{-2}$ & 3.8$\times$10$^{-2}$ & 1.1$\times$10$^{-1}$ \\
    \hline\hline
  \end{tabular}
  }
 \label{tab:match_HM}
\end{table}

\begin{figure}
  \centering 
    \includegraphics[width=0.45\textwidth]{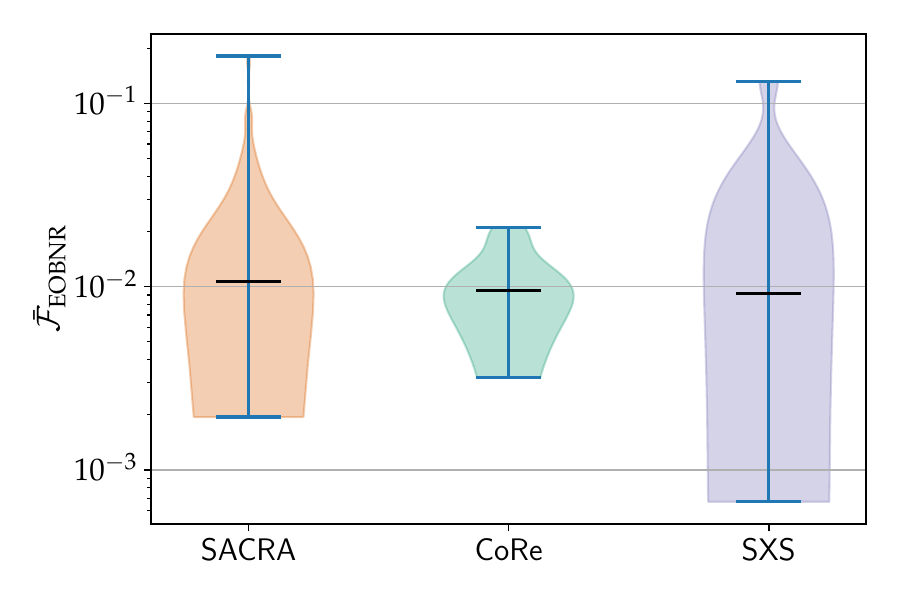}
    \caption{Mismatches for all NR simulations on the three available datasets: SACRA (orange), CoRe (green), and SXS (purple). The black line indicates the median for each case.}
    \label{fig:all_mm}
\end{figure}

Finally, we comment on the mismatches previously computed in~\cite{Albanesi:2025txj} for 184 \gls{BHNS} simulations across the available datasets including the CoRe data produced
in this work. These were obtained in a frequency range of $f\in[10,4096]$~Hz for the (2,2) mode. The results are shown in Fig.~\ref{fig:all_mm}. We mark the median values
for each dataset which lie around $\sim 1\%$ for all of them. The lowest mismatch corresponds to \texttt{SXS:BHNS:0001} with $\bar{\mathcal{F}}=0.07\%$, whereas the highest reaches 
$\bar{\mathcal{F}}=18\%$ for \texttt{2H-Q2M12a75} from the SACRA catalog. The latter is a configuration with a highly spinning \gls{BH} $\abh=0.75$ and $\Lambda=4392$.

\section{BHNS parameter space}
\label{sec:predict}

In this section, we discuss how \teobdali{} can guide the development of future \gls{NR} simulations for GW modeling.
We also demonstrate, for the first time, the capability of our model to generate eccentric and precessing waveforms for \gls{BHNS}.

\subsection{Where to further simulate?}

Despite having simulated 52 binaries, our work clearly highlights the necessity of more NR data in order to be able to develop faiuthful waveforms.
Just focusing on the non-precessing and quasicircular merger, a main issue is to identify the binaries that maximize the information required by EOB.
We are here able to address this issue by leveraging on the predictions of \teobdali{} and using a greedy algorithm~\cite{Field:2011mf,Canizares:2013ywa}.
The latter selects points in the parameter space by identifying a waveform basis among those that are ``most different'' according to a mismatch-based metric.
Points in the parameter space are sequentially added to the basis until either a desired number of points or a certain accuracy threshold is reached.

We run the greedy algorithm on a reduced parameter space described by mass ratio, effective
spin and tidal polarizability parameter $\{q,\chi_{\rm eff}, \Lambda
\}$, which adequately capture the relevant binary interactions and
effects impacting the waveform morphology. The parameter space sampled ranges from
$q\in[1,5]$ for masses $M_{\rm NS}\in[1,2]\Msun$; spins $\chi_1\in[-0.8,0.8]$, $\chi_2\in[-0.2,0.2]$,
and 10 different \gls{EoS}: 2B, 2H, ALF2, APR4, ENG, H4, MPA1,
MS1, MS1b, SLy~\cite{Read:2008iy} corresponding to a range of $\Lambda\in[1,10000]$.
The signals produced with \teobdali~start from a frequency of $f_0=0.0055$, corresponding to roughly $\sim 4000$M 
before reaching merger. The mismatch is obtained within a range of $f\in[0.007, 0.1]$ using a flat PSD.
The algorithm is stopped when the number of basis functions reaches $N_{\rm sim}=200$. 
We note these results have been anticipated in~\citet{Albanesi:2025txj}.

\begin{figure}
  \centering 
    \includegraphics[width=0.45\textwidth]{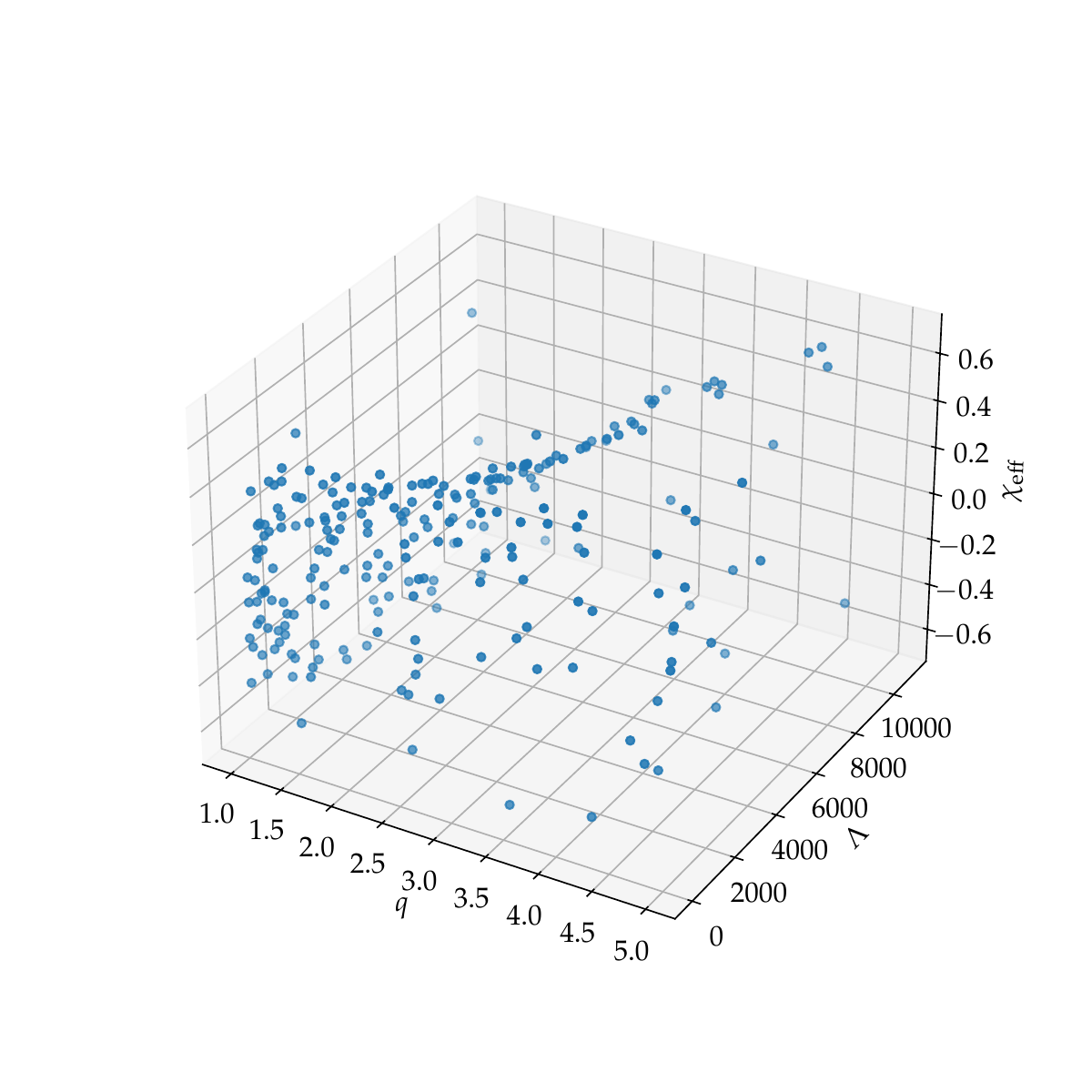}
    \caption{Greedy basis waveforms obtained in terms of $\Lambda$, $q$ and $\chi_{\rm eff}$ for non-precessing quasicircular orbits.}
    \label{fig:greedy}
\end{figure}

The result is shown in Fig.~\ref{fig:greedy}. The greedy algorithm
identifies 200 configurations, resulting in residuals with mismatches
below $\sim 0.05$, where $50\%$ are less than ${\sim} 0.02$.
The algorithm's results suggest that future \gls{NR} simulations should focus on
systems with $\Lambda>3000$ and mass ratio $q \leq
2$. These waveforms make up $\sim 50\%$ of the total points and
physically correspond to binaries where matter effects and tidal
disruption are significant. Therefore, they are more likely to produce
significant differences in the \gls{GW} signal. 
The distribution of the effective spin $\chi_{\rm eff}$ of the systems
selected by the greedy algorithm pushes towards the upper limit of the
parameter space considered.
Note that also in this case, binaries with large aligned spins are more likely
to produce significant tidal disruption.

\subsection{Eccentric and precessing waveforms}

\begin{figure*}
  \centering 
    \includegraphics[width=\textwidth]{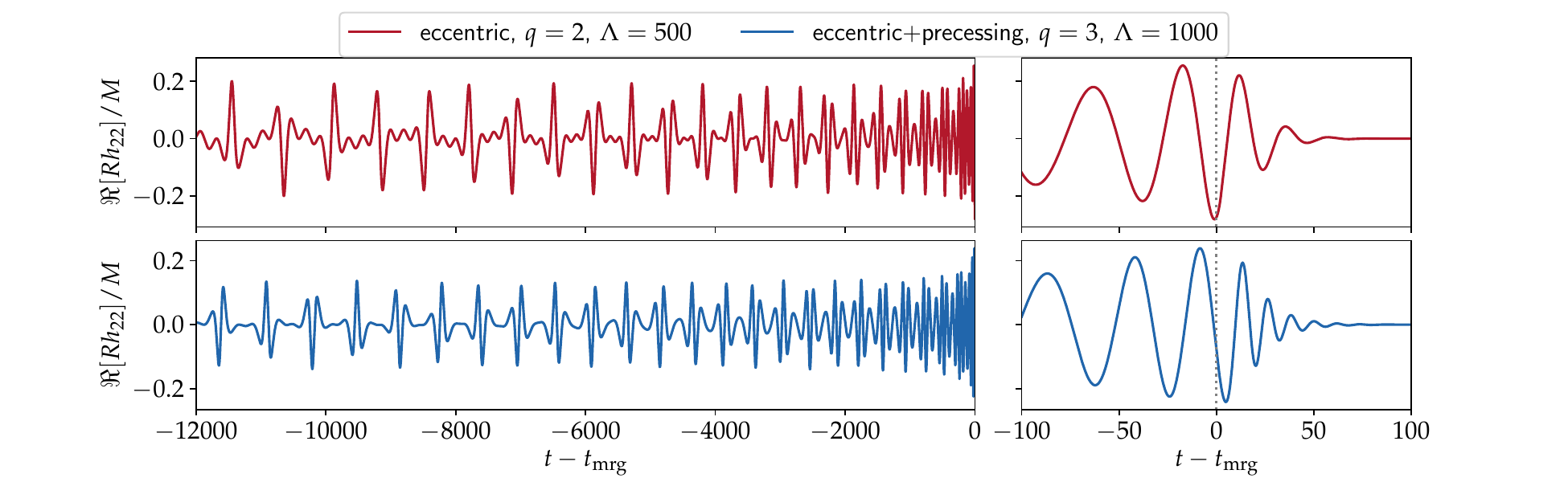}
    \caption{Gravitational waveforms produced by \teobdali~with eccentricity $e_0=0.5$ and no corresponding \gls{NR} data. Top (red line): 
    Binary in an eccentric orbit with nonspinning components, $q=2$ and $\Lambda=500$, (Type I). 
    Bottom (blue line): Binary in an eccentric orbit with $\vec{\chi}_1=(0.2,0.2,0.2)$, $q=3$ and $\Lambda=1000$, (Type III).}
    \label{fig:ecc_prec_wvfs}
\end{figure*}

\teobdali{} is a physically complete EOB model that allows to generate
BHNS waveforms from eccentric and precessing BHNS mergers \cite{Albanesi:2025txj}.
We have carefully checked the the robustness of our eccentric
\gls{BHNS} waveform by generating over 1000 different binaries
with $q\in[1,5]$,$\abh\in[-0.8,0.8]$, $\Lambda\in[1,5000]$ and $e_0\in[0.01,0.2]$. All the waveforms
generated are smooth and showed no evident unphysical features, see Appendix~\ref{app:wvf_check}. 
As an example, we show in Fig.~\ref{fig:ecc_prec_wvfs}, a gravitational waveform with initial eccentricity $e_0=0.5$, $q=2$ and $\Lambda=500$ (top, red line);
and a second one also including spin precessing effects, $q=3$, $\Lambda=1000$ (bottom, blue
line) with $\vec{\chi}_1=(0.2,0.2,0.2)$. Interestingly, one can notice
small oscillating features early in the inspiral which are not present for eccentric \gls{BBH} waveforms.
These appear for the $q=2$ (eccentric) configuration, where tidal disruption takes place.
These oscillations are still present but significantly damped for the $q=3$ (eccentric+precessing) waveform, 
corresponding to a Type III from our classification (intermediate case). 
Later in the inspiral, the waveforms show morphologies consistent with 
our predictions for \gls{BBH} but with the dampened ringdown that characterizes \gls{BHNS} binaries.

The early inspiral oscillations for eccentric configurations are consistently present throughout the parameter space 
of \teobdali, and get significantly more pronouced with higher positive spins and increasing $\Lambda$, corresponding to tidal disruption cases.
See Appendix~\ref{app:wvf_check} for more details on the behaviour of the eccentric model. 
Given the lack of publicly available \gls{NR} simulations in
elliptical orbits, we cannot directly assess the faithfulness of our
model against numerical data.

Recent work has suggested the event GW200105 shows evidence of both
eccentricity and precession~\cite{Morras:2025xfu}.
The analysis was carried out employing a post-Newtonian model, \texttt{pyEFPE}~\cite{Morras:2025nlp}, which 
incorporates eccentricity and precession, and considering a frequency range of $f\in[20,280]$~Hz.
They find an orbital eccentricity of $e_{20}=0.145^{+0.007}_{-0.097}$ at a \gls{GW} frequency of 20~Hz within the 90$\%$
credibility interval and evidence of precession with $\chi_p=0.06^{+0.13}_{-0.04}$ at the 95$\%$ credible upper limit.

\begin{figure*}
  \centering 
    \includegraphics[width=0.85\textwidth]{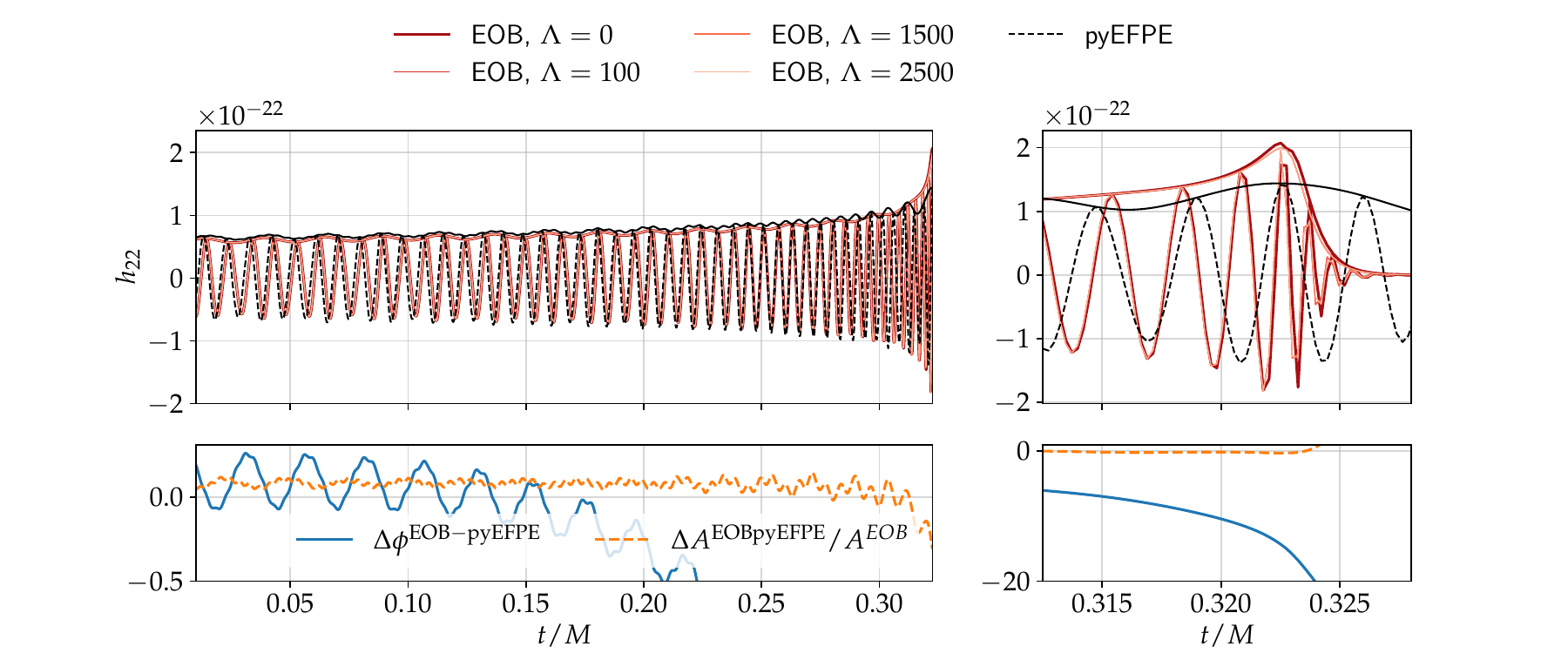}
    \caption{Alignment between the best waveform obtained in the analysis of~\cite{Morras:2025xfu} with the 
    \texttt{pyEFPE} model (black) and the equivalent waveform produced with
    \teobdali~(dark red). Their correponding phase and relative amplitude differences are shown at the bottom of the figure.
    As a reference, we show equivalent \gls{EOB} waveforms with different $\Lambda$ values on the top panel. }
    \label{fig:ecc_prec_gw200105}
\end{figure*}

Inspired by these claims, we put at test \teobdali{} on the best
waveform inferred in~\cite{Morras:2025xfu}.
In Figure~\ref{fig:ecc_prec_gw200105} we compare the best
\texttt{pyEFPE} waveform against a \teobdali~waveform produced with
the same parameters. 
We include in the plot \teobdali{} waveforms with different $\Lambda$
values as illustration, but perform the phase alignment between the
$\Lambda=0$ and the \texttt{pyEFPE} waveforms since the latter model
does not account for tides.
The PN based waveform aligns well in the early inspiral with the EOB
based one, but rapidly accumulate tidal dephasing starting at
GW frequencies of $f\sim760$~Hz (outside of the considered frequency range of the 
analysis in~\cite{Morras:2025xfu}).
By merger time, the PN approximant has dephased of several cycles in
comparison to \teobdali.

We compute mismatches among these waveforms in
the same way described in Sec.~\ref{sec:mismatch} but employing the
same frequency range as in~\cite{Morras:2025xfu} and considering 
the (2,2) mode only. To avoid complications arising from different definitions of eccentricity among models,
we optimize the \teobdali~waveform over the eccentricity to find the equivalent system to that of~\cite{Morras:2025xfu}.
We find a mismatch of $\bar{\mathcal{F}}=0.28$, indicating the PN model is affected by large systematics towards the upper limit of the analysis frequency range.
This result suggests that an analysis of GW200105 with \teobdali~ with eccentricity and precession is likely to deliver different parameters than the PN analysis. 
The mismatch computation with tidal effects ($\Lambda\neq0$) yields very similar mismatches, which is expected for this highly asymmetric event (Type II).
Therefore, using waveform models without tides for the inference of this event appears to be a robust choice.

As a further check, we compute the mismatches between the best waveform
obtained by the LVK
analysis~\cite{LIGOScientific:2021qlt} with \texttt{IMRPhenomXPHM}~\cite{Pratten:2020ceb} against the waveform produced with \teobdali, both with and
without tides, higher modes and same other parameters.
The mismatches stay at values $\sim 0.001$ and similarly as before, we find no difference
in the mismatch when including different values of $\Lambda$, which is expected from the high mass ratio of
this event.
This suggests that a \teobdali{} analysis of the same data (and without
tidal effects) is likely to
deliver source parameters compatible with the LIGO-Virgo-Kagra results.

\section{Conclusions} 
\label{sec:conclu}

In the first part of this paper, we presented 52 new NR simulations of
circularized BHNS with different configurations of mass ratios, spins,
and employing three different \gls{EoS}.
We validated our simulations with an extensive systematic
study including initial data quasiequilibrium sequences, grid setups
and convergence studies.
The simulation data were employed to quantitatively model BH remnants
and gravitational waveforms. About the former, we provided updated
formulas for the remnant mass and spin which smoothly deform 
NR-driven models for binary black holes. About the latter, we studied
for the first time the multipolar structure of the GW modes up to
$\ell=4$ and devised a quantitative estimate for the GW recoil. 

We found that the most relevant GW subdominant modes are
(2,1), (3,2), (3,3) and (4,4), as expected from the hierachy of binary
black holes. Contrary to the latter however, the (2,0) and
(3,0) amplitudes contribute more to the whole multipolar amplitude for \gls{BHNS}
and are related to the memory part of the \gls{GW}.
If these multipoles will be accurately modeled in future
waveform templates, \gls{GW} observations in these channels could
potentially help in distinguishing between BBH and BHNS, in addition to enhancing
the effects of \gls{GW} memory for these binaries.

With the numerical data, we develop a model to estimate the GW recoil of the remnant
 and find that tidal effects are more prominent for more comparable
masses and anti-aligned spins, i.e.~Type I (see
Fig.~\ref{fig:vk_fits}.) The net effect is to supress the kick velocity
with respect to BBH due to the correspondingly lower radiated momentum.
This is not the case for binaries with increasing spin magnitude, mass ratios $q\gtrsim2$ and
$\Lambda\lesssim500$, which approach more the estimates for BBH.

Further, we discussed the results from a 12 orbit spin precessing simulation
(\verb|BAM:0223|, Type I) compatible with the LIGO-Virgo-Kagra event
GW230529 and the implication for 
the interpretation of that event. Comparing the numerical simulation
to the best waveform obtained in \cite{LIGOScientific:2024elc}, we
find a mismatch of ${\sim}0.3$. The discrepancy is mainly ascribable
to tidal effects that are not modelled in the best inferred
waveform. 

In the second part of our work we presented \teobdali~for BHNS. The
latter EOB model is an extension of our previous work
\cite{Gonzalez:2022prs} which uses the improved NR-information
developed here to reduce systematic uncertainties.
\teobdali~ specifically models NQC and ringdown for multipoles (2,2),
(2,1), (3,2), (3,3) and (4,4). 
The (2,2) waveform amplitude at merger is improved by an order of
magnitude by the new NQC prescriptions. Such amplitude is employed to
phenomenologically classify Type I, II, III binaries and make a design
choice for the EOB waveforms.
The new multipolar ringdown allows to better capture the key
morphological features that distinguish BHNS from BBH waveforms.
Moreover, \teobdali~can make predictions for precessing and eccentric binaries.

We validated our model with \verb|BAM:0223| which was not used to
inform \teobdali. We obtain a phase difference throughout the
whole inspiral waveform below $\sim 0.5$ rad. The mismatches are the order of $\sim 0.01$
for low inclinations $\iota\leq\pi/4$, thus highlighting the model's 
capacity to accurately reproduce the waveforms of highly precessing configurations.
Additionally, we obtained a median mismatch of 0.01 against all available simulations,
demonstrating the accurate performance of the model.

Our work clearly emphasizes the necessity of more NR data in order to
be able to develop faiththful waveforms for advanced and next
generation observations. Focusing on quasicircular
non-precessing BHNS, we employ \teobdali~in a greedy search to
identify an optimal set of future simulations.
About 200 simulations with initial frequency of $f_0=0.0055$ appear
sufficient to describe BHNS waveforms with mismatches 
${\lesssim}0.02-0.05$. This results sets the task for the immediate 
future: new simulations should be performed for large tidal
polarizability parameters, mass ratio $q\sim2$ and large effective spins.
Similar studies also including precession
and eccentricity are current being conducted to identify
the most relevant parameters also for those cases.

Finally, we verified the robustness of \teobdali~ in producing 
the first eccentric (+precessing) \gls{BHNS} waveforms. In absence of 
\gls{NR} simulations, we verified the model produces sane waveforms
for over 1000 different binaries with parameters in ranges 
$q\in[1,5]$, $\abh\in[-0.8,0.8]$, $\Lambda\in[1,5000]$ and
$e_0\in[0.01,0.2]$. This is possibly the parameter space region which
more urgently needs specific BHNS waveforms.

Furthermore, we computed mismatches between \teobdali~and the best
waveforms for the event GW200105 as obtained by~\cite{Morras:2025xfu} and
LIGO-Virgo-KAGRA~\cite{LIGOScientific:2021qlt}. 
The former analysis was conducted with a \gls{PN} model incorporating
eccentricity and precession. For the best parameters, we found large 
mismatches corresponding to phase differences of more than 14 rad accumulated over the frequency
    interval analyzed. This suggests that the inference with
\teobdali~ including eccentricity and precession is likely to deliver
different source parameters.
The latter analysis was conducted with
\texttt{IMRPhenomXPHM} including higher modes and precession.
\teobdali~shows mismatches of order ${\sim}0.001$ for the same inferred parameters,
indicating consistency of the results independently on the
  inclusion of tidal effects.

The primary challenge facing contemporary BHNS modelling is the limited suite of NR simulations 
suitable for calibration. While the analytical description of the inspiral is robust, the strong-field dynamics 
represent a critical vulnerability across all existing models. Much of the available NR data lacks the convergence 
and length (often fewer than four orbits) necessary for high-fidelity modeling, underscoring the importance of the 
convergent simulations released in this study. In particular, the tidal disruption regime remains the most significant 
source of inaccuracy, despite its importance for multimessenger observations. Although recent NR efforts have integrated 
sophisticated microphysics, including temperature-dependent \gls{EoS} and neutrino transport, a gap remains in connecting
the complex merger dynamics and the resulting gravitational waveforms. Future progress will require a quantitative 
assessment of how matter and ejecta suppress the \gls{QNM} of the remnant black hole to better characterize 
the ringdown in these systems.

\begin{acknowledgments}
  A.G. acknowledges support by the Deutsche Forschungsgemeinschaft (DFG) under Grant No. 406116891 within the Research Training Group RTG 2522/1 and 
  by the Universitat de les Illes Balears (UIB); the Spanish Agencia Estatal de Investigación grants PID2022-138626NB-I00, RED2022-134204-E, RED2022-134411-T, 
  funded by MICIU/AEI/10.13039/501100011033 and the ERDF/EU; and the Comunitat Autònoma de les Illes Balears through the Conselleria d'Educació i Universitats with 
  funds from the European Union - NextGenerationEU/PRTR-C17.I1 (SINCO2022/6719) and from the European Union - European Regional Development Fund (ERDF) (SINCO2022/18146).
  S.B. acknowledges support by the EU Horizon under ERC Consolidator Grant, no.~InspiReM-101043372
  and from the DFG project ``GROOVHY'' (BE 6301/5-1 Projektnummer: 523180871).
  R.G. acknowledges support from NSF Grant PHY-2020275 (Network for Neutrinos, Nuclear Astrophysics, and Symmetries (N3AS)).
  AR was supported by NASA under Award No. 80NSSC21K1720.
  
  Simulations were performed on SuperMUC-NG at the Leibniz-Rechenzentrum (LRZ) Munich and and on the national HPE Apollo Hawk at the High Performance Computing Center Stuttgart (HLRS).
  The authors acknowledge the Gauss Centre for Supercomputing e.V. (\url{www.gauss-centre.eu}) for funding this project by providing computing time on the GCS Supercomputer SuperMUC-NG at LRZ (allocations {\tt pn36ge}, {\tt pn36jo}, {\tt pn68wi} and {\tt pn39go}) . The authors acknowledge HLRS for funding this project by providing access to the supercomputer HPE Apollo Hawk under the grant number INTRHYGUE/44215 and MAGNETIST/44288.
Postprocessing and development runs were performed on the ARA cluster at Friedrich Schiller University Jena. The ARA cluster is funded in part by DFG grants INST 275/334-1 FUGG and INST 275/363-1 FUGG, and ERC Starting Grant, grant agreement no. BinGraSp-714626.

  The data that support the findings of this article are openly available~\cite{teobresums-dali,CoRe:catalog}.

  \noindent
              \end{acknowledgments}

\appendix

\section{Numerical relativity data}
\label{app:nr}

In this Appendix we show the details of the configurations simulated for this work. 
In App.~\ref{app:nr_quality} we describe the accuracy of the waveforms extracted.
Appendix~\ref{app:final_bh} presents the obtained remnant properties and the coefficients necessary for the remnant \gls{BH} model.
Finally, resolution effects and errors on the measurement of the \gls{GW} kick velocity are discussed in App.~\ref{app:vkick}.

\subsection{Data quality}
\label{app:nr_quality}

In this section we assess the convergence of our simulations and justify our grid configuration choices for the production runs.
We consider four resolutions and three different refinement levels on the \gls{BH}, see Table~\ref{tab:grid_configurations}.
For these studies we employ the configuration of \verb|BAM:0206| to find the "cheapest", highest resolution configuration to 
employ for the rest of the simulations. 

\begin{table}[h!]
  \centering    
  \caption{ Grid configurations employed for the evolutions with the BAM code. }
  \begin{tabular}{ccccccc}        
    \hline\hline
    Name & $L$ & $l^{\rm{mv}}$ & $n^{\rm{mv}}$ & $h_{\rm{L-1}}$ & $n$ & $h_0$ \\
    \hline
    L6 & 6 & 2 & 64 & 0.209 & 96 & 13.38\\
    L8 & 8 & 2 & 64 & 0.052 & 128 & 13.38\\
    L9 & 9 & 2 & 64 & 0.026  & 128 & 13.38\\
    \hline
    M6 & 6 & 2 & 96 & 0.139 & 144 & 8.92\\
    M8 & 8 & 2 & 96 & 0.035 & 163 & 8.92\\
    M9 & 9 & 2 & 96 & 0.017 & 163 & 8.92\\
    \hline
    H6 & 6 & 2 & 128 & 0.104 & 192 & 6.69\\
    H8 & 8 & 2 & 128 & 0.026 & 218 & 6.69\\
    H9 & 9 & 2 & 128 & 0.013 & 384 & 6.69\\
    \hline
    F6 & 6 & 2 & 192 & 0.069 & 288 & 4.46\\
    F8 & 8 & 2 & 192 & 0.017 & 384 & 4.46\\
    F9 & 9 & 2 & 192 & 0.009 & 326 & 4.46\\
    \hline\hline
  \end{tabular}
 \label{tab:grid_configurations}
\end{table}

\begin{figure}[t]
  \centering 
    \includegraphics[width=0.45\textwidth]{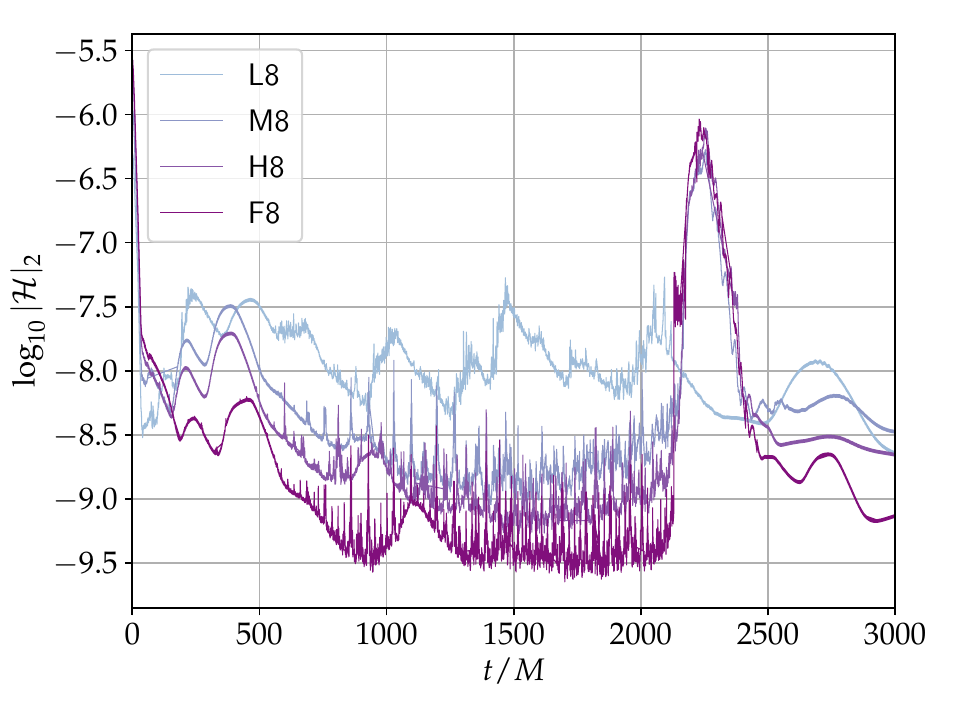}
    \cprotect\caption{Hamiltonian constraint of \verb|BAM:0206| with L8, M8, H8 and F8 settings.}
 \label{fig:ham_q2}
\end{figure}

Figure~\ref{fig:ham_q2} shows the norm of the Hamiltonian constraint for all the resolutions with eight refinement levels for the \gls{BH}. The norm stays at 
low values for most of the simulation time and decreases notably as we increase the resolution.

\begin{figure}[t]
  \centering 
    \includegraphics[width=0.45\textwidth]{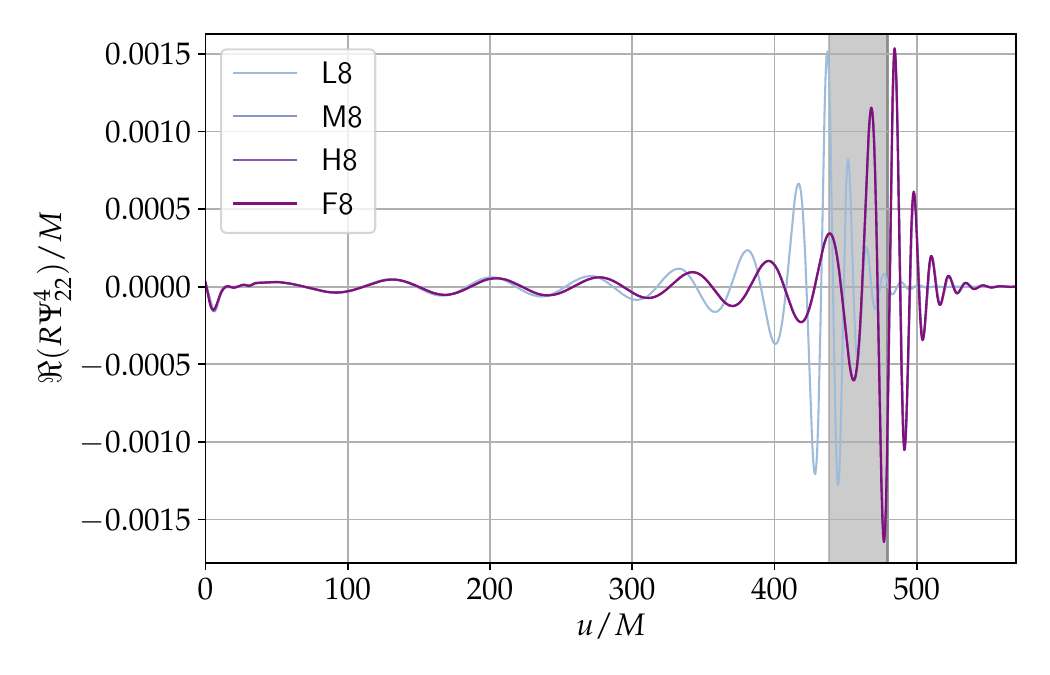}
    \cprotect\caption{Real part of $R\Psi^4_{22}/M$ of \verb|BAM:0206| with L8, M8, H8 and F8 settings. The grey area indicates the different merger times.}
 \label{fig:wvf_res_q2}
\end{figure}

The resulting waveforms are presented in Fig.~\ref{fig:wvf_res_q2}, where we also show the merger time difference (grey area). From this plot one can already 
tell that the set L8-M8-H8-F8 does not build a convergent series. Indeed, the following convergence study in Sec.~\ref{app:nr_conv} does show that the lowest 
resolution (L8), although the cheapest to simulate, doesn't produce accurate enough data.  

\begin{figure}[t]
  \centering 
    \includegraphics[width=0.45\textwidth]{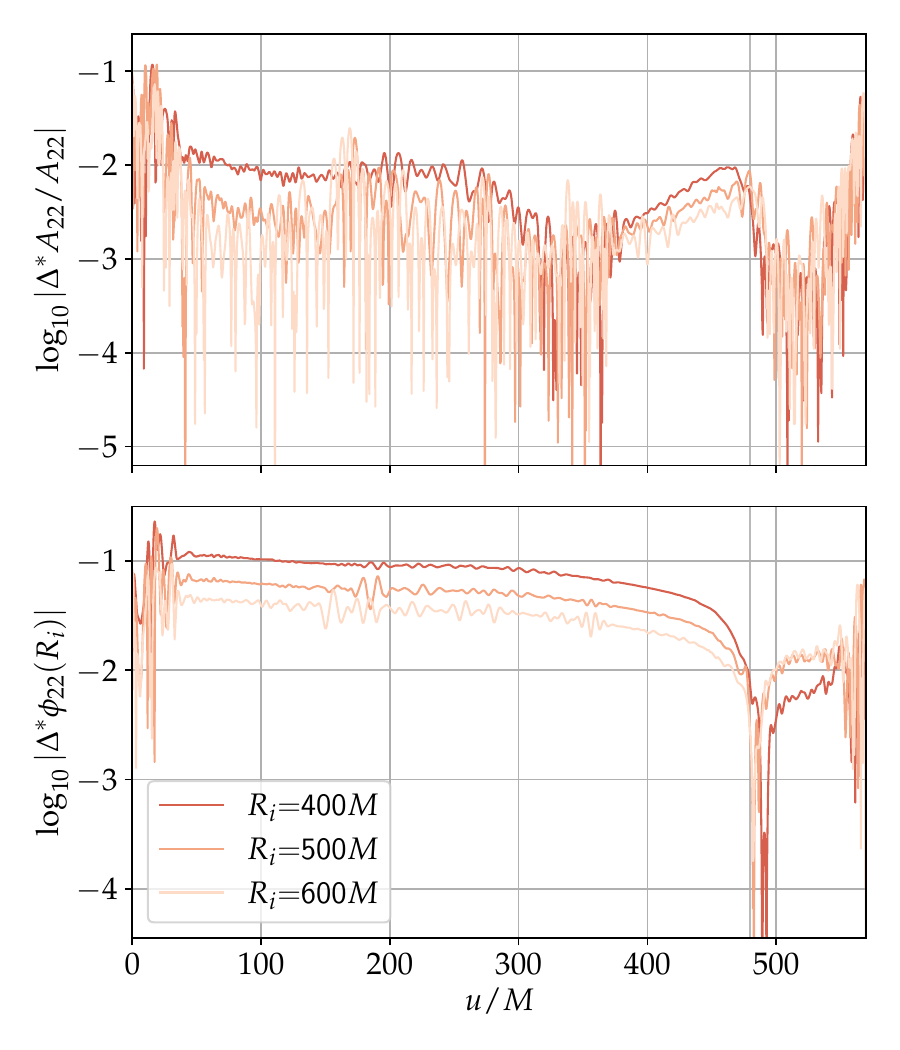}
    \cprotect\caption{Uncertainties due to finite extraction radii for \verb|BAM:0206| with H8 settings.}
 \label{fig:extradii_q2h}
\end{figure}

\begin{figure}[t]
  \centering 
    \includegraphics[width=0.45\textwidth]{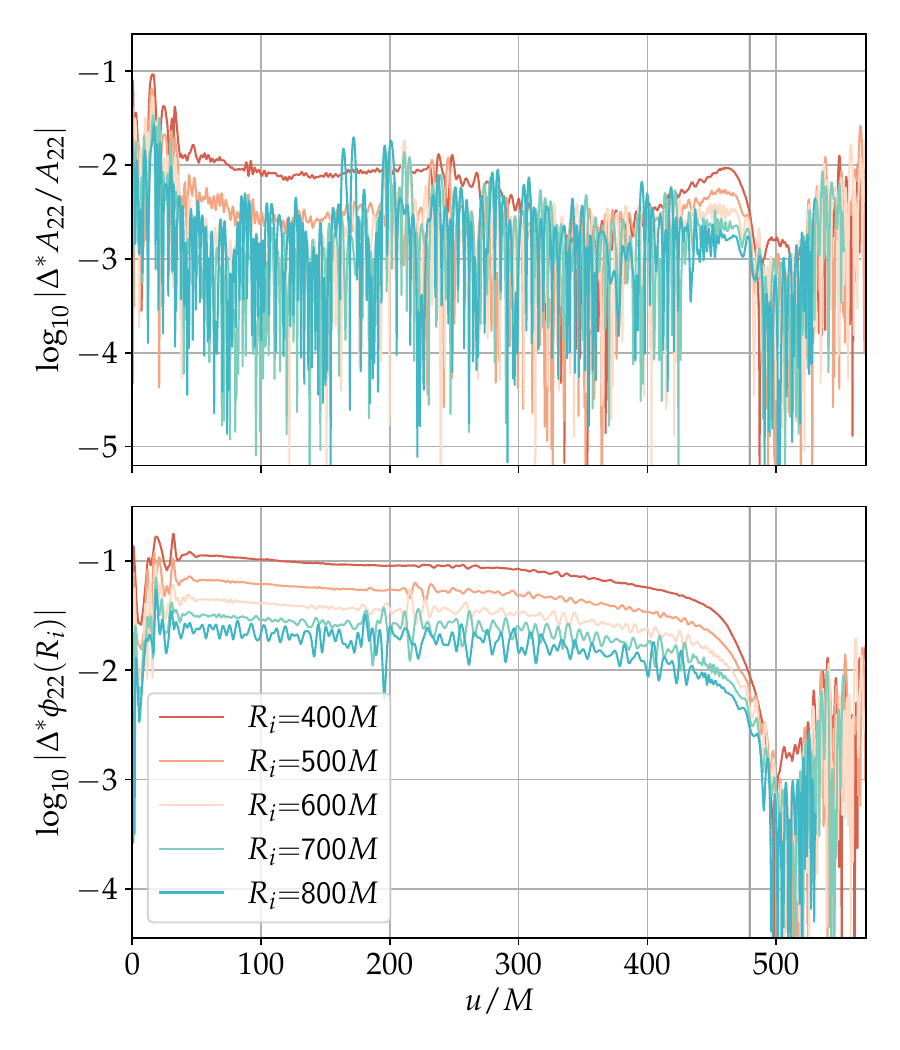}
    \cprotect\caption{Uncertainties due to finite extraction radii for \verb|BAM:0206| with F8 settings.}
    \label{fig:extradii_q2f}
\end{figure}

The first thing we want to check is the uncertainty due to extracting the waveform at a finite radius. These errors affect mostly the amplitude and phase, which are
critical quantities for waveform modelling. We show the differences between each extraction radius in Figures~\ref{fig:extradii_q2h} and~\ref{fig:extradii_q2f} for the 
two highest resolutions. Differences in both amplitude and phase are obtained as the difference between two consecutive radii, e.g. 
$\Delta^*\phi(R_i) = \phi(R_i) - \phi(R_{i-1})$. For both resolutions, differences in amplitude stay below $1\%$, whereas for the phase we see a clear 
decrease as we extract at higher radii. Furthermore, we notice how significant the phase errors are early in the evolution and are less relevant as we approach merger.
 
\subsubsection{Convergence tests}
\label{app:nr_conv}

\begin{figure}[t]
  \centering 
    \includegraphics[width=0.45\textwidth]{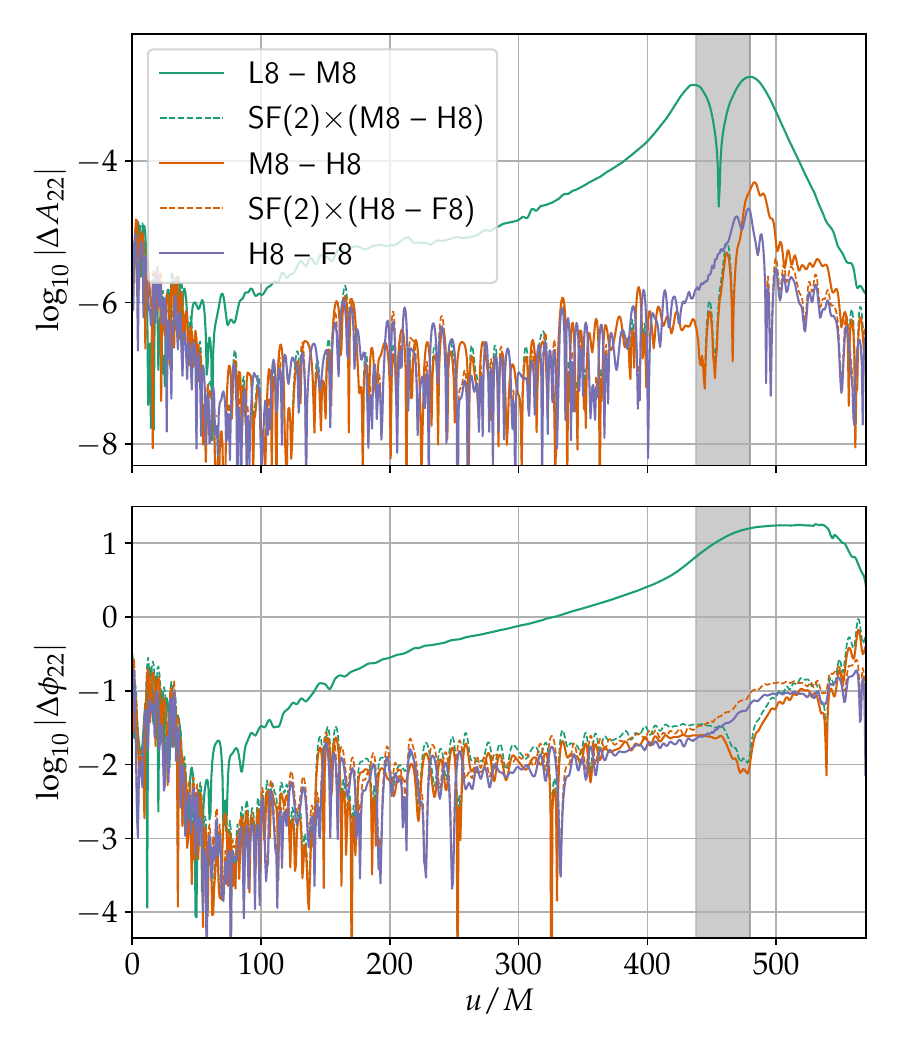}
    \cprotect\caption{Uncertainties due to grid resolution for \verb|BAM:0206|.
      Amplitude (top) and phase differences (bottom) of $R\Psi^4_{22}/M$ among the L8, M8, H8 and F8 settings.}
    \label{fig:diff_res_q2}
\end{figure}

We look again at the amplitude and phase difference among the different resolutions with eight refinement levels on the \gls{BH} in Fig.~\ref{fig:diff_res_q2}. 
As stated earlier, we note that the differences with the lowest resolution (green lines) are too high for it to belong to the convergent series M8-H8-F8. 
We obtain the convergence rate $r$ experimentally with the scaling factor $SF$,
\be 
SF=\frac{h^r_M - h^r_H}{h^r_H - h^r_F},
\ee
where $h$ corresponds to the minimum grid spacing for each resolution (M,H,F). As seen in the figure, our data shows second order convergence throughout the inspiral 
which decreases slightly around merger for the series M8-H8-F8. Truncation errors such as these tend to accumulate throughout the simulation, hence increasing 
with simulation time. At merger, uncertainties in the amplitude and phase stay below $\sim 0.01\%$ and $\sim10\%$ respectively for the convergent series.

\begin{figure}[t]
  \centering 
    \includegraphics[width=0.49\textwidth]{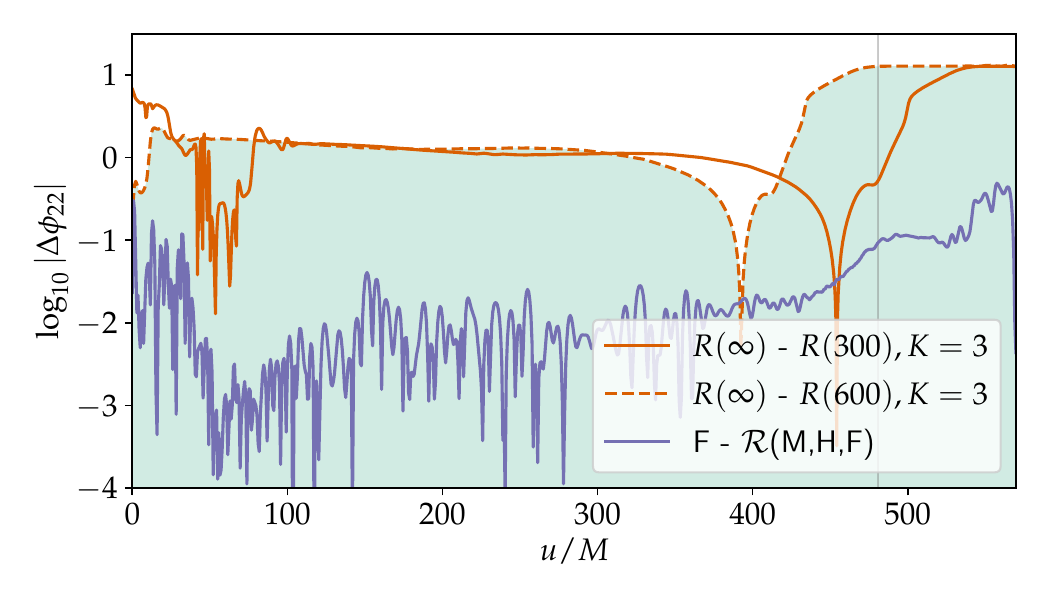}
    \cprotect\caption{Error budget for $R\Psi^4_{22}/M$ using the L8, M8, H8 and F8 convergent series for \verb|BAM:0206|.}
 \label{fig:errbud_res_q2}
\end{figure}

Aditionally, we obtain an error budget for our simulations as shown in Fig.~\ref{fig:errbud_res_q2}. To estimate the error from finite radius extraction $\delta\phi_{(r)}$, we 
compute an extrapolation to null infinity $R(\infty)$ by employing a polynomial in $1/R$ of order $K=3$ as described in~\cite{Bernuzzi:2011aq}. In the figure 
we compare the extrapolated waveform phase to that of the lowest and highest available extraction radius (orange solid and dashed lines respectively). We notice no significant
difference throughout the inspiral, but lower values towards merger for the highest radius. To account for resolution (or truncation) errors, we obtain an improved dataset 
$\mathcal{R}(M,H,F)$ through a Richardson extrapolation~\cite{Bernuzzi:2016pie,Bernuzzi:2011aq}. for this procedure we employ the convergence factor found experimentally and the convergent datasets 
we have (M8-H8-F8). The difference between the highest resolution and extrapolated values will provide an error estimate $\delta\phi_{(h)}$ for said value. Fig.~\ref{fig:errbud_res_q2} shows this estimate as a 
solid purple line, which typically increases as the evolution advances. The total error budget is obtained as the sum in quadrature of both error estimates (shaded green area in the figure), 
$\delta\phi = (\delta\phi^2_{(h)} + \delta\phi^2_{(r)})^{1/2}$. 

\subsubsection{Grid setups comparisons}
In the previous discussion, we presented the convergence of a simulation with three different simulations. Our goal is to select the most optimal grid configuration that both saves more computer time and 
still produces accurate results. We choose M8 to be our "base" configuration and in the following, we assess its accuracy by comparing its performance with higher resolutions and with more refinement levels.

\begin{figure}[t]
  \centering 
    \includegraphics[width=0.45\textwidth]{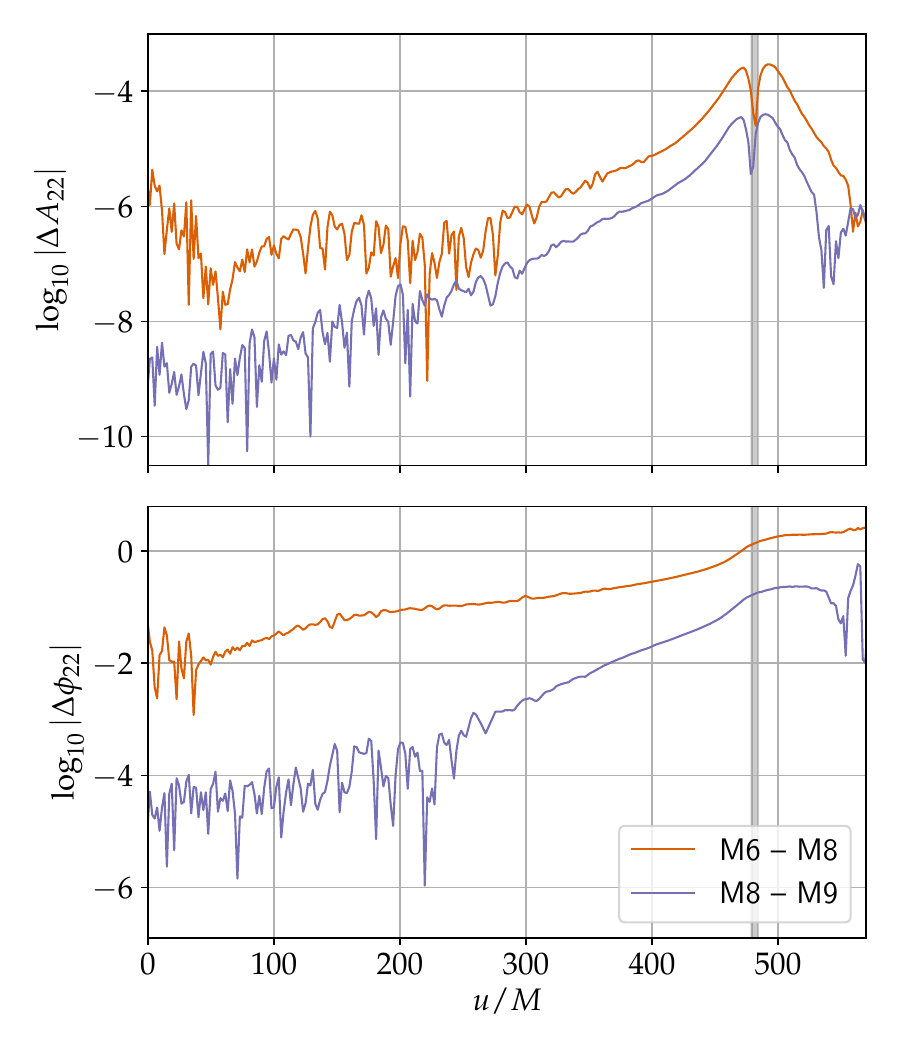}
    \caption{Amplitude and phase comparisons between the different refinement levels on the \gls{BH} for Medium resolution. 
    Differences between $L=6$ and $L=8$ (here M6 and M8) are shown in orange, whereas between $L=8$ and $L=9$ (M8 and M9) are shown in purple.
    The grey area indicates the different merger times.}
    \label{fig:reflevdiff_M}
\end{figure}

Figure~\ref{fig:reflevdiff_M} compares three refinement levels for the \gls{BH} at Medium resolution. Our base configuration (M8) has clearly lower differences in amplitude and phase with the more refined grid of M9
than with the less refined M6, with an order of magnitude difference at merger among the two.

\begin{figure}[t]
  \centering 
    \includegraphics[width=0.45\textwidth]{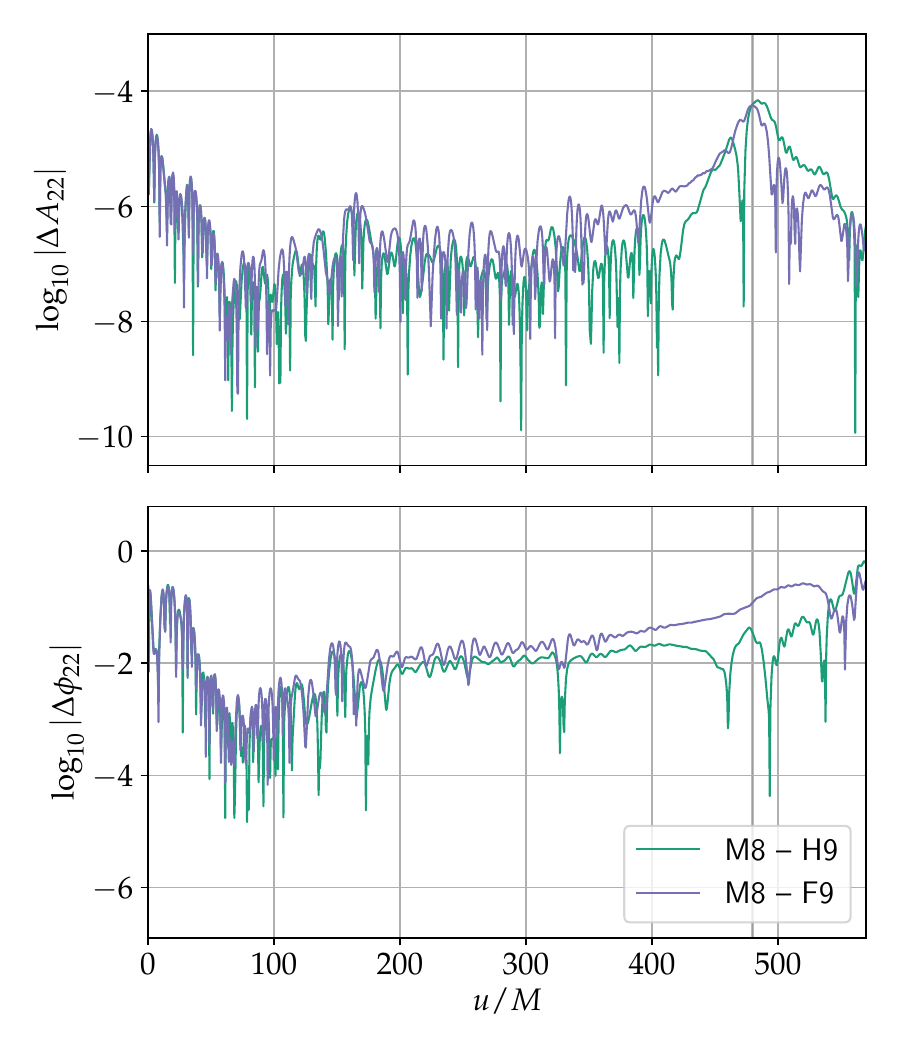}
    \caption{Amplitude and phase differences between M8 and the two highest resolutions H9 (green) and F9 (purple). The grey area indicates the different merger times.}
    \label{fig:resdiff_MH}
\end{figure}

If we now compare M8 with a higher resolution run with an extra refinement level, H9 (green line), we keep showing promising results as seen in Fig.~\ref{fig:resdiff_MH}. Overall in the inspiral, amplitude and phase differences stay around  
$\sim 0.00001\%$ and $\sim1\%$ respectively, with a phase difference at merger close to $1\%$, similarly as in the previous case with M9. Analogously, we show the same but comparing M8 with our highest 
simulation available, F9 (purple line). The amplitude differences show no considerable difference in comparison, however we do see a slight increase of the phase difference at merger, reaching almost $\sim10\%$. 

In general, we show good performance for the M8 configuration, presenting small enough differences with the most refined and highest resolutions. We therefore select this grid configuration as our minimal setup 
to produce high quality waveforms with the least computer resources.

\subsection{Remnant Black Hole}
\label{app:final_bh}

In this appendix we present the fitting parameters for the updated remnant \gls{BH} model. Table~\ref{tab:remnant_coef} shows the best coefficients
obtained for $\afbh$ and $\Mfbh$ as a deviation from the \gls{BBH} case.

 \begin{table*}[t]
	\centering    
	\caption{Fitting parameters for $\afbh$ and $\Mfbh$ with $R^2=0.914$ and $R^2=0.930$ respectively. 
  Here, we make a fit of a quantity $\mathcal{F}$ as $\mathcal{F}^{\rm BHNS}/\mathcal{F}^{\rm BBH}$.}
\begin{tabular}{ccccccccccc}        
	\hline\hline
	$\mathcal{F}$ & $k$ & $c_{k12}$ & $c_{k11}$ & $c_{k10}$ & $c_{k22}$ & $c_{k21}$ & $c_{k20}$ & $c_{k32}$ & $c_{k31}$ & $c_{k30}$\\
	\hline
	
	\multirow{ 4}{*}{$\afbh$} & 1 & $1.4 \times 10^{-3}$ & $4.6 \times 10^{-3}$ & $9.1 \times 10^{-4}$ & $-5.0 \times 10^{-2}$ & $-2.3 \times 10^{-2}$ & $-2.5 \times 10^{-2}$ & $1.8 \times 10^{-1}$ & $1.5 \times 10^{-2}$ & $1.3 \times 10^{-1}$\\ 
	& 2 & $2.1 \times 10^{-5}$ & $-3.1 \times 10^{-5}$ & $8.1 \times 10^{-6}$ & $-1.7 \times 10^{-4}$ & $2.8 \times 10^{-4}$ & $-4.4 \times 10^{-5}$ & $3.9 \times 10^{-4}$ & $-6.3 \times 10^{-4}$ & $2.1 \times 10^{-5}$\\ 
	& 3 & $-1.6 \times 10^{-8}$ & $1.7 \times 10^{-8}$ & $-3.1 \times 10^{-9}$ & $1.5 \times 10^{-7}$ & $-1.6 \times 10^{-7}$ & $2.4 \times 10^{-8}$ & $-3.9 \times 10^{-7}$ & $4.0 \times 10^{-7}$ & $-2.5 \times 10^{-8}$\\
  & 4 & $7.9 \times 10^{-7}$ & - & - & - & - & - & - & - & -\\
	\hline 
	
	\multirow{ 3}{*}{$\Mfbh$} & 1 & $-9.4 \times 10^{-4}$ & $1.7 \times 10^{-3}$ & $-5.9 \times 10^{-4}$ & $1.9 \times 10^{-3}$ & $-9.0 \times 10^{-3}$ & $4.2 \times 10^{-3}$ & - & - & -\\ 
	& 2 & $1.6 \times 10^{-7}$ & $-1.5 \times 10^{-6}$ & $1.0 \times 10^{-6}$ & $9.4 \times 10^{-7}$ & $6.6 \times 10^{-6}$ & $4.0 \times 10^{-5}$ & - & - & -\\ 
  & 3 & $4.7 \times 10^{-5}$ & - & - & - & - & - & - & - & -\\
	\hline\hline
\end{tabular}
\label{tab:remnant_coef}
\end{table*} 

\subsection{Kick velocity}
\label{app:vkick}
In the following, we describe the effects of resolution and integration errors on the computation of $v^{\rm GW}_{\rm kick}$.
The coefficients of the kick velocity fits presented in Sec.~\ref{sec:nr_vkick} are shown in Table~\ref{tab:vkick_coef}.

\begin{table}[t]
  \centering    
  \caption{Fitting parameters for $v^{\rm GW}_{\rm kick}$ yielding $R^2=0.926$.}
  \begin{tabular}{ccc}        
    \hline\hline
    Coefficient & $k=1$ & $k=2$ \\
    \hline
    $c_{k13}$ & 2.143 & - \\
    $c_{k12}$ & 0.096 & - \\
    $c_{k11}$ & -0.820 & - \\
    $c_{k10}$ & -0.051 & - \\
    $c_{k23}$ & -11.668 & - \\
    $c_{k22}$ & 0.339 & - \\
    $c_{k21}$ & 4.109 & 1.272 \\
    $c_{k20}$ & 0.400 & -2.297 \\
    \hline\hline
  \end{tabular}
\label{tab:vkick_coef}
\end{table}

\subsubsection{Effects due to resolution}
\label{app:vkick_res}
\begin{figure}[t]
  \centering 
    \includegraphics[width=0.45\textwidth]{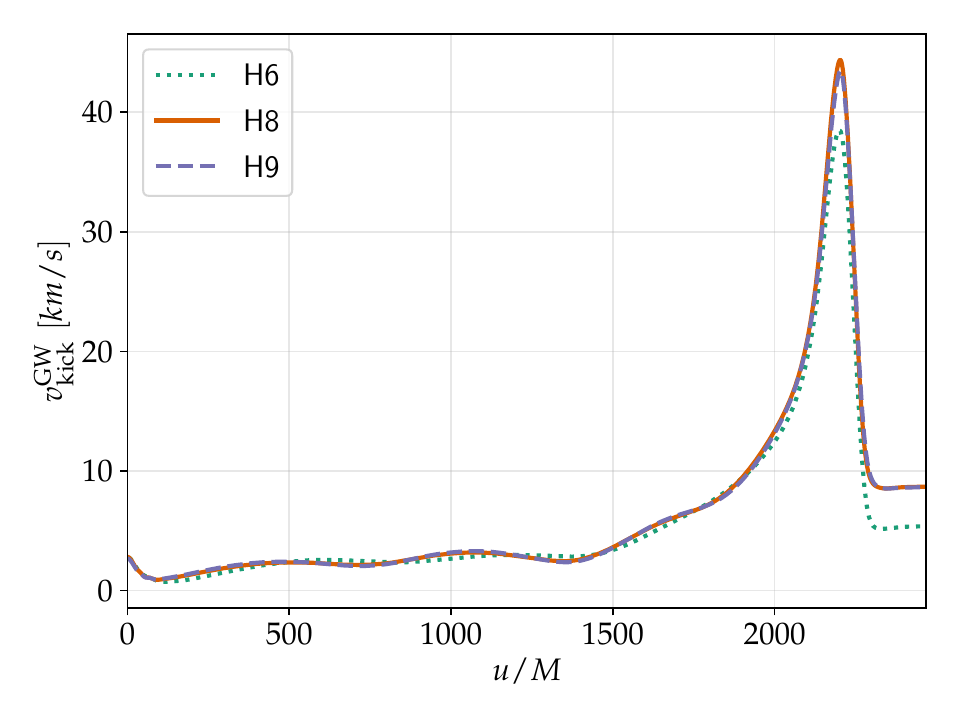} 
    \cprotect\caption{Computed recoil velocity of \verb|BAM:0206| using the (H)igh resolution and different refinement levels for the 
    \gls{BH}.}
 \label{fig:vk_reflev}
\end{figure}

In this subsection we discuss the effects of resolution on the measurement of the recoil velocity. Firstly, we focus on the different refinement levels 
for the \gls{BH} considered in this work. Fig.~\ref{fig:vk_reflev} shows the obtained kick velocity for \verb|BAM:0206| produced 
with the (H)igh resolution (see Table~\ref{tab:grid_configurations}) and different refinement levels. We notice a considerable gap between the velocity 
computed employing 6 refinement levels ($\sim 5~km/s$) with the others ($\sim 8~km/s$), thus showcasing the importance of resolving the \gls{BH} for accurate 
measurements of the remnant \gls{BH}.  On the other hand, choosing between 8 or 9 refinement levels doesn't seem to make a significant difference in the final 
kick velocity of the remnant.

\begin{figure}[t]
  \centering 
    \includegraphics[width=0.45\textwidth,height=5.5cm]{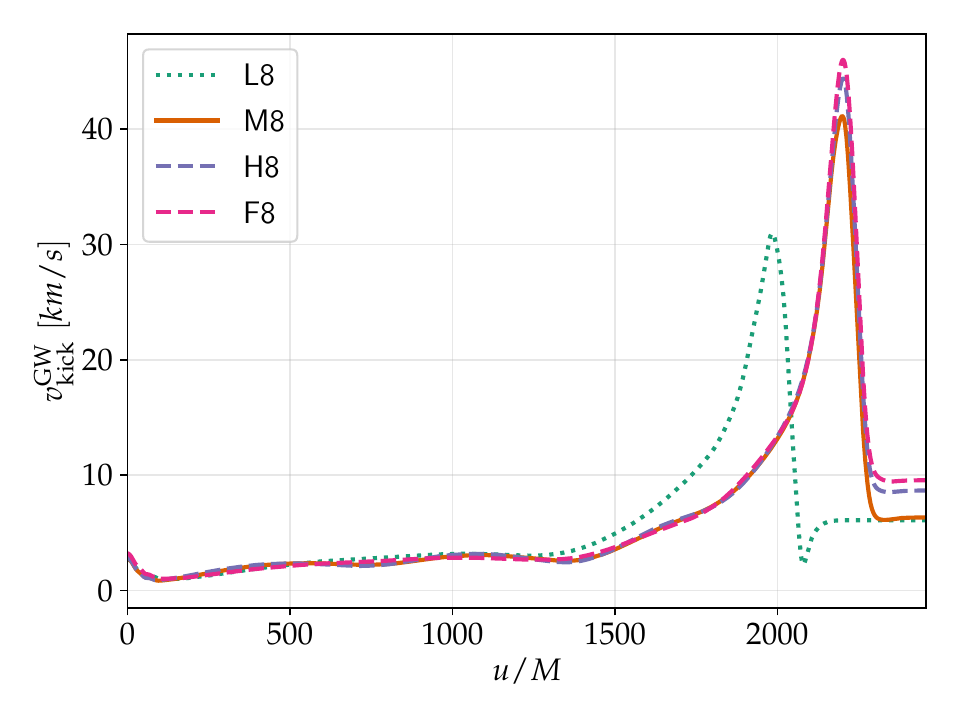}
    \includegraphics[width=0.45\textwidth,height=5.5cm]{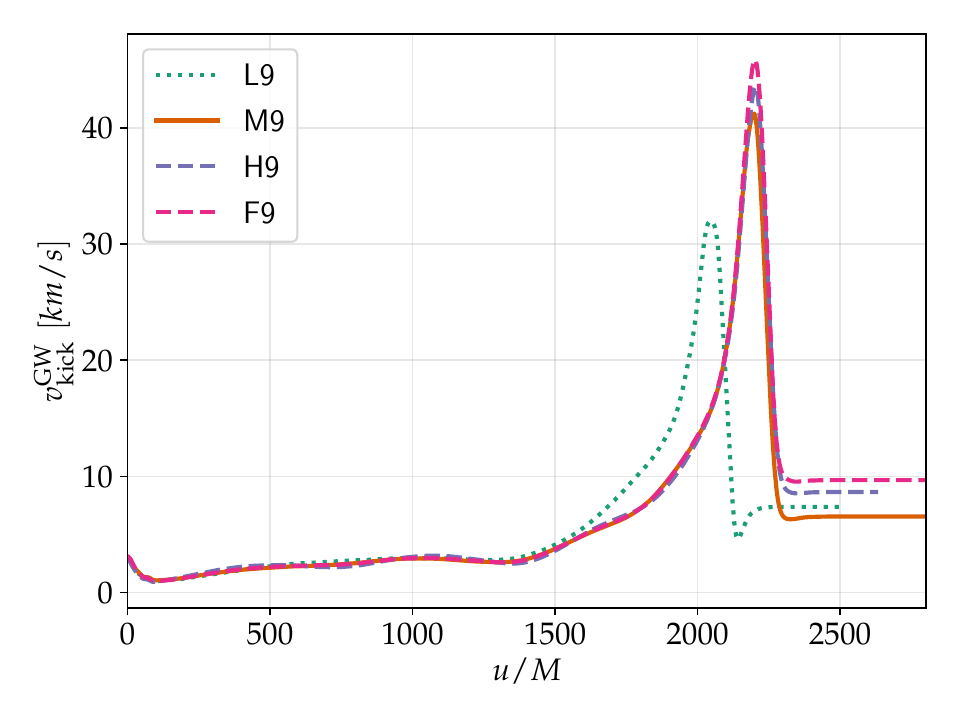}
    \caption{Same as Fig.~\ref{fig:vk_reflev} but for different resolutions and employing 8 (top) and 9 (bottom) refinement levels on the \gls{BH}.}
 \label{fig:vk_resos}
\end{figure}

Aditionally, we also see effect that resolution has in Fig.~\ref{fig:vk_resos} with 8 refinement levels (top) and using 9 (bottom). For both figures we see 
similar effects as was expected from our previous discussion. The configurations L8 ad L9 stand out from the rest as there are not part of the convergent series,
however their resulting $v^{\rm GW}_{\rm kick}$ lies within the values obtained with higher resolutions. From the figure, we notice that the recoil velocity 
tends to converge to a value of $\sim 10~km/s$. We remind the reader that for the results presented in this work, our chosen configuration for most simulations 
(with the exception of binaries with higher mass ratios) is M8. It is therefore expected to obtain errors of $\sim 3-4 km/s$ in the results presented in this paper 
according to what is shown in Fig.~\ref{fig:vk_resos}.

\subsubsection{Reducing measurement error}
\label{app:vkick_err}
\begin{figure}
  \centering 
    \includegraphics[width=0.45\textwidth]{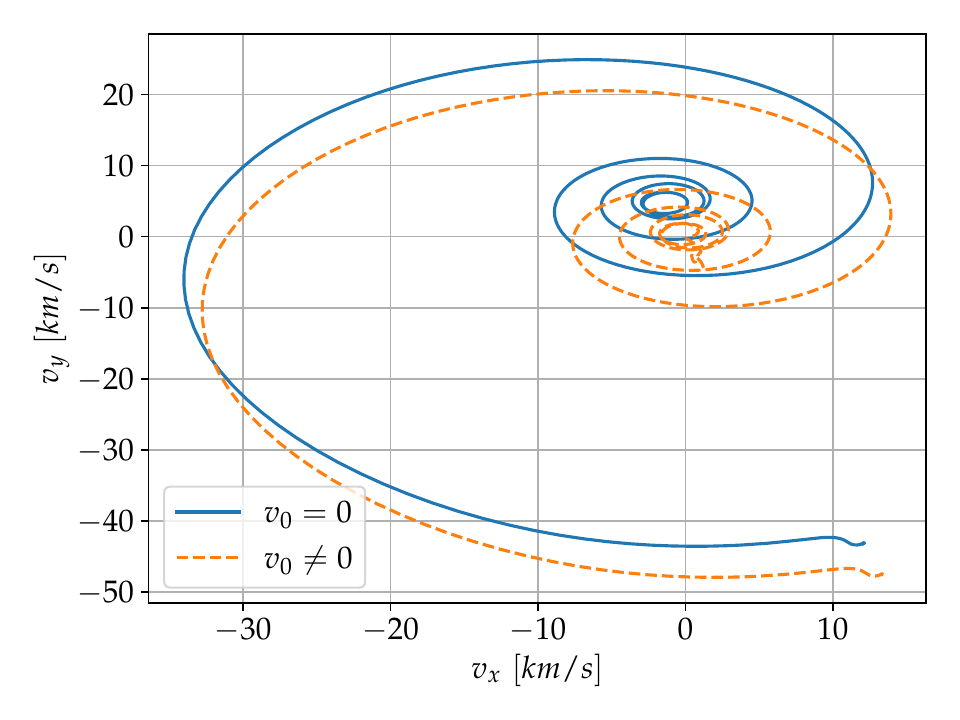}
    \includegraphics[width=0.45\textwidth]{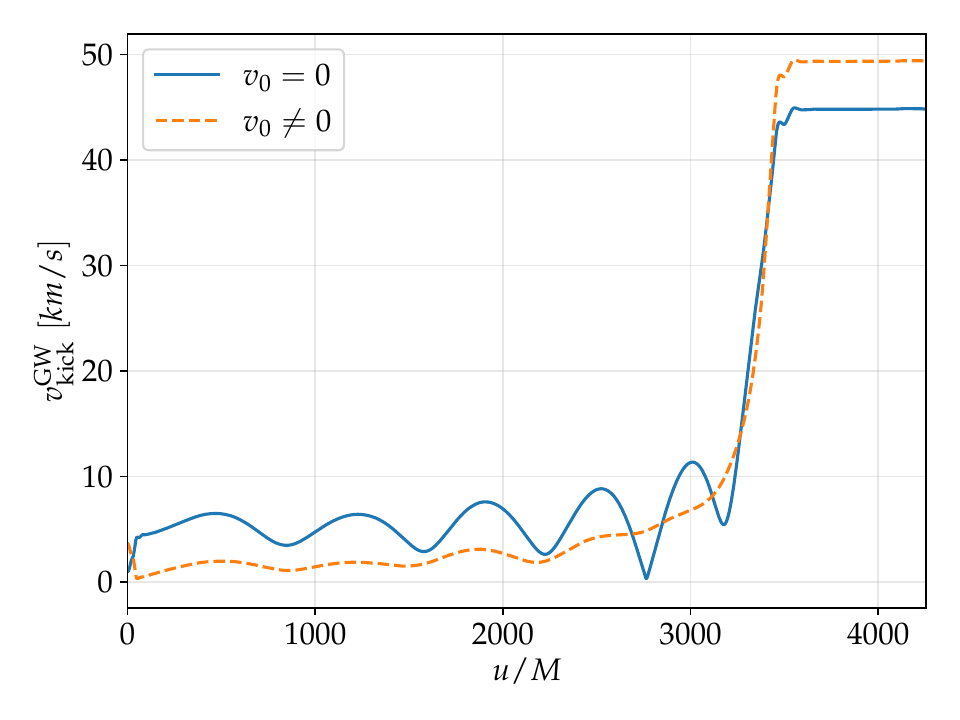}
    \caption{Left: hodograph of the component recoil velocities $v_x$ and $v_y$. Right: evolution in time of the kick velocity magnitude. The blue lines ($v_0=0$) are computed by integrating 
    from a finite $t_0$, and the orange ones by accouting for a non-zero integration constant ($v_0\neq 0$), see text.}
 \label{fig:kickvel_example}
\end{figure}

As described in the main text, the linear momentum of the binary is not taken into account for earlier times when integrating the kick velocity vector. This effect adds a substantial
amount of error on the measurement of $v^{\rm GW}_{\rm kick}$, which can be reduced by estimating an adequate integration constant,
i.e.
\be
v = v_0 -\frac{1}{M}\int^t_{t_0} \left( \dot{P}_x + i\dot{P}_y \right)dt.
\ee
We obtain $v_0$ by means of a hodograph as shown in the left panel of Fig.~\ref{fig:kickvel_example}, where we want to shift the center of the spiral to the origin. The resulting correction 
is presented on the right panel, showing a more monotonic increase of the recoil velocity. There is a significant difference of $9\%$ between the integration done from $t_0$ ($v_0=0$) and the one 
accounting for $v_0\neq 0$. This procedure is thus applied to all of our simulations. 

\subsection{Long simulation accuracy}
\label{app:errorbud_prec}

\begin{figure}
  \centering 
    \includegraphics[width=0.5\textwidth]{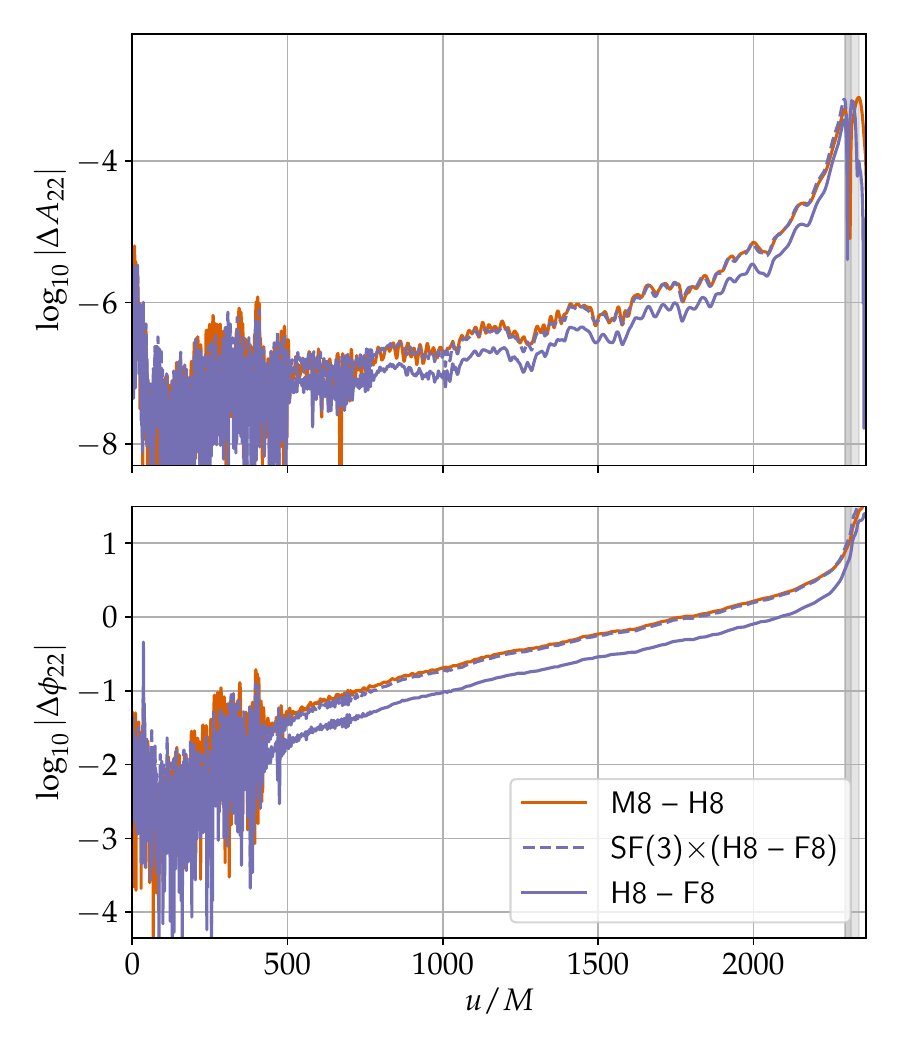}
    \cprotect\caption{Same as Fig.~\ref{fig:diff_res_q2} but considering the M8, H8 and F8 settings of 
      \verb|BAM:0223|. The vertical grey area shows the merger time differences among the 3 resolutions.}
 \label{fig:ap_diff}
\end{figure}

In Sec.~\ref{sec:validation} we compared our model to a new 12 orbit precessing simulation,
\verb|BAM:0223|, performed in 3 different resolutions: M8, H8, 
F8 (see Table~\ref{tab:grid_configurations}). We inspect the self convergence of the configuration in
Fig.~\ref{fig:ap_diff}, where we show the amplitude and phase differences between the available datasets.
The amplitude differences for all resolutions stay well below $10^{-5}$ before reaching merger, whereas the 
phase differences lie below 1 rad up until $\sim 2000M$ for the finest grids and until $~\sim 1700M$ for the
coarser ones.
The medium and high resolutions show a clear $\sim 3$rd order convergence throughout all evolution. 
Although computationally expensive,
additional simulations with even higher resolution would help us determine the convergence behaviour of the 
highest resolutions presented here.

\begin{figure}
  \centering 
    \includegraphics[width=0.49\textwidth]{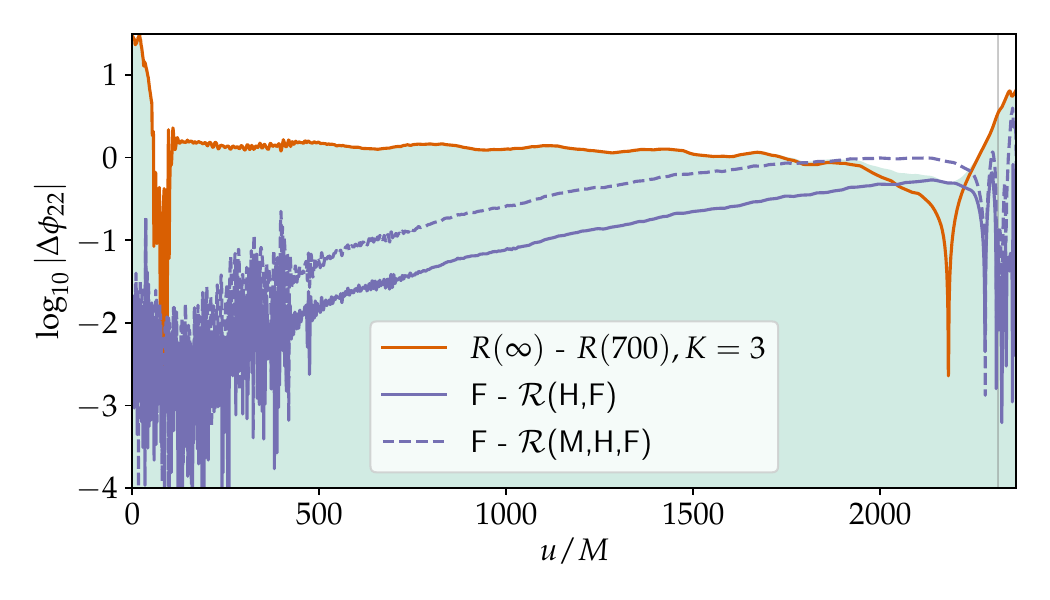}
    \cprotect\caption{Same as Fig.~\ref{fig:errbud_res_q2} but considering the M8, H8 and F8 settings of 
    \verb|BAM:0223| used for validation of the model. The vertical grey line indicated the merger time difference between the extrapolated
    $\mathcal{R}(H,F)$ and the finest resolution}
 \label{fig:errbud_prec}
\end{figure}

Here we present the total phase error budget on the 
phase, as a crucial quantity sensitive to numerical errors. This error needs to be considered 
when employing this simulation for calibration or validation of a model. We use the
same strategy as in App.~\ref{app:nr_conv}, see Fig.~\ref{fig:errbud_prec}. Furthermore, we show
two different Richardson extrapolated datasets, one including all resolutions and another 
only employing the two highest ones. As seen in the figure, we find the most optimal dataset to be
$\mathcal{R}(M,F)$ resulting in a significant decrease in the truncation errors. To account for finite extraction 
radii we extrapolate to infinity employing a $(1/R)^K$ polynomial with $K=3$ since other K orders resulted in high phase variations
over time. Nevertheless, although errors due to resolution slowly increase over simulation time and dominate briefly before merger, we notice
that extrapolation errors are the ones that contribute the most to the total phase error throughout most of the evolution.

 \section{Waveform fit models}
 \label{app:fit_models}

 In this Appendix we discuss the fitting models developed for the waveform model presented in this work for each individual mode. Appendix~\ref{app:fitmodels_ring} collects
 the fitting functions and coefficients for all the quantities related to building the ringdown waveform, i.e. ($A^{\rm peak}_{\lm}$, $\omega^{\rm peak}_{\lm}$, $\alpha_{\lm 1}$, $\omega_{\lm 1}$),
 and in App.~\ref{app:fitmodels_nqc} one can find the models for all the parameters needed for the \gls{NQC}, 
 ($A^{\rm NQC}_{\lm}$, $\dot{A}^{\rm NQC}_{\lm}$, $\omega^{\rm NQC}_{\lm}$, $\dot{\omega}^{\rm NQC}_{\lm}$).
 
 Due to the low quantity of higher modes data, some fits present higher order terms. In order to mitigate extreme values, we include artificial \gls{BBH} data in the relevant regions of
 the parameter space and ensure a physical behaviour of the fits. Results out of the data sample region could reach up to a 10$\%$ error.
 
 \subsection{Ringdown model parameters}
 \label{app:fitmodels_ring}
 
 Here we present the fit models and corresponding coefficients for the quantities needed to build the ringdown model described in Sec.~\ref{sec:eob_ringdown}.
All fitting models for each quantity $\mathcal{F}$, are based on a deviation from the \gls{BBH} fits from~\cite{Nagar:2020pcj} as 
$\mathcal{F}^{\rm BHNS}/\mathcal{F}^{\rm BBH}$ and modelled using a Pade approximant function with dependence on masses, spins and tides. 
For the multipolar amplitude peaks we employ the rescaling shown in Eq.~\ref{eq:amplm} (see Sec.~\ref{sec:nr_gw}) and are represented as $\hat{A}_{\lm}$.

The ringdown part of the waveform requires for each multipole the quantities ($A^{\rm peak}_{\lm}$, $\omega^{\rm peak}_{\lm}$, $\alpha_{\lm 1}$, $\omega_{\lm 1}$)
extracted from \gls{NR} data. We evaluate the reliability of our fits by computing the coefficient of determination $R^2$, which we report below for every quantity.

 \subsubsection{$(\ell,m)$ $=$ $(2,2)$}
We first develop the fits for the dominant (2,2) mode and model the peak of the amplitude as 
\be
\label{eq:apeak22}
\hat{A}^{\rm BHNS}_{22}/\hat{A}^{\rm BBH}_{22} = \frac{1 + \Lambda p^{(3)}_{1}(\nu,\abh) + \Lambda^2 p^{(3)}_{2}(\nu,\abh)}{1 + \Lambda p^{(2)}_{3}(\nu)}
\ee
with the following polynomials
\begin{subequations}
  \begin{align}
  p^{(3)}_{k}(\nu,\abh) &= p^{(3)}_{k1}(\abh)\nu + p^{(3)}_{k2}(\abh)\nu^2 \nonumber \\
  &\quad + p^{(3)}_{k3}(\abh)\nu^3,\\
  p^{(3)}_{kj}(\abh) &= c_{kj2}\abh^2 + c_{kj1}\abh + c_{kj0}, \\
  p^{(2)}_{3}(\nu) &= c_{311}\nu + c_{312}\nu^2.
  \end{align}
\end{subequations}
With this fitting model for $\hat{A}^{\rm BHNS}_{22}/\hat{A}^{\rm BBH}_{22}$ we achieve $R^2=0.97$.

Next, we extract the frequency at merger from our numerical data and use the form
\be
\label{eq:opeak22}
\hat{\omega}^{\rm BHNS}_{22}/\hat{\omega}^{\rm BBH}_{22} = \frac{1 + \Lambda p^{(3)}_{1}(\nu,\abh) + \Lambda^2 p^{(3)}_{2}(\nu,\abh)}{(1 + \Lambda p^{(2)}_{3}(\nu))^2}
\ee
to model it using
\begin{subequations}
  \begin{align}
  p^{(3)}_{k}(\nu,\abh) &= p^{(2)}_{k1}(\abh)\nu + p^{(2)}_{k2}(\abh)\nu^2 \nonumber \\
  &\quad + p^{(2)}_{k3}(\abh)\nu^3,\\
  p^{(2)}_{kj}(\abh) &= c_{kj2}\abh^2 + c_{kj1}\abh + c_{kj0}, \\
  p^{(2)}_{3}(\nu) &= c_{311}\nu^2,
  \end{align}
\end{subequations}
thus also reaching $R^2=0.97$.

To model the characteristic ringdown form, we need to fit the inverse damping time (Eq.~\ref{eq:idt22}) and the frequency (Eq.~\ref{eq:omg22}) of the \gls{QNM}.
For our $\alpha^{\rm BHNS}_{221}/\alpha^{\rm BBH}_{221}$ model we obtain a $R^2=0.95$, whereas for the \gls{QNM} frequency the 
coefficient of determination is $R^2=0.98$
\begin{widetext}
\be 
\label{eq:idt22}
\frac{\alpha^{\rm BHNS}_{221}}{\alpha^{\rm BBH}_{221}} = \frac{1 + \Lambda p^{(2)}_{1}(\nu,\abh) + \Lambda^2 p^{(2)}_{2}(\nu,\abh) + \Lambda^3 p^{(2)}_{3}(\nu,\abh)}{1 + \Lambda p^{(1)}_{41}(\abh)\nu}
\ee
\end{widetext}
\begin{subequations}
  \begin{align}
  p^{(2)}_{k}(\nu,\abh) &= p^{(1)}_{k1}(\abh)\nu + p^{(1)}_{k2}(\abh)\nu^2,\\
  p^{(1)}_{kj}(\abh) &= c_{kj1}\abh + c_{kj0}.
  \end{align}
\end{subequations}
\be 
\label{eq:omg22}
\omega^{\rm BHNS}_{221}/\omega^{\rm BBH}_{221} = \frac{1 + \Lambda p^{(2)}_{1}(\nu,\abh) + \Lambda^2 p^{(2)}_{2}(\nu,\abh)}{\left[1 + \Lambda \left(p^{(2)}_{3}(\nu,\abh)\right)^2\right]^2}
\ee
\begin{subequations}
  \begin{align}
  p^{(2)}_{k}(\nu,\abh) &= p^{(1)}_{k1}(\abh)\nu + p^{(1)}_{k2}(\abh)\nu^2,\\
  p^{(1)}_{kj}(\abh) &= c_{kj1}\abh + c_{kj0}.
  \end{align}
\end{subequations}

The coefficients for the (2,2) mode fit models are shown in Table~\ref{tab:ringdown_coef22}.

\subsubsection{$(\ell,m)$ $=$ $(2,1)$}
Contrary to the dominant mode, we employ the \gls{BBH} fits from~\cite{Pompili:2023tna} to model the amplitude and frequency at merger for the (2,1) mode 
(see Sec.~\ref{sec:eob_ringdown}). The amplitude peak is fitted in the form
\be
\label{apeak21}
\hat{A}^{\rm BHNS}_{21}/\hat{A}^{\rm BBH}_{21} = \frac{1 + \Lambda p^{(3)}_{1}(\nu,\abh) + \Lambda^2 p^{(3)}_{2}(\nu,\abh)}{1 + \Lambda \left(p^{(1)}_{31}(\abh)\nu\right)^2}
\ee
with the polynomials
\begin{subequations}
  \begin{align}
  p^{(3)}_{k}(\nu,\abh) &= p^{(2)}_{k1}(\abh)\nu + p^{(2)}_{k2}(\abh)\nu^2 \nonumber \\
  &\quad + p^{(2)}_{k3}(\abh)\nu^3,\\
  p^{(1)}_{31}(\abh) &= c_{311}\abh + c_{310},\\
  p^{(2)}_{kj}(\abh) &= c_{kj2}\abh^2 + c_{kj1}\abh + c_{kj0}.
  \end{align}
\end{subequations}
Thus obtaining a determination coefficient of $R^2=0.94$. On the other hand, for the frequency at merger we get $R^2=0.93$ employing the model
\begin{widetext}
\be
\label{opeak21}
\frac{\hat{\omega}^{\rm BHNS}_{21}}{\hat{\omega}^{\rm BBH}_{21}} = \frac{1 + \Lambda p^{(3)}_1(\nu,\abh) + \Lambda^2 p^{(3)}_2(\nu,\abh) + \Lambda^3 p^{(3)}_3(\nu,\abh)}{1 + \Lambda p^{(2)}_4(\nu,\abh)}
\ee
\end{widetext}
with the following expressions
\begin{subequations}
  \begin{align}
  p^{(3)}_{k}(\nu,\abh) &= p^{(3)}_{k1}(\abh)\nu + p^{(3)}_{k2}(\abh)\nu^2 + p^{(3)}_{k3}(\abh)\nu^3,\\
  p^{(2)}_{4}(\nu,\abh) &= p^{(2)}_{41}(\abh)\nu + p^{(2)}_{42}(\abh)\nu^2,\\
  p^{(2)}_{kj}(\abh) &= c_{4j1}\abh + c_{4j0}.
  \end{align}
\end{subequations}

The \gls{QNM} quantities $\alpha_{211}$ and $\omega_{211}$ are modelled with Eq.~\ref{idt21} and Eq.~\ref{omg21} respectively.
\be 
\label{idt21}
\alpha^{\rm BHNS}_{211}/\alpha^{\rm BBH}_{211} = \frac{1 + \Lambda p^{(2)}_{1}(\nu,\abh) + \Lambda^2 p^{(2)}_{2}(\nu,\abh)}{\left[1 + \Lambda \left(p^{(2)}_{3}(\nu,\abh)\right)^2\right]^2}
\ee
\begin{subequations}
  \begin{align}
    p^{(2)}_{k}(\nu,\abh) &= p^{(3)}_{k1}(\abh)\nu + p^{(3)}_{k2}(\abh)\nu^2,\\
    p^{(3)}_{kj}(\abh) &= c_{kj3}\abh^3 + c_{kj2}\abh^2 + c_{kj1}\abh \nonumber \\
    &\quad + c_{kj0},\\
    p^{(2)}_{3}(\nu,\abh) &= p^{(1)}_{31}(\abh)\nu + p^{(1)}_{32}(\abh)\nu^2,\\
    p^{(1)}_{3j}(\abh) &= c_{kj1}\abh + c_{kj0}.
  \end{align}
\end{subequations}
\be 
\label{omg21}
\omega^{\rm BHNS}_{211}/\omega^{\rm BBH}_{211} =  \frac{1 + \Lambda p^{(3)}_{1}(\nu,\abh) + \Lambda^2 p^{(3)}_{2}(\nu,\abh)}{\left[1 + \Lambda^2 \left(p^{(3)}_{3}(\nu,\abh)\right)^2\right]^2}
\ee 
\begin{subequations}
  \begin{align}
    p^{(2)}_{k}(\nu,\abh) &= p^{(4)}_{k1}(\abh)\nu + p^{(4)}_{k2}(\abh)\nu^2,\\
    p^{(3)}_{kj}(\abh) &= c_{kj3}\abh^3 + c_{kj2}\abh^2 + c_{kj1}\abh \nonumber \\
    &\quad + c_{kj0},\\
    p^{(2)}_{3}(\nu,\abh) &= p^{(2)}_{31}(\abh)\nu + p^{(2)}_{32}(\abh)\nu^2,\\
    p^{(1)}_{3j}(\abh) &= c_{3j1}\abh + c_{3j0}.
  \end{align}
\end{subequations}
For $\alpha^{\rm BHNS}_{211}/\alpha^{\rm BBH}_{211}$ we get $R^2=0.90$ and for the \gls{QNM} frequency rational between the \gls{BHNS} and \gls{BBH} cases 
one obtains $R^2=0.98$. All the corresponding coefficients for the (2,1) mode models are reported in Table~\ref{tab:ringdown_coef21}.

\subsubsection{$(\ell,m)$ $=$ $(3,2)$}
Similarly to the dominant (2,2), we fit the amplitude peak of the (3,2) mode using the expression
\begin{widetext}
\be
\label{apeak32}
\frac{\hat{A}^{\rm BHNS}_{32}}{\hat{A}^{\rm BBH}_{32}} = \frac{1 + \Lambda p^{(3)}_{1}(\nu,\abh) + \Lambda^2 p^{(3)}_{2}(\nu,\abh) + \Lambda^3 p^{(3)}_{3}(\nu,\abh)}{1 + \Lambda \left( p^{(2)}_{4}(\nu,\abh) \right)^2}
\ee
\end{widetext}
with
\begin{subequations}
  \begin{align}
    p^{(3)}_{k}(\nu,\abh) &= p^{(2)}_{k1}(\abh)\nu + p^{(2)}_{k2}(\abh)\nu^2 + p^{(2)}_{k3}(\abh)\nu^3,\\
    p^{(2)}_{kj}(\abh) &= c_{kj2}\abh^2 + c_{kj1}\abh + c_{kj0},\\
    p^{(2)}_{4}(\nu,\abh) &= p^{(1)}_{41}(\abh)\nu + p^{(1)}_{42}(\abh)\nu^2,\\
    p^{(1)}_{4j}(\abh) &= c_{4j1}\abh + c_{4j0}.
  \end{align}
\end{subequations}

The frequency at merger is instead modelled with
\be
\label{opeak32}
\hat{\omega}^{\rm BHNS}_{32}/\hat{\omega}^{\rm BBH}_{32} = \frac{1 + \Lambda p^{(2)}_{1}(\nu,\abh) + \Lambda^2 p^{(2)}_{2}(\nu,\abh)}{1 + \Lambda \left(p^{(2)}_{3}(\nu,\abh)\right)^2}
\ee
where the polynomials are expressed as
\begin{subequations}
  \begin{align}
    p^{(2)}_{k}(\nu,\abh) &= p^{(1)}_{k1}(\abh)\nu + p^{(1)}_{k2}(\abh)\nu^2,\\
    p^{(1)}_{kj}(\abh) &= c_{kj1}\abh + c_{kj0}.
  \end{align}
\end{subequations}
The quality of the amplitude and frequency peak fits is proved by the coefficients of determination $R^2=0.92$ and $R^2=0.91$ respectively. However,
given the high order nature of the resulting amplitude fit, we perform an out-of-sample test to discard overfitting. Employing a simulation not used for calibration, \gls{BHNS} data on
Fig. 1 of~\cite{Albanesi:2025txj}, we obtain a relative error of 9.92$\%$. 

The \gls{QNM} inverse damping time of the (3,2) mode is fitted with a template in the form
\begin{widetext}
\be 
\label{idt32}
\frac{\alpha^{\rm BHNS}_{321}}{\alpha^{\rm BBH}_{321}} =  \frac{1 + \Lambda p^{(3)}_{1}(\nu,\abh) + \Lambda^2 p^{(3)}_{2}(\nu,\abh) + \Lambda^3 p^{(3)}_{3}(\nu,\abh)}{\left[1 + \Lambda \left(p^{(2)}_{4}(\nu,\abh)\right)^2\right]^2}
\ee
\end{widetext}
using the polynomials
\begin{subequations}
  \begin{align}
    p^{(3)}_{k}(\nu,\abh) &= p^{(3)}_{k2}(\abh)\nu^2 + p^{(3)}_{k3}(\abh)\nu^3,\\
    p^{(2)}_{kj}(\abh) &= c_{kj2}\abh^2 + c_{kj1}\abh + c_{kj0},\\
    p^{(2)}_{4}(\nu,\abh) &= p^{(1)}_{41}(\abh)\nu + p^{(1)}_{42}(\abh)\nu^2,\\
    p^{(1)}_{4j}(\abh) &= c_{4j1}\abh + c_{4j0},
  \end{align}
\end{subequations}
which give $R^2=0.97$. For the \gls{QNM} frequency we obtain as well $R^2=0.97$ employing the expression
\begin{widetext}
\be 
\label{idt32}
\frac{\omega^{\rm BHNS}_{321}}{\omega^{\rm BBH}_{321}} =  \frac{1 + \Lambda p^{(2)}_{1}(\nu,\abh) + \Lambda^2 p^{(2)}_{2}(\nu,\abh) + \Lambda^3 p^{(2)}_{3}(\nu,\abh)}{1 + \Lambda \left(p^{(1)}_{42}(\abh)\nu^2\right)^2}
\ee
\end{widetext}
with 
\begin{subequations}
  \begin{align}
    p^{(2)}_{k}(\nu,\abh) &= p^{(1)}_{k1}(\abh)\nu + p^{(1)}_{k2}(\abh)\nu^2, \\
    p^{(1)}_{kj}(\abh) &= c_{kj1}\abh + c_{kj0}.
  \end{align}
\end{subequations}

considering that we make use of the fits from~\cite{Pompili:2023tna} for (2,1), (3,3) and (4,4), where the matching time is set to be the amplitude peak of (2,2), 
we only need to fit the time shift $\Delta t_{\lm}$ for the (3,2) mode. We do this with the model
\be 
\label{dt32}
\Delta t^{\rm BHNS}_{32}/\Delta t^{\rm BBH}_{32} =  \frac{1 + \Lambda^2 p^{(3)}_1(\nu, \abh) + \Lambda^4 p^{(3)}_2(\nu, \abh)}{\left(1 + \Lambda^2 \left(p^{(2)}_3(\nu, \abh)\right)^2\right)^2}
\ee
with $R^2=0.91$, and the polynomials are expressed as
\begin{subequations}
  \begin{align}
    p^{(3)}_k(\nu, \abh) &= p^{(3)}_{k1}(\abh) \nu + p^{(3)}_{k2}(\abh) \nu^2 \nonumber \\
    &\quad + p^{(3)}_{k3}(\abh) \nu^3, \\
    p^{(3)}_{kj}(\abh) &= c_{kj3} \abh^3 + c_{kj2} \abh^2 + c_{kj1} \abh \nonumber \\
    &\quad + c_{kj0}, \\
    p^{(2)}_3(\nu, \abh) &= p^{(1)}_{31}(\abh) \nu + p^{(1)}_{32}(\abh) \nu^2, \\
    p^{(1)}_{3j}(\abh) &= c_{3j1} \abh + c_{3j0}.
  \end{align}
\end{subequations}
Table~\ref{tab:ringdown_coef32} presents the coefficients of all the fit models made for the (3,2) mode.

\subsubsection{$(\ell,m)$ $=$ $(3,3)$}

Equations~\ref{apeak33} and~\ref{opeak33} below with their corresponding polynomials are used as templates to fit the amplitude and frequency at merger of 
the (3,3) mode.
\be
\label{apeak33}
\hat{A}^{\rm BHNS}_{33}/\hat{A}^{\rm BBH}_{33} = \frac{1 + \Lambda p^{(3)}_1(\nu, \abh) + \Lambda^3 p^{(3)}_2(\nu, \abh)}{\left[1 + \Lambda^2 \left( p^{(2)}_3(\nu, \abh) \right)^2 \right]^2}
\ee
\begin{subequations}
  \begin{align}
    p^{(2)}_k(\nu,\abh) &= p^{(4)}_{k1}(\abh)\nu + p^{(4)}_{k2}(\abh)\nu^2 \nonumber \\
    &\quad + p^{(4)}_{k3}(\abh)\nu^3, \\
    p^{(4)}_{kj}(\abh) &= c_{kj4} \abh^4 + c_{kj3} \abh^3 + c_{kj2} \abh^2 \nonumber \\
    &\quad + c_{kj1} \abh + c_{kj0}, \\
    p^{(2)}_3(\nu,\abh) &= p^{(1)}_{k1}(\abh)\nu + p^{(1)}_{k2}(\abh)\nu^2, \\
    p^{(1)}_{3j}(\abh) &= c_{kj1}\abh + c_{kj0}.
  \end{align}
\end{subequations}
\be
\label{opeak33}
\hat{\omega}^{\rm BHNS}_{33}/\hat{\omega}^{\rm BBH}_{33} = \frac{1 + \Lambda p^{(2)}_1(\nu, \abh) + \Lambda^2 p^{(2)}_2(\nu, \abh)}{\left[ 1 + \Lambda^2 p^{(3)}_3(\nu) \right]^2}
\ee
with
\begin{subequations}
  \begin{align}
    p^{(2)}_k(\nu,\abh) &= p^{(1)}_{k1}(\abh)\nu + p^{(1)}_{k2}(\abh)\nu^2, \\
    p^{(1)}_{kj}(\abh) &= c_{kj3} \abh^3 + c_{kj2} \abh^2 + c_{kj1} \abh \nonumber \\
    &\quad + c_{kj0}, \\
    p^{(2)}_3(\nu) &= c_{310}\nu + c_{320}\nu^2.
  \end{align}
\end{subequations}
The fit of these two quantities result in determination coefficients of $R^2=0.89$ and $R^2=0.91$ respectively. 

To model the \gls{BBH} to \gls{BHNS} deviation of the \gls{QNM}'s inverse damping time, we use the template
\be 
\label{idt33}
\alpha^{\rm BHNS}_{331}/\alpha^{\rm BBH}_{331} =  \frac{1 + \Lambda p^{(2)}_{1}(\nu,a_{BH}) + \Lambda^2 p^{(2)}_{2}(\nu,a_{BH})  }{(1 + \Lambda (p^{(1)}_{31} (a_{BH})\nu)^2) }
\ee
with the polynomials
\begin{subequations}
  \begin{align}
    p^{(2)}_{k}(\nu,a_{BH}) &= p^{(3)}_{k1}(a_{BH})\nu + p^{(3)}_{k2}(a_{BH})\nu^2, \\
    p^{(3)}_{kj}(a_{BH}) &=   c_{kj3}a^3_{BH} + c_{kj2}a^2_{BH} + c_{kj1}a_{BH} \nonumber \\
    &\quad + c_{kj0}, \\
    p^{(1)}_{31}(a_{BH}) &=  c_{311}a_{BH} + c_{310}.
  \end{align}
\end{subequations}

For the \gls{QNM} frequency of the (3,3) mode we employ instead
\be 
\label{omg33}
\omega^{\rm BHNS}_{331}/\omega^{\rm BBH}_{331} =  \frac{1 + \Lambda p^{(2)}_1(\nu, \abh) + \Lambda^2 p^{(2)}_2(\nu, \abh)}{ 1 + \Lambda^2 \left(p^{(1)}_{32}(\abh)\nu^2\right)^2 }
\ee
with
\begin{subequations}
  \begin{align}
    p^{(2)}_1(\nu,\abh) &= p^{(1)}_{11}(\abh)\nu + p^{(1)}_{12}(\abh)\nu^2, \\
    p^{(3)}_{kj}(\abh) &= c_{kj3} \abh^3 + c_{kj2} \abh^2 + c_{kj1} \abh \nonumber \\
    &\quad + c_{kj0}, \\
    p^{(1)}_{32}(\abh) &= c_{321} \abh + c_{320}.
  \end{align}
\end{subequations}
With these \gls{QNM} fit models we obtain $R^2=0.90$ and $R^2=0.95$ respectively. The fitting paramaters for all the (3,3) mode models are reported in 
Table~\ref{tab:ringdown_coef33}.  

\subsubsection{$(\ell,m)$ $=$ $(4,4)$}

The amplitude peak of the (4,4) mode is obtained by fitting the following expression
\begin{widetext}
\be
\label{apeak44}
\frac{\hat{A}^{\rm BHNS}_{44}}{\hat{A}^{\rm BBH}_{44}} = \frac{1 + \Lambda p^{(3)}_1(\nu, \abh) + \Lambda^2 p^{(3)}_2(\nu, \abh) + \Lambda^3 p^{(3)}_3(\nu, \abh)}{1 + \Lambda \left(p^{(2)}_3(\nu, \abh)\right)^2}
\ee
\end{widetext}
with the polynomials
\begin{subequations}
  \begin{align}
    p^{(3)}_k(\nu, \abh) &= p^{(2)}_{k1}(\abh) \nu + p^{(2)}_{k2}(\abh) \nu^2 \nonumber \\
    &\quad + p^{(2)}_{k3}(\abh) \nu^3, \\
    p^{(2)}_{kj}(\abh) &= c_{kj2} \abh^2 + c_{kj1} \abh + c_{kj0}, \\
    p^{(2)}_4(\nu, \abh) &= p^{(1)}_{41}(\abh) \nu + p^{(1)}_{42}(\abh) \nu^2, \\
    p^{(1)}_{4j}(\abh) &= c_{4j1} \abh + c_{4j0}.
  \end{align}
\end{subequations}
Using this template to fit our model we obtain a determination coefficient of $R^2=0.93$.

For the frequency at merger, we employ a template of the form
\be
\label{opeak44}
\hat{\omega}^{\rm BHNS}_{44}/\hat{\omega}^{\rm BBH}_{44} = \frac{1 + \Lambda \left( p^{(2)}_1(\nu, \abh) + \Lambda p^{(2)}_2(\nu, \abh) \right)}{1 + \Lambda \left( p^{(2)}_3(\nu, \abh) \right)^2}
\ee
where
\begin{subequations}
  \begin{align}
    p^{(2)}_k(\nu, \abh) &= p^{(3)}_{k1}(\abh) \nu + p^{(3)}_{k2}(\abh) \nu^2, \\
    p^{(3)}_{kj}(\abh) &= c_{kj3} \abh^3 + c_{kj2} \abh^2 + c_{kj1} \abh \nonumber \\
    &\quad + c_{kj0}, \\
    p^{(2)}_3(\nu, \abh) &= p^{(1)}_{31}(\abh) \nu + p^{(1)}_{32}(\abh) \nu^2, \\
    p^{(1)}_{3j}(\abh) &= c_{3j1} \abh + c_{3j0}.
  \end{align}
\end{subequations}
We obtain $R^2=0.94$ as a measure of the fit model's performance.

Finally, the \gls{QNM} quantities $\alpha_{441}$ and $\omega_{441}$ are modelled as Equations~\ref{idt44} and~\ref{omg44} below with their respective polynomials.
\be 
\label{idt44}
\alpha^{\rm BHNS}_{441}/\alpha^{\rm BBH}_{441} =  \frac{1 + \Lambda p^{(3)}_1(\nu, \abh) + \Lambda^2 p^{(3)}_2(\nu, \abh)}{\left[1 + \Lambda \left(p^{(2)}_3(\nu, \abh)\right)^2\right]^2}
\ee
\begin{subequations}
  \begin{align}
    p^{(3)}_k(\nu, \abh) &= p^{(2)}_{k1}(\abh) \nu + p^{(2)}_{k2}(\abh) \nu^2 \nonumber \\
    &\quad + p^{(2)}_{k3}(\abh) \nu^3, \\
    p^{(2)}_{kj}(\abh) &= c_{kj2} \abh^2 + c_{kj1} \abh + c_{kj0}, \\
    p^{(2)}_3(\nu, \abh) &= p^{(1)}_{31}(\abh) \nu + p^{(1)}_{32}(\abh) \nu^2, \\
    p^{(1)}_{3j}(\abh) &= c_{3j1} \abh + c_{3j0}.
  \end{align}
\end{subequations}
\be 
\label{omg44}
\omega^{\rm BHNS}_{441}/\omega^{\rm BBH}_{441} =  \frac{1 + \Lambda p^{(3)}_1(\nu, \abh) + \Lambda^2 p^{(3)}_2(\nu, \abh)}{\left[1 + \Lambda \left(p^{(1)}_{32}(\abh)\nu^2\right)^2\right]^2}
\ee
\begin{subequations}
  \begin{align}
    p^{(3)}_k(\nu, \abh) &= p^{(2)}_{k1}(\abh) \nu + p^{(2)}_{k2}(\abh) \nu^2 \nonumber \\
    &\quad + p^{(2)}_{k3}(\abh) \nu^3, \\
    p^{(1)}_{kj}(\abh) &= c_{kj1} \abh + c_{kj0}.
  \end{align}
\end{subequations}
These result in a coefficient of determination of $R^2=0.95$ and $R^2=0.99$ correspondingly. The coefficients for all the fitted quantities of the (4,4)
mode are presented in Table~\ref{tab:ringdown_coef44}.

\subsection{\gls{NQC} extraction points}
\label{app:fitmodels_nqc}

In this subsection we present the fits developed for the \gls{NQC} extraction points for all multipoles available (not available for the (3,2) mode), namely 
the quantities ($A^{\rm NQC}_{\lm}$, $\dot{A}^{\rm NQC}_{\lm}$, $\omega^{\rm NQC}_{\lm}$, $\dot{\omega}^{\rm NQC}_{\lm}$), see Sec.~\ref{sec:eob_imr} for more 
details on how these fits are used. Note that both the amplitude $A^{\rm NQC}_{\lm}$ and its time derivative  $\dot{A}^{\rm NQC}_{\lm}$ are normalized by 
$1/\sqrt{(\ell+2)(\ell+1)\ell(\ell-1)}$ in all cases.

\subsubsection{$(\ell,m)$ $=$ $(2,2)$}

The (2,2) mode amplitude extracted at $t^{\rm NQC}_{22}$ is fitted with the template
\be
\label{anqc22}
A^{\rm BHNS}_{22} / A^{\rm BBH}_{22} = \frac{1 + \Lambda p^{(3)}_1(\nu, \abh) + \Lambda^2 p^{(3)}_2(\nu, \abh)}{1 + \Lambda p^{(2)}_3(\nu, \abh)}
\ee
with the polynomials
\begin{subequations} 
\begin{align} 
p^{(3)}_k(\nu, \abh) &= p^{(2)}_{k1}(\abh) \nu + p^{(2)}_{k2}(\abh) \nu^2 \nonumber \\
&\quad + p^{(2)}_{k3}(\abh) \nu^3, \\
p^{(2)}_{kj}(\abh) &= c_{kj2} \abh^2 + c_{kj1} \abh + c_{kj0}, \\
p^{(2)}_3(\nu) &= c_{310} \nu + c_{320} \nu^2. 
\end{align} 
\end{subequations}
This results in a value of $R^2=0.99$ for the determination coefficient.

For the first time derivative of the \gls{NQC} amplitude we obtain instead $R^2=0.87$ using the expression
\be
\label{danqc22}
\dot{A}^{\rm BHNS}_{22} / \dot{A}^{\rm BBH}_{22} = \frac{1 + \Lambda p^{(2)}_1(\nu, \abh) + \Lambda^2 p^{(2)}_2(\nu, \abh)}{\left[1 + \Lambda \left(p^{(2)}_3(\nu, \abh)\right)^2\right]^2}
\ee
as fitting model, where the polynomials are defined as 
\begin{subequations}
  \begin{align}
    p^{(2)}_k(\nu, \abh) &= p^{(1)}_{k1}(\abh) \nu + p^{(1)}_{k2}(\abh) \nu^2, \\
    p^{(1)}_{kj}(\abh) &= c_{kj1} \abh + c_{kj0}.
  \end{align}
\end{subequations}

Additionally, the (2,2) mode frequency and its time derivative extracted at the NQC point are modelled employing Eq.~\ref{onqc22} and~\ref{donqc22} respectively.
\be
\label{onqc22}
\omega^{\rm BHNS}_{22} / \omega^{\rm BBH}_{22} = \frac{1 + \Lambda p^{(3)}_1(\nu, \abh) + \Lambda^2 p^{(3)}_2(\nu, \abh)}{\left(1 + \Lambda \, p^{(1)}_3(\nu)\right)^2}
\ee
\begin{subequations}
  \begin{align}
    p^{(3)}_k(\nu, \abh) &= p^{(2)}_{k1}(\abh) \nu + p^{(2)}_{k2}(\abh) \nu^2 \nonumber \\
    &\quad + p^{(2)}_{k3}(\abh) \nu^3, \\
    p^{(2)}_{kj}(\abh) &= c_{kj2} \abh^2 + c_{kj1} \abh + c_{kj0}, \\
    p^{(1)}_3(\nu) &= c_{310} \nu.
  \end{align}
\end{subequations}
\be
\label{donqc22}
\dot{\omega}^{\rm BHNS}_{22} / \dot{\omega}^{\rm BBH}_{22} = \frac{1 + \Lambda p^{(3)}_1(\nu, \abh) + \Lambda^2 p^{(3)}_2(\nu, \abh)}{\left(1 + \Lambda c_{310} \nu \right)^2}
\ee
\begin{subequations}
  \begin{align}
    p^{(3)}_k(\nu, \abh) &= p^{(2)}_{k1}(\abh) \nu + p^{(2)}_{k2}(\abh) \nu^2 \nonumber \\
    &\quad + p^{(2)}_{k3}(\abh) \nu^3, \\
    p^{(2)}_{kj}(\abh) &= c_{kj2} \abh^2 + c_{kj1} \abh + c_{kj0}.
  \end{align}
\end{subequations}
Thus resulting in the following corresponding determination coefficients: $R^2=0.96$ and $R^2=0.98$. The fit paramaters for all the \gls{NQC} quantities 
modelled for the (2,2) are reported in Table~\ref{tab:nqc_coef22}.

\subsubsection{$(\ell,m)$ $=$ $(2,1)$}

For the (2,1) mode, we extract all \gls{NQC} quantities at $t^{\rm NQC}_{21}=t^{\rm peak}_{22}$ according to the fits from~\cite{Pompili:2023tna} (see 
Sec.~\ref{sec:eob_imr}). We model the \gls{NQC} amplitude of the (2,1) mode with the following expression
\be
\label{anqc21}
A^{\rm BHNS}_{21} / A^{\rm BBH}_{21} = \frac{1 + \Lambda p^{(2)}_1(\nu, \abh) + \Lambda^2 p^{(2)}_2(\nu, \abh)}{1 + \Lambda \left(p^{(2)}_3(\nu, \abh)\right)^2}
\ee
where
\begin{subequations}
  \begin{align}
    p^{(2)}_k(\nu, \abh) &= p^{(3)}_{k1}(\abh) \nu + p^{(3)}_{k2}(\abh) \nu^2, \\
    p^{(3)}_{kj}(\abh) &= c_{kj3} \abh^3 + c_{kj2} \abh^2 + c_{kj1} \abh \nonumber \\
    &\quad + c_{kj0}, \\
    p^{(2)}_3(\nu, \abh) &= p^{(1)}_{32}(\abh) \nu^2, \\
    p^{(1)}_{31}(\abh) &= c_{311} \abh + c_{310}.
  \end{align}
\end{subequations}
With this fit model we obtain $R^2=0.95$.

For the time derivative of the \gls{NQC} amplitude we employ instead
\be
\label{danqc21}
\dot{A}^{\rm BHNS}_{21} / \dot{A}^{\rm BBH}_{21} = \frac{1 + \Lambda  p^{(3)}_1(\nu, \abh) + \Lambda^2 p^{(3)}_2(\nu, \abh) }{1 + \Lambda^2 \left(p^{(1)}_3(\abh)\nu\right)^2}
\ee
with 
\begin{subequations}
  \begin{align}
    p^{(3)}_k(\nu, \abh) &= p^{(2)}_{k1}(\abh) \nu + p^{(2)}_{k2}(\abh) \nu^2 \nonumber \\
    &\quad + p^{(2)}_{k3}(\abh) \nu^3, \\
    p^{(2)}_{kj}(\abh) &= c_{kj2} \abh^2 + c_{kj1} \abh + c_{kj0}, \\
    p^{(1)}_{32}(\abh) &= c_{311} \abh + c_{310}.
  \end{align}
\end{subequations}
which gives us a determination coefficient of $R^2=0.99$.

Moreover, for the \gls{NQC} frequency and its time derivative we employ Eq.~\ref{onqc21} and~\ref{donqc21} which result correspondingly in $R^2=0.85$ and 
$R^2=0.91$,
\begin{widetext}
\be
\label{onqc21}
\frac{\omega^{\rm BHNS}_{21}}{\omega^{\rm BBH}_{21}}  = \frac{1 + \Lambda p^{(3)}_1(\nu, \abh) + \Lambda^2 p^{(3)}_2(\nu, \abh) + \Lambda^3 p^{(3)}_3(\nu, \abh)}{ \left[1 + \Lambda \left(p^{(1)}_{41}(\abh)\nu \right)^2 \right]^2 }
\ee
\end{widetext}
with 
\begin{subequations}
  \begin{align}
    p^{(3)}_k(\nu, \abh) &= p^{(3)}_{k1}(\abh) \nu + p^{(3)}_{k2}(\abh) \nu^2 \nonumber \\
    &\quad + p^{(3)}_{k3}(\abh) \nu^3 \\
    p^{(3)}_{kj}(\abh) &= c_{kj3} \abh^3 + c_{kj2} \abh^2 + c_{kj1} \abh \nonumber \\
    &\quad + c_{kj0} \\
    p^{(2)}_{41}(\abh) &= c_{411} \abh + c_{410} 
  \end{align}
\end{subequations}
\be
\label{donqc21}
\dot{\omega}^{\rm BHNS}_{21} / \dot{\omega}^{\rm BBH}_{21} = \frac{1 + \Lambda p^{(2)}_1(\nu, \abh) + \Lambda^2 p^{(2)}_2(\nu, \abh)}{\left(1 + \Lambda \left( p^{(2)}_3(\nu, \abh)\right)^2 \right)^2}
\ee
where
\begin{subequations}
  \begin{align}
    p^{(2)}_k(\nu, \abh) &= p^{(3)}_{k1}(\abh) \nu + p^{(3)}_{k2}(\abh) \nu^2 \\
    p^{(3)}_{kj}(\abh) &= c_{kj3} \abh^3 + c_{kj2} \abh^2 + c_{kj1} \abh \nonumber \\
    &\quad + c_{kj0}
  \end{align}
\end{subequations}
Table~\ref{tab:nqc_coef21} summarizes all fit parameters obtained for each \gls{NQC} extraction point of the (2,1) mode.

\subsubsection{$(\ell,m)$ $=$ $(3,3)$}

Similarly as before, we set the \gls{NQC} extraction point at $t^{\rm NQC}_{33}=t^{\rm peak}_{22}$ for the (3,3) mode, and model the amplitude as
\be
\label{anqc33}
A^{\rm BHNS}_{33} / A^{\rm BBH}_{33} = \frac{1 + \Lambda p^{(3)}_1(\nu, \abh) + \Lambda^2 p^{(3)}_2(\nu, \abh)}{\left[1 + \Lambda \left(p^{(2)}_3(\nu, \abh)\right)^2 \right]^2}
\ee
with the expressions detailed as 
\begin{subequations}
  \begin{align}
    p^{(3)}_k(\nu, \abh) &= p^{(3)}_{11}(\abh) \nu + p^{(3)}_{12}(\abh) \nu^2 \nonumber \\
    &\quad + p^{(3)}_{13}(\abh) \nu^3, \\
    p^{(3)}_{kj}(\abh) &= c_{kj3} \abh^3 + c_{kj2} \abh^2 + c_{kj1} \abh \nonumber \\
    &\quad + c_{kj0}, \\
    p^{(2)}_3(\nu, \abh) &= p^{(1)}_{31}(\abh) \nu + p^{(1)}_{32}(\abh) \nu^2, \\
    p^{(1)}_{3j}(\abh) &= c_{3j1} \abh + c_{3j0}.
  \end{align}
\end{subequations}
We thus obtain a coefficient of determination of $R^2=0.89$. On the other hand, for the amplitude's time derivative we obtain $R^2=0.90$ with the template in 
Eq.~\ref{danqc33},
\be
\label{danqc33}
\dot{A}^{\rm BHNS}_{33} / \dot{A}^{\rm BBH}_{33} = \frac{1 + \Lambda p^{(3)}_1(\nu, \abh) + \Lambda^2 p^{(3)}_2(\nu, \abh)}{\left[1 + \Lambda \left(p^{(1)}_{32}(\abh) \nu^2\right)^2 \right]^2}
\ee
where the polynomials $p^{(o)}_{kj}$ are
\begin{subequations}
  \begin{align}
    p^{(3)}_k(\nu, \abh) &= p^{(3)}_{k1}(\abh) \nu + p^{(3)}_{k2}(\abh) \nu^2 \nonumber \\
    &\quad + p^{(3)}_{k3}(\abh) \nu^3, \\
    p^{(4)}_{kj}(\abh) &= c_{kj3} \abh^3 + c_{kj2} \abh^2 + c_{kj1} \abh \nonumber \\
    &\quad + c_{kj0}, \\
    p^{(1)}_{32}(\abh) &= c_{321} \abh + c_{320}.
  \end{align}
\end{subequations}

We fit the \gls{NQC} frequency with the template,
\be
\label{onqc33}
\omega^{\rm BHNS}_{33} / \omega^{\rm BBH}_{33} = \frac{1 + \Lambda p^{(2)}_1(\nu, \abh) + \Lambda^2 p^{(2)}_2(\nu, \abh)}{\left(1 + \Lambda \left(p^{(1)}_{32}(\abh) \nu^2 \right)^2\right)^2}
\ee
where
\begin{subequations}
  \begin{align}
    p^{(2)}_k(\nu, \abh) &= p^{(3)}_{k1}(\abh) \nu + p^{(3)}_{k2}(\abh) \nu^2, \\
    p^{(3)}_{kj}(\abh) &= c_{kj3} \abh^3 + c_{kj2} \abh^2 + c_{kj1} \abh \nonumber \\
    &\quad + c_{kj0}, \\
    p^{(1)}_{32}(\abh) &= c_{321} \abh + c_{320}.
  \end{align}
\end{subequations}

For the time derivative of the \gls{NQC} frequency we employ instead,
\begin{widetext}
\be
\label{donqc33}
\frac{\dot{\omega}^{\rm BHNS}_{33}}{\dot{\omega}^{\rm BBH}_{33}}  = \frac{1 + \Lambda p^{(2)}_1(\nu, \abh) + \Lambda^2 p^{(2)}_2(\nu, \abh) + \Lambda^3 p^{(2)}_3(\nu, \abh)}{\left[1 + \Lambda^2 \left(p^{(1)}_{42}(\abh) \nu^2 \right)^2\right]^2}
\ee
\end{widetext}
with 
\begin{subequations}
  \begin{align}
    p^{(2)}_k(\nu, \abh) &= p^{(3)}_{k1}(\abh) \nu + p^{(3)}_{k2}(\abh) \nu^2, \\
    p^{(3)}_{kj}(\abh) &= c_{kj3} \abh^3 + c_{kj2} \abh^2 + c_{kj1} \abh\nonumber \\
    &\quad  + c_{kj0}, \\
    p^{(1)}_{42}(\abh) &= c_{421} \abh + c_{420}.
  \end{align}
\end{subequations}
For both fitting models we obtain a determination coefficient of $R^2=0.89$. The fit paramaters for all \gls{NQC} quantities of the (3,3) are listed
in Table~\ref{tab:nqc_coef33}.

\subsubsection{$(\ell,m)$ $=$ $(4,4)$}

In the case of the (4,4) mode we also make use of the \gls{NQC} fits of~\cite{Pompili:2023tna} and extract at $t^{\rm NQC}_{44}=t^{\rm peak}_{22}$.
The amplitude is modelled by employing the following expression with its corresponding polynomials below,
\begin{widetext}
\be
\label{anqc44}
\frac{A^{\rm BHNS}_{44}}{A^{\rm BBH}_{44}}  = \frac{1 + \Lambda p^{(2)}_1(\nu, \abh) + \Lambda^2 p^{(2)}_2(\nu, \abh) + \Lambda^3 p^{(2)}_3(\nu, \abh)}{\left(1 + \Lambda^2 p^{(2)}_4(\nu, \abh) \right)^2}
\ee
\end{widetext}
where
\begin{subequations}
  \begin{align}
    p^{(2)}_k(\nu, \abh) &= p^{(2)}_{k1}(\abh) \nu + p^{(2)}_{k2}(\abh) \nu^2, \\
    p^{(2)}_{kj}(\abh) &= c_{kj2} \abh^2 + c_{kj1} \abh + c_{kj0}, \\
    p^{(2)}_4(\nu, \abh) &= p^{(1)}_{41}(\abh) \nu + p^{(1)}_{42}(\abh) \nu^2, \\
    p^{(1)}_{4j}(\abh) &= c_{4j1} \abh + c_{4j0},
  \end{align}
\end{subequations}
By employing Eq.~\ref{anqc44} to fit our \gls{NQC} amplitude with our \gls{NR} data, we get $R^2=0.91$.
For its time derivative we use
\be
\label{danqc44}
\dot{A}^{\rm BHNS}_{44} / \dot{A}^{\rm BBH}_{44} = \frac{1 + \Lambda p^{(3)}_1(\nu, \abh) + \Lambda^2 p^{(3)}_2(\nu, \abh)}{\left(1 + \Lambda^2 \left(p^{(2)}_3(\nu, \abh) \right)^2 \right)^2}
\ee
with
\begin{subequations}
  \begin{align}
    p^{(3)}_k(\nu, \abh) &= p^{(2)}_{k1}(\abh) \nu + p^{(2)}_{k2}(\abh) \nu^2\nonumber \\
    &\quad  + p^{(2)}_{k3}(\abh) \nu^3, \\
    p^{(2)}_3(\nu, \abh) &= p^{(2)}_{31}(\abh) \nu + p^{(2)}_{32}(\abh) \nu^2, \\
    p^{(2)}_{kj}(\abh) &= c_{kj2} \abh^2 + c_{kj1} \abh + c_{kj0},
  \end{align}
\end{subequations}
thus giving us a coefficient of determination of $R^2=0.87$.

The \gls{NQC} frequency and its time derivative are fitted with the templates in Eq.~\ref{onqc44} and~\ref{donqc44} below, where we obtain $R^2=0.94$
and $R^2=0.95$ respectively. 

Table~\ref{tab:nqc_coef44} shows the best fit coefficients for all the fitted \gls{NQC} quantities of the (4,4) mode.
\be
\label{onqc44}
\omega^{\rm BHNS}_{44} / \omega^{\rm BBH}_{44} = \frac{1 + \Lambda p^{(2)}_1(\nu, \abh) + \Lambda^2 p^{(2)}_2(\nu, \abh)}{\left( 1 + \Lambda \left(p^{(1)}_{32}(\abh)\nu^2 \right)^2 \right)^2}
\ee
with
\begin{subequations}
  \begin{align}
    p^{(2)}_k(\nu, \abh) &= p^{(3)}_{k1}(\abh) \nu + p^{(3)}_{k2}(\abh) \nu^2, \\
    p^{(3)}_{kj}(\abh) &= c_{kj3} \abh^3 + c_{kj2} \abh^2 + c_{kj1} \abh \nonumber \\
    &\quad + c_{kj0}, \\
    p^{(2)}_{32}(\abh) &= c_{321} \abh + c_{320}.
  \end{align}
\end{subequations}
\be
\label{donqc44}
\dot{\omega}^{\rm BHNS}_{44} / \dot{\omega}^{\rm BBH}_{44} = \frac{1 + \Lambda p^{(2)}_1(\nu, \abh) + \Lambda^2 p^{(2)}_2(\nu, \abh)}{1 + \Lambda \left( p^{(1)}_{32}(\abh)\nu^2 \right)^2}
\ee
where the polynomials are expressed as
\begin{subequations}
  \begin{align}
    p^{(2)}_k(\nu, \abh) &= p^{(3)}_{k1}(\abh) \nu + p^{(3)}_{k2}(\abh) \nu^2, \\
    p^{(2)}_{kj}(\abh) &= c_{kj2} \abh^2 + c_{kj1} \abh + c_{kj0}, \\
    p^{(1)}_{32}(\abh) &= c_{320} \abh + c_{321}.
  \end{align}
\end{subequations}

\section{Waveform robustness for generic orbits}
\label{app:wvf_check}

\begin{figure}
  \centering 
    \includegraphics[width=0.5\textwidth]{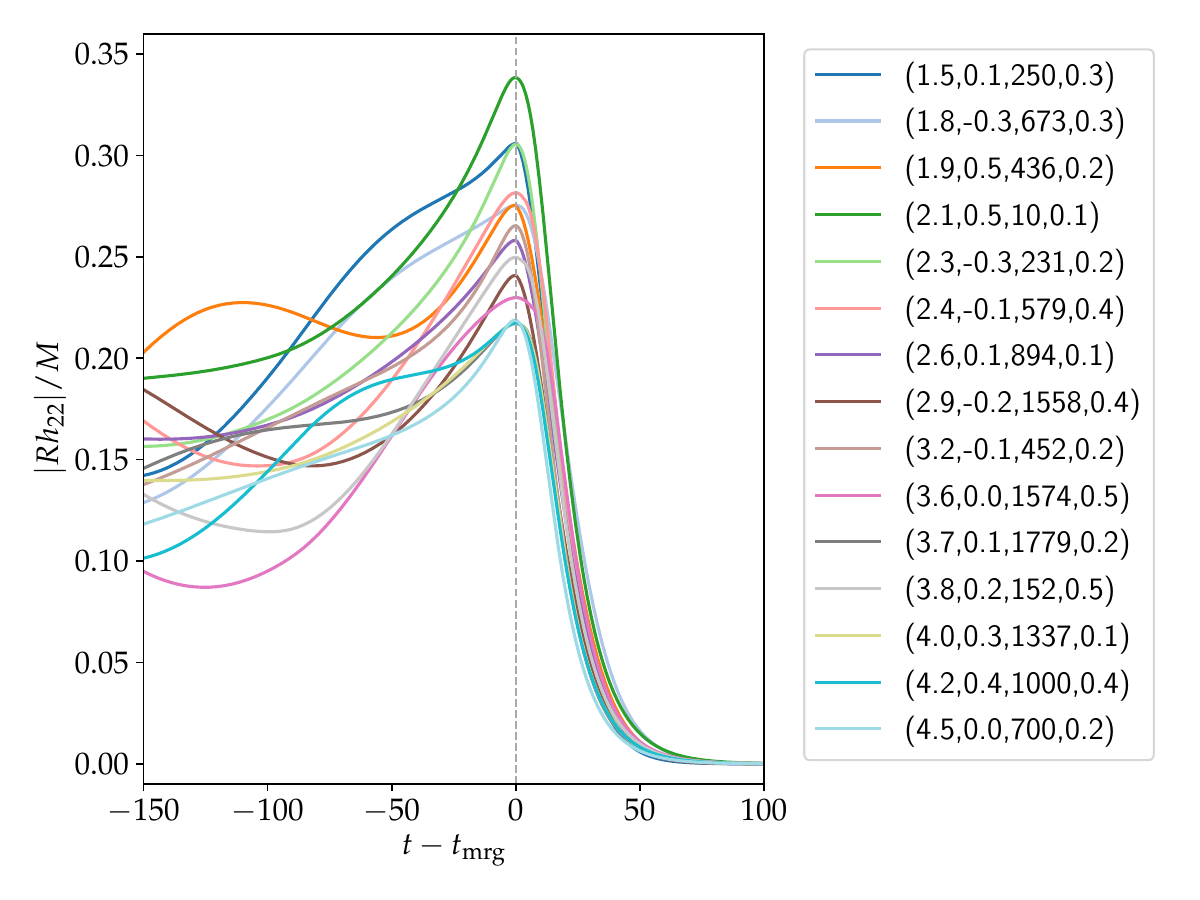}
    \caption{Sanity check for eccentric waveform amplitudes employing different parameters $(q,\abh,\Lambda,e)$. The grey dashed line indicates the moment of merger for all 
    configurations.}
    \label{fig:ecc_wvfs}
\end{figure}

In this Appendix, we highlight the model's robustness and consistency for a variety of configurations with eccentricity. We generate waveform from  over 1000 different binaries
with parameters $q\in[1,5]$,$\abh\in[-0.8,0.8]$, $\Lambda\in[1,5000]$ and $e_0\in[0.01,0.2]$.
In Fig.~\ref{fig:ecc_wvfs} we show the amplitude of
the model's waveforms produced with different $q$, $\abh$, $\Lambda$ and eccentricity, $e$. This figure demonstrates the model's capacity to produce a smooth amplitude and proper ringdown attachment for eccentricities as high as $e \sim 0.5$.

\begin{figure}
  \centering 
    \includegraphics[width=0.5\textwidth]{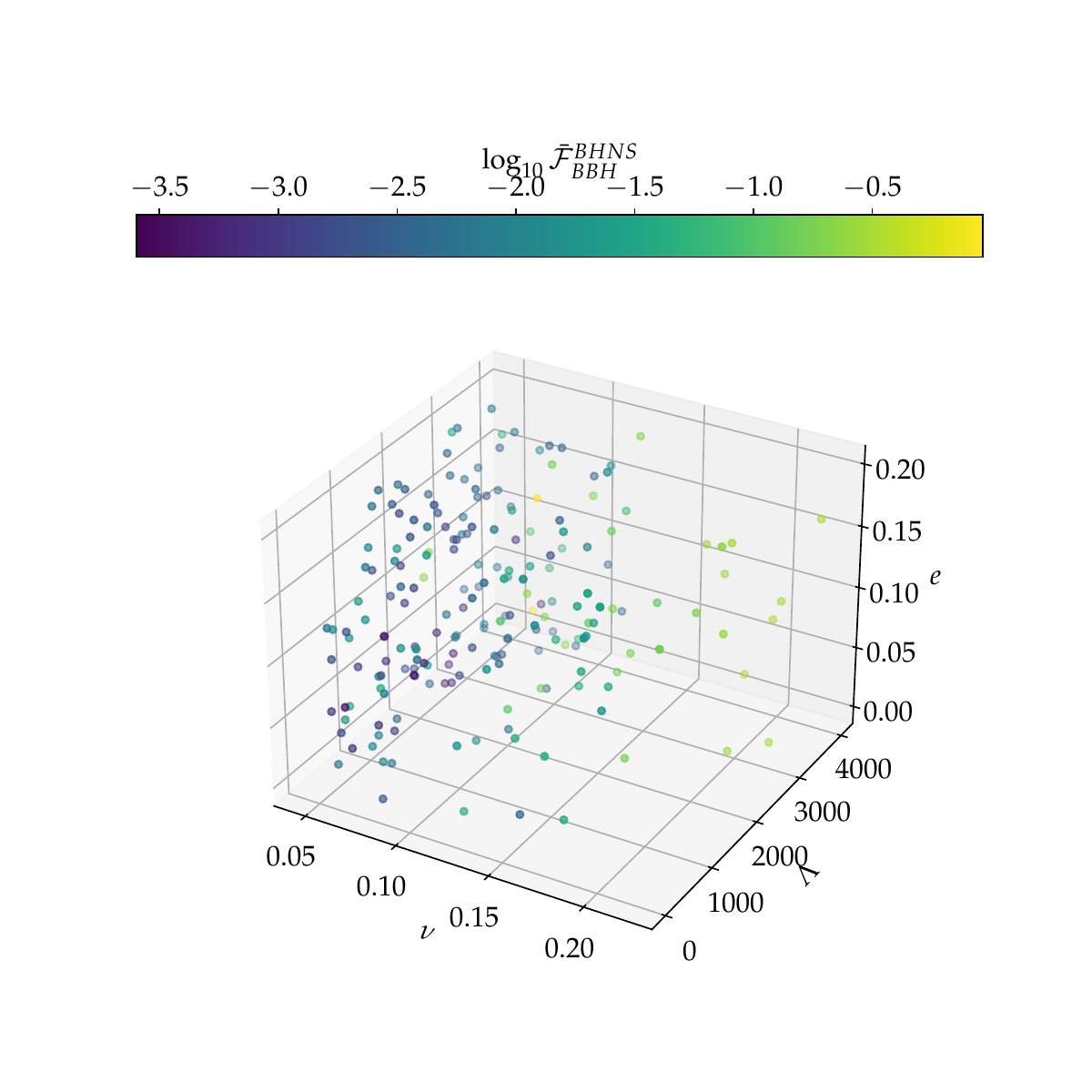}
    \caption{Mismatches between \gls{BHNS} and its corresponding \gls{BBH} waveform, $\bar{\mathcal{F}}^{\rm BHNS}_{\rm BBH}$, in the parameter
    space of ($\nu$, $\Lambda$, $e$).}
    \label{fig:ecc_tides_mm}
\end{figure}

As a second consistency check, we computed mismatches between the \gls{BHNS} waveforms against their corresponding \gls{BBH} cases in order to verify the behaviour that we expect
between the two models: almost identical waveforms towards $\nu\rightarrow0$ and $\Lambda\rightarrow0$, and increasing deviations with higher values of $\Lambda$, $\nu\rightarrow\frac{1}{4}$, 
together with increasing eccentricities. This behaviour is shown by the mismatches in Fig.~\ref{fig:ecc_tides_mm}, demonstrating the robustness of our new model across the vast parameter space.
Overall, the model is sufficiently stable to serve in future parameter estimation studies.

\begin{figure*}[ht]
  \centering 
    \includegraphics[width=0.9\textwidth]{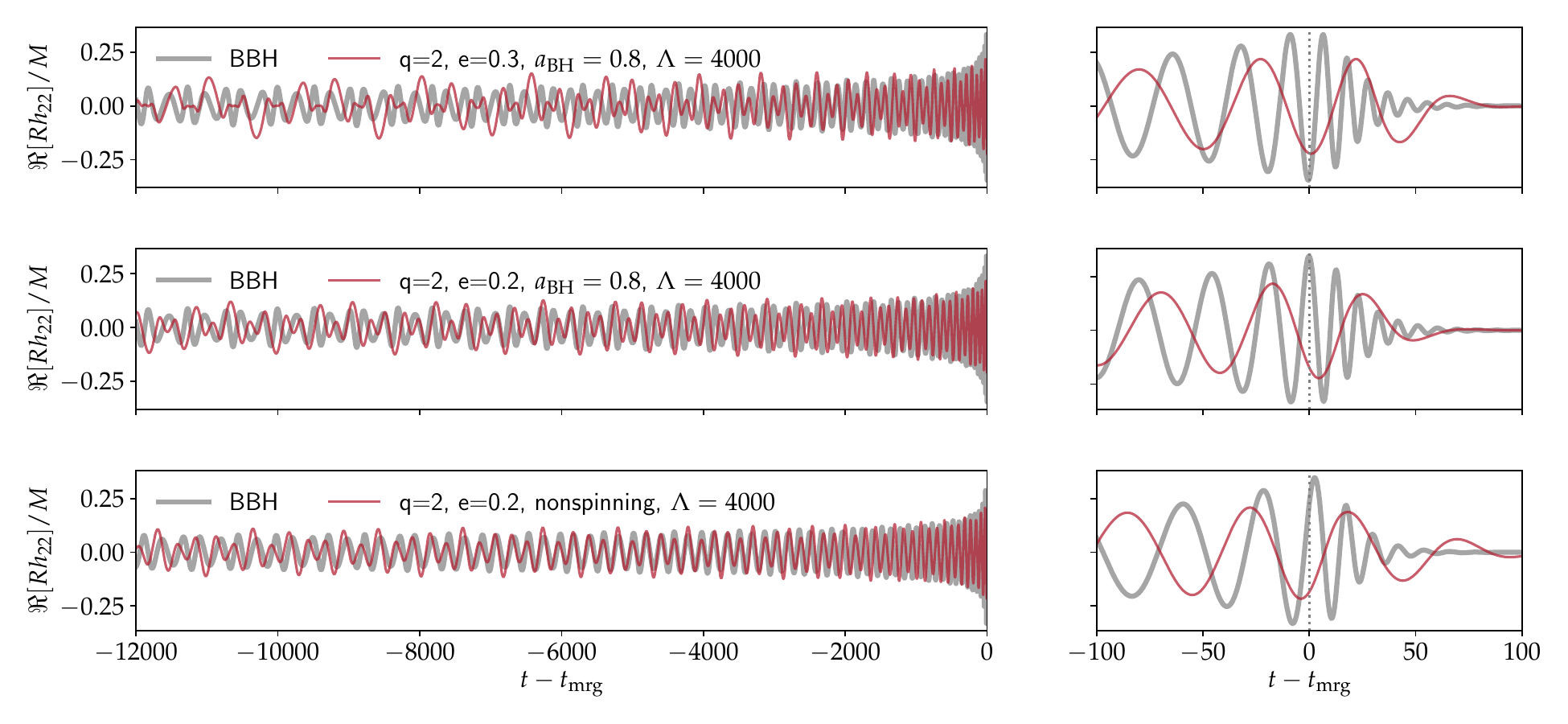}
    \caption{Gravitational waveforms from eccentric binary configurations with increasing initial eccentricity and spin. The corresponding \gls{BBH} ($\Lambda=0$) waveform is 
    added as reference as a gray solid line.}
    \label{fig:eccspins_wvfs}
\end{figure*}

Finally, we discuss further the effect of eccentricity, spins, and tides in our model.
In comparison to the effect from the spins aligned to the orbital plane, precessing spins have little to no noticeable effect on the eccentric-tidal waveforms. 
The combination of increasing $\Lambda$ and $\abh$ has a significant effect on the eccentric waveform, as seen in Fig.~\ref{fig:eccspins_wvfs}.
The oscillating features in the early inspiral get more pronounced with higher spin magnitudes and are specially sensible to increasing eccentricity. On the other hand,
amplitude deviations from the \gls{BBH} case on the inspiral are already significant when tides are turned on (bottom panel of Fig.~\ref{fig:eccspins_wvfs}). A further analytical investigation would be necessary to
understand the interplay between eccentricity, tides, and spin.

\begin{table*}[t]
	\centering    
	\caption{Fitting coefficients for ringdown paramters from the (2,2) waveform. Here, we make a fit of a quantity $\mathcal{F}$ as $\mathcal{F}^{\rm BHNS}/\mathcal{F}^{\rm BBH}$.}
\begin{tabular}{ccccccccccc}        
	\hline\hline
	$\mathcal{F}$ & $k$ & $c_{k12}$ & $c_{k11}$ & $c_{k10}$ & $c_{k22}$ & $c_{k21}$ & $c_{k20}$ & $c_{k32}$ & $c_{k31}$ & $c_{k30}$\\
	\hline
	
	\multirow{ 3}{*}{$\hat{A}_{22}$} & 1 & $2.5 \times 10^{-2}$ & $1.1 \times 10^{-2}$ & $-2.5 \times 10^{-3}$ & $-3.1 \times 10^{-1}$ & $-1.3 \times 10^{-1}$ & $1.4 \times 10^{-1}$ & $8.7 \times 10^{-1}$ & $3.1 \times 10^{-1}$ & $-5.1 \times 10^{-1}$\\ 
	& 2 & $-1.9 \times 10^{-5}$ & $-2.8 \times 10^{-6}$ & $7.7 \times 10^{-6}$ & $2.2 \times 10^{-4}$ & $2.5 \times 10^{-5}$ & $-9.7 \times 10^{-5}$ & $-5.8 \times 10^{-4}$ & $-5.1 \times 10^{-5}$ & $2.8 \times 10^{-4}$\\ 
	& 3 & $-1.6 \times 10^{-2}$ & $8.7 \times 10^{-3}$ & - & - & - & - & - & - & -\\
	\hline 
	
	\multirow{ 3}{*}{$\hat{\omega}_{22}$} & 1 & $1.1 \times 10^{-1}$ & $3.1 \times 10^{-2}$ & $-6.2 \times 10^{-2}$ & $-1.5 \times 10^{0}$ & $-2.3 \times 10^{-1}$ & $9.0 \times 10^{-1}$ & $4.4 \times 10^{0}$ & $2.3 \times 10^{-1}$ & $-2.4 \times 10^{0}$\\ 
	& 2 & $-1.3 \times 10^{-4}$ & $4.7 \times 10^{-5}$ & $3.1 \times 10^{-5}$ & $1.6 \times 10^{-3}$ & $-6.8 \times 10^{-4}$ & $-3.1 \times 10^{-4}$ & $-4.4 \times 10^{-3}$ & $2.1 \times 10^{-3}$ & $9.1 \times 10^{-4}$\\ 
	& 3 & - & $5.1 \times 10^{-2}$ & - & - & - & - & - & - & -\\
  \hline

  \multirow{ 4}{*}{$\alpha_{221}$} & 1 & - & $7.0 \times 10^{-2}$ & $7.1 \times 10^{-3}$ & - & $-2.9 \times 10^{-1}$ & $7.2 \times 10^{-2}$ & - & - & -\\
  & 2 & - & $-5.2 \times 10^{-5}$ & $5.0 \times 10^{-5}$ & - & $1.8 \times 10^{-4}$ & $-2.4 \times 10^{-4}$ & - & - & -\\ 
	& 3 & - & $6.4 \times 10^{-9}$ & $-1.1 \times 10^{-8}$ & - & $-1.2 \times 10^{-8}$ & $5.2 \times 10^{-8}$ & - & - & -\\
  & 4 & - & $-2.7 \times 10^{-3}$ & $5.2 \times 10^{-2}$ & - & - & - & - & - & -\\
  \hline

  \multirow{ 3}{*}{$\omega_{221}$} & 1 & - & $-4.8 \times 10^{6}$ & $4.0 \times 10^{6}$ & - & $1.5 \times 10^{7}$ & $-1.2 \times 10^{7}$ & - & - & -\\
  & 2 & - & $6.3 \times 10^{3}$ & $-4.3 \times 10^{3}$ & - & $-2.8 \times 10^{4}$ & $2.0 \times 10^{4}$ & - & - & -\\ 
	& 3 & - & $6.2 \times 10^{1}$ & $-1.6 \times 10^{1}$ & - & $-3.0 \times 10^{2}$ & $1.8 \times 10^{2}$ & - & - & -\\
	\hline\hline
\end{tabular}
\label{tab:ringdown_coef22}
\end{table*} 

\begin{table*}[t]
	\centering    
	\caption{Fitting coefficients for ringdown parameters from the (2,1) waveform. Here, we make a fit of a quantity $\mathcal{F}$ as $\mathcal{F}^{\rm BHNS}/\mathcal{F}^{\rm BBH}$.}
\begin{tabular}{ccccccccccc}        
	\hline\hline
	$\mathcal{F}$ & $k$ & $c_{k12}$ & $c_{k11}$ & $c_{k10}$ & $c_{k22}$ & $c_{k21}$ & $c_{k20}$ & $c_{k32}$ & $c_{k31}$ & $c_{k30}$\\
	\hline
	
	\multirow{ 3}{*}{$\hat{A}_{21}$} & 1 & $8.0 \times 10^{1}$ & $-6.4 \times 10^{1}$ & $4.8 \times 10^{2}$ & $1.4 \times 10^{4}$ & $-1.7 \times 10^{4}$ & $1.9 \times 10^{2}$ & $-4.7 \times 10^{4}$ & $5.6 \times 10^{4}$ & $-3.7 \times 10^{3}$\\ 
	& 2 & $-3.7 \times 10^{0}$ & $4.1 \times 10^{0}$ & $-1.8 \times 10^{0}$ & $4.5 \times 10^{1}$ & $-4.2 \times 10^{1}$ & $1.6 \times 10^{1}$ & $-1.4 \times 10^{2}$ & $1.1 \times 10^{2}$ & $-3.7 \times 10^{1}$\\ 
	& 3 & - & $-7.2 \times 10^{1}$ & $4.4 \times 10^{1}$ & - & - & - & - & - & -\\
	\hline 
	
	\multirow{ 3}{*}{$\hat{\omega}_{21}$} & 1 & - & $-1.6 \times 10^{7}$ & $2.9 \times 10^{7}$ & - & $1.9 \times 10^{8}$ & $-3.0 \times 10^{8}$ & - & $-5.6 \times 10^{8}$ & $7.8 \times 10^{8}$\\ 
	& 2 & - & $6.6 \times 10^{4}$ & $-2.9 \times 10^{4}$ & - & $-7.2 \times 10^{5}$ & $2.0 \times 10^{5}$ & - & $1.9 \times 10^{6}$ & $-2.0 \times 10^{5}$\\ 
	& 3 & - & $-8.0 \times 10^{1}$ & $4.2 \times 10^{1}$ & - & $8.7 \times 10^{2}$ & $-4.1 \times 10^{2}$ & - & $-2.3 \times 10^{3}$ & $9.9 \times 10^{2}$\\
  & 4 & - & $3.5 \times 10^{1}$ & $-2.1 \times 10^{1}$ & - & $-1.7 \times 10^{2}$ & $2.3 \times 10^{2}$ & - & - & -\\
  \hline

  \multirow{ 3}{*}{$\alpha_{211}$} & & $c_{k13}$ & $c_{k12}$ & $c_{k11}$ & $c_{k10}$ & $c_{k23}$ & $c_{k22}$ & $c_{k21}$ & $c_{k20}$ &  \\
  \hline
  & 1 & $1.0 \times 10^{0}$ & $-6.6 \times 10^{-2}$ & $-1.8 \times 10^{-1}$ & $-4.1 \times 10^{-2}$ & $-4.8 \times 10^{0}$ & $6.4 \times 10^{-1}$ & $1.2 \times 10^{0}$ & $2.5 \times 10^{-1}$ & \\
  & 2 & $-6.5 \times 10^{-4}$ & $2.0 \times 10^{-4}$ & $4.0 \times 10^{-5}$ & $-1.5 \times 10^{-5}$ & $3.6 \times 10^{-3}$ & $-3.7 \times 10^{-4}$ & $5.6 \times 10^{-5}$ & $1.1 \times 10^{-4}$ & \\ 
	& 3 & - & - & $8.3 \times 10^{-1}$ & $7.7 \times 10^{-2}$ & - & - & $-2.4 \times 10^{0}$ & $4.9 \times 10^{-1}$ & \\
  \hline

  \multirow{ 3}{*}{$\omega_{211}$} & 1 & $3.3 \times 10^{0}$ & $1.7 \times 10^{0}$ & $6.4 \times 10^{-1}$ & $1.5 \times 10^{-1}$ & $-1.7 \times 10^{1}$ & $-9.6 \times 10^{0}$ & $-4.0 \times 10^{0}$ & $-9.2 \times 10^{-1}$ & \\
  & 2 & $-4.3 \times 10^{-3}$ & $-6.7 \times 10^{-3}$ & $-4.1 \times 10^{-3}$ & $-6.3 \times 10^{-4}$ & $2.5 \times 10^{-2}$ & $3.9 \times 10^{-2}$ & $2.3 \times 10^{-2}$ & $3.8 \times 10^{-3}$ & \\ 
	& 3 & - & - & $-3.7 \times 10^{-2}$ & $-7.5 \times 10^{-3}$ & - & - & $2.6 \times 10^{-1}$ & $9.2 \times 10^{-2}$ & \\
	\hline\hline
\end{tabular}
\label{tab:ringdown_coef21}
\end{table*} 

\begin{table*}[t]
	\centering    
	\caption{Fitting coefficients for ringdown paramters from the (3,2) waveform. Here, we make a fit of a quantity $\mathcal{F}$ as $\mathcal{F}^{\rm BHNS}/\mathcal{F}^{\rm BBH}$.}
\begin{tabular}{ccccccccccc}        
	\hline\hline
	$\mathcal{F}$ & $k$ & $c_{k12}$ & $c_{k11}$ & $c_{k10}$ & $c_{k22}$ & $c_{k21}$ & $c_{k20}$ & $c_{k32}$ & $c_{k31}$ & $c_{k30}$\\
	\hline
	
	\multirow{ 3}{*}{$\hat{A}_{32}$} & 1 & $1.6 \times 10^{1}$ & $6.4 \times 10^{0}$ & $-4.0 \times 10^{0}$ & $-1.6 \times 10^{2}$ & $-6.8 \times 10^{1}$ & $4.0 \times 10^{1}$ & $4.2 \times 10^{2}$ & $1.9 \times 10^{2}$ & $-9.8 \times 10^{1}$\\ 
	& 2 & $-1.3 \times 10^{-2}$ & $-5.0 \times 10^{-3}$ & $5.1 \times 10^{-3}$ & $1.2 \times 10^{-1}$ & $4.5 \times 10^{-2}$ & $-5.0 \times 10^{-2}$ & $-2.6 \times 10^{-1}$ & $-1.0 \times 10^{-1}$ & $1.2 \times 10^{-1}$\\ 
	& 3 & $2.2 \times 10^{-6}$ & $6.0 \times 10^{-7}$ & $-1.1 \times 10^{-6}$ & $-1.1 \times 10^{-5}$ & $-3.0 \times 10^{-7}$ & $9.9 \times 10^{-6}$ & $2.8 \times 10^{-6}$ & $-1.1 \times 10^{-5}$ & $-2.2 \times 10^{-5}$\\
	& 4 & - & $2.8 \times 10^{0}$ & $1.9 \times 10^{-1}$ & - & $-1.9 \times 10^{1}$ & $-4.3 \times 10^{0}$ & - & - & -\\
  \hline 
	
	\multirow{ 3}{*}{$\hat{\omega}_{32}$} & 1 & - & $-1.9 \times 10^{-1}$ & $1.7 \times 10^{-1}$ & - & $8.0 \times 10^{-1}$ & $-7.4 \times 10^{-1}$ & - & - & -\\ 
	& 2 & - & $3.5 \times 10^{-5}$ & $-4.5 \times 10^{-5}$ & - & $-1.5 \times 10^{-4}$ & $2.0 \times 10^{-4}$ & - & - & -\\ 
	& 3 & - & $1.1 \times 10^{0}$ & $-1.2 \times 10^{0}$ & - & $-4.4 \times 10^{0}$ & $4.3 \times 10^{0}$ & - & - & -\\
  \hline

  \multirow{ 3}{*}{$\alpha_{321}$} & 1 & - & - & - & $1.4 \times 10^{1}$ & $1.1 \times 10^{1}$ & $-4.0 \times 10^{0}$ & $-6.1 \times 10^{1}$ & $-5.1 \times 10^{1}$ & $1.8 \times 10^{1}$\\
  & 2 & - & - & - & $-6.6 \times 10^{-2}$ & $-5.2 \times 10^{-2}$ & $2.1 \times 10^{-2}$ & $3.0 \times 10^{-1}$ & $2.4 \times 10^{-1}$ & $-9.6 \times 10^{-2}$\\ 
	& 3 & - & - & - & $5.5 \times 10^{-5}$ & $4.2 \times 10^{-5}$ & $-1.6 \times 10^{-5}$ & $-2.5 \times 10^{-4}$ & $-2.0 \times 10^{-4}$ & $7.4 \times 10^{-5}$\\
  & 4 & - & $2.4 \times 10^{-1}$ & $7.7 \times 10^{-1}$ & - & $-3.9 \times 10^{-1}$ & $-3.0 \times 10^{0}$ & - & - & -\\
  \hline

  \multirow{ 3}{*}{$\omega_{321}$} & 1 & - & $2.7 \times 10^{-2}$ & $-2.9 \times 10^{-2}$ & - & $-1.6 \times 10^{-1}$ & $1.0 \times 10^{-1}$ & - & - & -\\
  & 2 & - & $4.4 \times 10^{-6}$ & $5.8 \times 10^{-5}$ & - & $2.7 \times 10^{-5}$ & $-2.4 \times 10^{-4}$ & - & - & -\\ 
	& 3 & - & $-2.8 \times 10^{-8}$ & $-3.0 \times 10^{-8}$ & - & $1.2 \times 10^{-7}$ & $1.3 \times 10^{-7}$ & - & - & -\\
  & 4 & - & - & - & - & $5.3 \times 10^{-1}$ & $6.1 \times 10^{-1}$ & - & - & -\\
  \hline
  $\Delta t^{\rm BHNS}_{32}$ & & $c_{k13}$ & $c_{k12}$ & $c_{k11}$ & $c_{k10}$ &  &  &  &  & \\
  \hline
  & 1 & $-1.5 \times 10^{0}$ & $-1.5 \times 10^{0}$ & $1.2 \times 10^{-1}$ & $2.8 \times 10^{-1}$ &  &  &  &  & \\
  & 2 & $8.3 \times 10^{-6}$ & $2.0 \times 10^{-5}$ & $1.4 \times 10^{-5}$ & $2.9 \times 10^{-6}$ &  &  &  &  & \\ 
	& 3 & - & - & $-1.4 \times 10^{-1}$ & $-1.3 \times 10^{-1}$ &  &  &  &  & \\
  \hline
  & & $c_{k23}$ & $c_{k22}$ & $c_{k21}$ & $c_{k20}$ &  &  &  &  & \\
  \hline
  & 1 & $1.4 \times 10^{1}$ & $1.4 \times 10^{1}$ & $-1.9 \times 10^{0}$ & $-2.9 \times 10^{0}$ &  &  &  &  & \\
  & 2 & $-9.9 \times 10^{-5}$ & $-2.3 \times 10^{-4}$ & $-1.6 \times 10^{-4}$ & $-3.4 \times 10^{-5}$ &  &  &  &  & \\ 
	& 3 & - & - & $1.1 \times 10^{0}$ & $8.9 \times 10^{-1}$ &  &  &  &  & \\
  \hline
  & & $c_{k33}$ & $c_{k32}$ & $c_{k31}$ & $c_{k30}$ &  &  &  &  & \\
  \hline
  & 1 & $-3.3 \times 10^{1}$ & $-2.9 \times 10^{1}$ & $7.1 \times 10^{0}$ & $7.7 \times 10^{0}$ &  &  &  &  & \\
  & 2 & $3.0 \times 10^{-4}$ & $6.9 \times 10^{-4}$ & $4.7 \times 10^{-4}$ & $1.0 \times 10^{-4}$ &  &  &  &  & \\ 
	\hline\hline
\end{tabular}
\label{tab:ringdown_coef32}
\end{table*}

\begin{table*}[t]
	\centering    
	\caption{Fitting coefficients for ringdown paramters from the (3,3) waveform. Here, we make a fit of a quantity $\mathcal{F}$ as $\mathcal{F}^{\rm BHNS}/\mathcal{F}^{\rm BBH}$.}
\begin{tabular}{ccccccccccc}        
	\hline\hline
	$\hat{A}_{33}$ & $k$ & $c_{k14}$ & $c_{k13}$ & $c_{k12}$ & $c_{k11}$ & $c_{k10}$ &  &  &  & \\
	\hline
	
	 & 1 & $9.9 \times 10^{1}$ & $2.3 \times 10^{1}$ & $-2.9 \times 10^{1}$ & $1.5 \times 10^{0}$ & $1.5 \times 10^{-1}$ &  &  &  & \\ 
	& 2 & $-1.0 \times 10^{3}$ & $-2.5 \times 10^{2}$ & $3.0 \times 10^{2}$ & $-1.4 \times 10^{1}$ & $-1.4 \times 10^{0}$ &  &  &  & \\ 
  & 3 & - & - & - & $-5.4 \times 10^{-2}$ & $9.6 \times 10^{-3}$ &  &  &  & \\
  \hline
  &  & $c_{k24}$ & $c_{k23}$ & $c_{k22}$ & $c_{k21}$ & $c_{k20}$ &  &  &  & \\
  \hline
   & 1 & $2.6 \times 10^{3}$ & $6.6 \times 10^{2}$ & $-7.7 \times 10^{2}$ & $3.5 \times 10^{1}$ & $3.2 \times 10^{0}$ &  &  &  & \\
  & 2 & $-2.3 \times 10^{-4}$ & $-8.6 \times 10^{-5}$ & $7.1 \times 10^{-5}$ & $1.3 \times 10^{-5}$ & $6.8 \times 10^{-7}$ &  &  &  & \\ 
  & 3 & - & - & - & $2.9 \times 10^{-1}$ & $9.8 \times 10^{-3}$ &  &  &  & \\
  \hline
  &  & $c_{k34}$ & $c_{k33}$ & $c_{k32}$ & $c_{k31}$ & $c_{k30}$ &  &  &  & \\
  \hline
   & 1 & $2.4 \times 10^{-3}$ & $8.9 \times 10^{-4}$ & $-7.5 \times 10^{-4}$ & $-1.5 \times 10^{-4}$ & $-6.7 \times 10^{-6}$ &  &  &  & \\
  & 2 & $-6.2 \times 10^{-3}$ & $-2.3 \times 10^{-3}$ & $2.0 \times 10^{-3}$ & $4.2 \times 10^{-4}$ & $1.9 \times 10^{-5}$ &  &  &  & \\ 
	\hline 
	\hline
	$\mathcal{F}$ & $k$ & $c_{k13}$ & $c_{k12}$ & $c_{k11}$ & $c_{k10}$ & $c_{k23}$ & $c_{k22}$ & $c_{k21}$ & $c_{k20}$ & \\
  \hline
  \multirow{ 3}{*}{$\hat{\omega}_{33}$} & 1 & $-1.8 \times 10^{-1}$ & $-6.3 \times 10^{-2}$ & $3.9 \times 10^{-2}$ & $-1.3 \times 10^{-2}$ & $8.6 \times 10^{-1}$ & $3.1 \times 10^{-1}$ & $-1.9 \times 10^{-1}$ & $-2.2 \times 10^{-4}$ & \\ 
	& 2 & $2.6 \times 10^{-4}$ & $7.3 \times 10^{-5}$ & $-4.7 \times 10^{-5}$ & $7.8 \times 10^{-5}$ & $-1.3 \times 10^{-3}$ & $-3.9 \times 10^{-4}$ & $2.2 \times 10^{-4}$ & $-2.9 \times 10^{-4}$ & \\ 
	& 3 & - & - & $3.2 \times 10^{-5}$ & $-1.2 \times 10^{-4}$ & - & - & - & - & \\
  \hline

  \multirow{ 3}{*}{$\alpha_{331}$} & 1 & $1.6 \times 10^{0}$ & $9.1 \times 10^{-1}$ & $1.1 \times 10^{-1}$ & $-3.4 \times 10^{-2}$ & $-7.7 \times 10^{0}$ & $-1.7 \times 10^{0}$ & $1.1 \times 10^{0}$ & $3.6 \times 10^{-1}$ & \\
  & 2 & $-1.6 \times 10^{-3}$ & $9.1 \times 10^{-5}$ & $3.8 \times 10^{-4}$ & $9.3 \times 10^{-5}$ & $7.1 \times 10^{-3}$ & $-1.4 \times 10^{-3}$ & $-2.1 \times 10^{-3}$ & $-4.3 \times 10^{-4}$ & \\ 
	& 3 & - & - & $2.1 \times 10^{0}$ & $6.6 \times 10^{-1}$ & - & - & - & - & \\
  \hline

  \multirow{ 3}{*}{$\omega_{331}$} & 1 & $-2.0 \times 10^{-2}$ & $-2.4 \times 10^{-2}$ & $2.7 \times 10^{-2}$ & $8.1 \times 10^{-3}$ & $-7.6 \times 10^{-1}$ & $-1.8 \times 10^{-1}$ & $4.6 \times 10^{-1}$ & $2.1 \times 10^{-1}$ & \\
  & 2 & $4.5 \times 10^{-6}$ & $5.3 \times 10^{-6}$ & $-8.2 \times 10^{-6}$ & $-2.8 \times 10^{-6}$ & $-1.4 \times 10^{-4}$ & $1.8 \times 10^{-3}$ & $2.4 \times 10^{-3}$ & $7.3 \times 10^{-4}$ & \\ 
	& 3 & - & - & $3.9 \times 10^{-1}$ & $2.3 \times 10^{-1}$ & - & - & - & - & \\
	\hline\hline
\end{tabular}
\label{tab:ringdown_coef33}
\end{table*} 

\begin{table*}[t]
	\centering    
	\caption{Fitting coefficients for ringdown paramters from the (4,4) waveform. Here, we make a fit of a quantity $\mathcal{F}$ as $\mathcal{F}^{\rm BHNS}/\mathcal{F}^{\rm BBH}$.}
\begin{tabular}{ccccccccccc}        
	\hline\hline
	$\mathcal{F}$ & $k$ & $c_{k12}$ & $c_{k11}$ & $c_{k10}$ & $c_{k22}$ & $c_{k21}$ & $c_{k20}$ & $c_{k32}$ & $c_{k31}$ & $c_{k30}$\\
	\hline
	
	\multirow{ 3}{*}{$\hat{A}_{44}$} & 1 & $3.5 \times 10^{3}$ & $3.3 \times 10^{3}$ & $8.1 \times 10^{2}$ & $-4.1 \times 10^{4}$ & $-3.7 \times 10^{4}$ & $-8.6 \times 10^{3}$ & $1.2 \times 10^{5}$ & $1.1 \times 10^{5}$ & $2.4 \times 10^{4}$\\ 
	& 2 & $2.6 \times 10^{-1}$ & $-5.6 \times 10^{0}$ & $-2.7 \times 10^{0}$ & $-5.5 \times 10^{0}$ & $5.7 \times 10^{1}$ & $2.8 \times 10^{1}$ & $1.9 \times 10^{1}$ & $-1.4 \times 10^{2}$ & $-7.1 \times 10^{1}$\\ 
	& 3 & $-1.6 \times 10^{-3}$ & $2.3 \times 10^{-3}$ & $1.8 \times 10^{-3}$ & $1.9 \times 10^{-2}$ & $-2.2 \times 10^{-2}$ & $-1.8 \times 10^{-2}$ & $-5.5 \times 10^{-2}$ & $5.4 \times 10^{-2}$ & $4.7 \times 10^{-2}$\\
  & 4 & - & $-1.2 \times 10^{2}$ & $-1.0 \times 10^{1}$ & - & $7.4 \times 10^{2}$ & $1.3 \times 10^{2}$ & - & - & -\\
	\hline 
	
	\multirow{ 3}{*}{$\hat{\omega}_{44}$} &  & $c_{k13}$ & $c_{k12}$ & $c_{k11}$ & $c_{k10}$ & $c_{k23}$ & $c_{k22}$ & $c_{k21}$ & $c_{k20}$ & \\
  \hline
  & 1 & $-4.8 \times 10^{-1}$ & $-2.7 \times 10^{-2}$ & $7.4 \times 10^{-1}$ & $3.7 \times 10^{-1}$ & $2.2 \times 10^{0}$ & $5.3 \times 10^{-1}$ & $-2.8 \times 10^{0}$ & $-1.5 \times 10^{0}$ & \\ 
	& 2 & $7.5 \times 10^{-5}$ & $9.0 \times 10^{-5}$ & $-7.8 \times 10^{-5}$ & $-3.1 \times 10^{-5}$ & $-5.9 \times 10^{-4}$ & $-8.8 \times 10^{-4}$ & $8.4 \times 10^{-5}$ & $8.2 \times 10^{-5}$ & \\ 
	& 3 & - & - & $2.3 \times 10^{0}$ & $2.5 \times 10^{0}$ & - & - & $-6.3 \times 10^{0}$ & $-7.5 \times 10^{0}$ & \\
  \hline

  \multirow{ 3}{*}{$\alpha_{441}$} &  & $c_{k12}$ & $c_{k11}$ & $c_{k10}$ & $c_{k22}$ & $c_{k21}$ & $c_{k20}$ & $c_{k32}$ & $c_{k31}$ & $c_{k30}$\\
	\hline
  & 1 & $1.1 \times 10^{1}$ & $-1.7 \times 10^{1}$ & $2.7 \times 10^{0}$ & $-9.8 \times 10^{1}$ & $1.8 \times 10^{2}$ & $-2.6 \times 10^{1}$ & $2.2 \times 10^{2}$ & $-4.5 \times 10^{2}$ & $6.2 \times 10^{1}$\\
  & 2 & $8.8 \times 10^{-2}$ & $-1.2 \times 10^{-2}$ & $1.9 \times 10^{-2}$ & $-8.9 \times 10^{-1}$ & $7.2 \times 10^{-2}$ & $-1.7 \times 10^{-1}$ & $2.3 \times 10^{0}$ & $-7.5 \times 10^{-2}$ & $3.9 \times 10^{-1}$\\ 
	& 3 & - & $-3.4 \times 10^{0}$ & $3.7 \times 10^{0}$ & - & $1.8 \times 10^{1}$ & $-1.7 \times 10^{1}$ & - & - & -\\
  \hline

  \multirow{ 3}{*}{$\omega_{441}$} & 1 & - & $-8.3 \times 10^{-1}$ & $-3.7 \times 10^{-1}$ & - & $8.4 \times 10^{0}$ & $3.8 \times 10^{0}$ & - & $-2.1 \times 10^{1}$ & $-9.8 \times 10^{0}$\\
  & 2 & - & $8.9 \times 10^{-4}$ & $4.9 \times 10^{-4}$ & - & $-9.3 \times 10^{-3}$ & $-5.0 \times 10^{-3}$ & - & $2.4 \times 10^{-2}$ & $1.3 \times 10^{-2}$\\ 
	& 3 & - & - & - & - & $-6.5 \times 10^{-2}$ & $6.5 \times 10^{-1}$ & - & - & -\\
	\hline\hline
\end{tabular}
\label{tab:ringdown_coef44}
\end{table*}

\begin{table*}[t]
	\centering    
	\caption{Fitting coefficients for \gls{NQC} extraction points from the (2,2) waveform. Here, we make a fit of a quantity $\mathcal{F}$ as $\mathcal{F}^{\rm BHNS}/\mathcal{F}^{\rm BBH}$.}
\begin{tabular}{ccccccccccc}        
	\hline\hline
	$\mathcal{F}$ & $k$ & $c_{k12}$ & $c_{k11}$ & $c_{k10}$ & $c_{k22}$ & $c_{k21}$ & $c_{k20}$ & $c_{k32}$ & $c_{k31}$ & $c_{k30}$\\
	\hline
	
	\multirow{ 3}{*}{$A^{\rm NQC}_{22}$} & 1 & $8.5 \times 10^{-4}$ & $1.1 \times 10^{-3}$ & $6.2 \times 10^{-2}$ & $2.1 \times 10^{-1}$ & $-2.6 \times 10^{0}$ & $1.7 \times 10^{0}$ & $-1.3 \times 10^{-1}$ & $7.1 \times 10^{0}$ & $-3.5 \times 10^{0}$\\ 
	& 2 & $-4.6 \times 10^{-7}$ & $3.0 \times 10^{-7}$ & $3.3 \times 10^{-7}$ & $2.4 \times 10^{-4}$ & $-1.1 \times 10^{-4}$ & $-1.7 \times 10^{-5}$ & $-1.2 \times 10^{-3}$ & $1.1 \times 10^{-3}$ & $-6.2 \times 10^{-4}$\\ 
	& 3 & - & - & $6.4 \times 10^{-2}$ & - & - & $1.1 \times 10^{0}$ & - & - & -\\
	\hline 
	
	\multirow{ 3}{*}{$\dot{A}^{\rm NQC}_{22}$} & 1 & - & $6.5 \times 10^{-3}$ & $-2.9 \times 10^{-3}$ & - & $9.1 \times 10^{-2}$ & $1.5 \times 10^{-1}$ & - & - & -\\ 
	& 2 & - & $-1.7 \times 10^{-6}$ & $8.7 \times 10^{-7}$ & - & $1.0 \times 10^{-5}$ & $-2.2 \times 10^{-5}$ & - & - & -\\ 
	& 3 & - & $6.3 \times 10^{-1}$ & $1.9 \times 10^{-1}$ & - & $-1.7 \times 10^{0}$ & $4.3 \times 10^{-1}$ & - & - & -\\
  \hline

  \multirow{ 3}{*}{$\omega^{\rm NQC}_{22}$} & 1 & $3.1 \times 10^{-3}$ & $-1.3 \times 10^{-3}$ & $4.4 \times 10^{-2}$ & $-8.5 \times 10^{-1}$ & $6.5 \times 10^{-1}$ & $3.4 \times 10^{-1}$ & $3.8 \times 10^{0}$ & $-3.5 \times 10^{0}$ & $-5.7 \times 10^{-1}$\\
  & 2 & $-8.7 \times 10^{-7}$ & $7.7 \times 10^{-7}$ & $6.7 \times 10^{-7}$ & $2.2 \times 10^{-4}$ & $-1.3 \times 10^{-3}$ & $1.1 \times 10^{-3}$ & $-7.0 \times 10^{-4}$ & $5.6 \times 10^{-3}$ & $-4.2 \times 10^{-3}$\\ 
	& 3 & - & - & $2.3 \times 10^{-2}$ & - & - & - & - & - & -\\
  \hline

  \multirow{ 3}{*}{$\dot{\omega}^{\rm NQC}_{22}$} & 1 & $-2.1 \times 10^{-3}$ & $-1.4 \times 10^{-4}$ & $3.6 \times 10^{-2}$ & $-3.3 \times 10^{-1}$ & $-5.0 \times 10^{-1}$ & $4.8 \times 10^{-1}$ & $2.0 \times 10^{0}$ & $1.1 \times 10^{0}$ & $-2.3 \times 10^{0}$\\
  & 2 & $1.3 \times 10^{-7}$ & $6.0 \times 10^{-7}$ & $-3.1 \times 10^{-7}$ & $1.2 \times 10^{-3}$ & $-1.3 \times 10^{-3}$ & $2.5 \times 10^{-4}$ & $-5.3 \times 10^{-3}$ & $6.4 \times 10^{-3}$ & $-1.3 \times 10^{-3}$\\ 
	& 3 & - & - & $1.7 \times 10^{-2}$ & - & - & - & - & - & -\\
	\hline\hline
\end{tabular}
\label{tab:nqc_coef22}
\end{table*} 

\begin{table*}[t]
	\centering    
	\caption{Fitting coefficients for \gls{NQC} extraction points from the (2,1) waveform. Here, we make a fit of a quantity $\mathcal{F}$ as $\mathcal{F}^{\rm BHNS}/\mathcal{F}^{\rm BBH}$.}
\begin{tabular}{ccccccccccc}        
	\hline\hline
	$\mathcal{F}$ & $k$ & $c_{k13}$ & $c_{k12}$ & $c_{k11}$ & $c_{k10}$ & $c_{k23}$ & $c_{k22}$ & $c_{k21}$ & $c_{k20}$ & \\
	\hline
	
	\multirow{ 3}{*}{$A^{\rm NQC}_{21}$} & 1 & $-9.3 \times 10^{1}$ & $-1.2 \times 10^{3}$ & $1.1 \times 10^{3}$ & $-1.8 \times 10^{2}$ & $1.9 \times 10^{3}$ & $9.0 \times 10^{3}$ & $-9.5 \times 10^{3}$ & $2.1 \times 10^{3}$ & \\ 
	& 2 & $8.7 \times 10^{-1}$ & $1.4 \times 10^{0}$ & $-1.4 \times 10^{0}$ & $1.5 \times 10^{-1}$ & $-7.1 \times 10^{0}$ & $-6.6 \times 10^{0}$ & $8.6 \times 10^{0}$ & $-1.3 \times 10^{0}$ & \\ 
	& 3 & - & - & $-3.0 \times 10^{2}$ & $1.8 \times 10^{2}$ & - & - & - & - & \\
	\hline 
	
	 &  & $c_{k12}$ & $c_{k11}$ & $c_{k10}$ & $c_{k22}$ & $c_{k21}$ & $c_{k20}$ & $c_{k32}$ & $c_{k31}$ & $c_{k30}$\\
  \hline
  \multirow{ 3}{*}{$\dot{A}^{\rm NQC}_{21}$}& 1 & $-3.7 \times 10^{5}$ & $3.3 \times 10^{5}$ & $-1.3 \times 10^{5}$ & $3.7 \times 10^{6}$ & $-3.4 \times 10^{6}$ & $1.4 \times 10^{6}$ & $-9.1 \times 10^{6}$ & $8.5 \times 10^{6}$ & $-3.5 \times 10^{6}$\\ 
	& 2 & $-7.4 \times 10^{2}$ & $-1.7 \times 10^{2}$ & $4.2 \times 10^{2}$ & $8.9 \times 10^{3}$ & $9.9 \times 10^{2}$ & $-4.2 \times 10^{3}$ & $-2.6 \times 10^{4}$ & $-1.1 \times 10^{3}$ & $1.1 \times 10^{4}$\\ 
	& 3 & - & - & - & - & $-9.3 \times 10^{0}$ & $5.3 \times 10^{0}$ & - & - & -\\
  \hline
  \hline
  $\omega^{\rm NQC}_{21}$ &  & $c_{k13}$ & $c_{k12}$ & $c_{k11}$ & $c_{k10}$ &  &  &  &  & \\
  \hline
  & 1 & $3.5 \times 10^{2}$ & $3.7 \times 10^{2}$ & $4.0 \times 10^{1}$ & $2.5 \times 10^{0}$ &  &  &  &  & \\
  & 2 & $-4.3 \times 10^{-1}$ & $-1.5 \times 10^{0}$ & $-4.6 \times 10^{-1}$ & $-1.5 \times 10^{-2}$ &  &  &  &  & \\ 
	& 3 & $-1.8 \times 10^{-4}$ & $1.4 \times 10^{-3}$ & $6.6 \times 10^{-4}$ & $1.4 \times 10^{-5}$ &  &  &  &  & \\
  & 4 & - & - & $-7.1 \times 10^{-1}$ & $3.8 \times 10^{-1}$ &  &  &  &  & \\
  \hline
  &  & $c_{k23}$ & $c_{k22}$ & $c_{k21}$ & $c_{k20}$ &  &  &  &  & \\
  \hline
  & 1 & $-3.8 \times 10^{3}$ & $-3.9 \times 10^{3}$ & $-3.7 \times 10^{2}$ & $-2.3 \times 10^{1}$ &  &  &  &  & \\
  & 2 & $5.1 \times 10^{0}$ & $1.6 \times 10^{1}$ & $4.6 \times 10^{0}$ & $1.4 \times 10^{-1}$ &  &  &  &  & \\ 
	& 3 & $1.4 \times 10^{-3}$ & $-1.5 \times 10^{-2}$ & $-6.7 \times 10^{-3}$ & $-1.3 \times 10^{-4}$ &  &  &  &  & \\
  \hline
  &  & $c_{k33}$ & $c_{k32}$ & $c_{k31}$ & $c_{k30}$ &  &  &  &  & \\
  \hline
  & 1 & $1.0 \times 10^{4}$ & $1.0 \times 10^{4}$ & $8.6 \times 10^{2}$ & $5.1 \times 10^{1}$ &  &  &  &  & \\
  & 2 & $-1.5 \times 10^{1}$ & $-4.1 \times 10^{1}$ & $-1.1 \times 10^{1}$ & $-3.1 \times 10^{-1}$ &  &  &  &  & \\ 
	& 3 & $-2.3 \times 10^{-3}$ & $3.9 \times 10^{-2}$ & $1.7 \times 10^{-2}$ & $3.0 \times 10^{-4}$ &  &  &  &  & \\
  \hline
   & $k$ & $c_{k13}$ & $c_{k12}$ & $c_{k11}$ & $c_{k10}$ & $c_{k23}$ & $c_{k22}$ & $c_{k21}$ & $c_{k20}$ & \\
  \hline
  \multirow{ 3}{*}{$\dot{\omega}^{\rm NQC}_{21}$} & 1 & $1.1 \times 10^{2}$ & $2.8 \times 10^{2}$ & $1.8 \times 10^{2}$ & $3.0 \times 10^{1}$ & $7.5 \times 10^{2}$ & $2.7 \times 10^{2}$ & $-1.8 \times 10^{2}$ & $-4.8 \times 10^{1}$ & \\
  & 2 & $-5.6 \times 10^{0}$ & $-3.5 \times 10^{0}$ & $4.3 \times 10^{-1}$ & $3.2 \times 10^{-1}$ & $2.7 \times 10^{1}$ & $1.7 \times 10^{1}$ & $-1.9 \times 10^{0}$ & $-1.5 \times 10^{0}$ & \\ 
	& 3 & $1.8 \times 10^{2}$ & $4.6 \times 10^{1}$ & $-7.7 \times 10^{1}$ & $-2.1 \times 10^{1}$ & $-8.3 \times 10^{2}$ & $-2.0 \times 10^{2}$ & $3.8 \times 10^{2}$ & $1.0 \times 10^{2}$ & \\
	\hline\hline
\end{tabular}
\label{tab:nqc_coef21}
\end{table*} 

\begin{table*}[t]
	\centering    
	\caption{Fitting coefficients for \gls{NQC} extraction points from the (3,3) waveform. Here, we make a fit of a quantity $\mathcal{F}$ as $\mathcal{F}^{\rm BHNS}/\mathcal{F}^{\rm BBH}$.}
\begin{tabular}{ccccccccccc}        
	\hline\hline
	$A^{\rm NQC}_{33}$ & $k$ & $c_{k13}$ & $c_{k12}$ & $c_{k11}$ & $c_{k10}$ &  &  &  &  & \\
	\hline
	& 1 & $-8.6 \times 10^{1}$ & $-1.1 \times 10^{1}$ & $1.7 \times 10^{1}$ & $5.1 \times 10^{0}$ &  &  &  &  & \\ 
	& 2 & $3.7 \times 10^{-2}$ & $-2.3 \times 10^{-2}$ & $-3.2 \times 10^{-2}$ & $1.7 \times 10^{-2}$ &  &  &  &  & \\ 
  & 3 & - & - & $-1.1 \times 10^{0}$ & $2.4 \times 10^{0}$ &  &  &  &  & \\ 
	\hline 
  &  & $c_{k23}$ & $c_{k22}$ & $c_{k21}$ & $c_{k20}$ &  &  &  &  & \\
  \hline
	& 1 & $8.5 \times 10^{2}$ & $8.8 \times 10^{1}$ & $-1.7 \times 10^{2}$ & $-4.4 \times 10^{1}$ &  &  &  &  & \\ 
	& 2 & $-3.3 \times 10^{-1}$ & $2.7 \times 10^{-1}$ & $3.2 \times 10^{-1}$ & $-1.5 \times 10^{-1}$ &  &  &  &  & \\ 
  & 3 & - & - & $6.9 \times 10^{0}$ & $-9.8 \times 10^{0}$ &  &  &  &  & \\ 
  \hline 
  &  & $c_{k33}$ & $c_{k32}$ & $c_{k31}$ & $c_{k30}$ &  &  &  &  & \\
  \hline
	& 1 & $-2.1 \times 10^{3}$ & $-1.8 \times 10^{2}$ & $4.1 \times 10^{2}$ & $9.6 \times 10^{1}$ &  &  &  &  & \\ 
	& 2 & $7.2 \times 10^{-1}$ & $-7.7 \times 10^{-1}$ & $-8.0 \times 10^{-1}$ & $3.4 \times 10^{-1}$ &  &  &  &  & \\ 
  \hline
	$\dot{A}^{\rm NQC}_{33}$ &  & $c_{k13}$ & $c_{k12}$ & $c_{k11}$ & $c_{k10}$ &  &  &  &  & \\
  \hline
  & 1 & $-1.0 \times 10^{8}$ & $-2.8 \times 10^{7}$ & $-1.0 \times 10^{7}$ & $7.8 \times 10^{6}$ &  &  &  &  & \\ 
	& 2 & $1.2 \times 10^{5}$ & $7.2 \times 10^{4}$ & $3.3 \times 10^{4}$ & $-2.0 \times 10^{4}$ &  &  &  &  & \\ 
  \hline
  &  & $c_{k23}$ & $c_{k22}$ & $c_{k21}$ & $c_{k20}$ &  &  &  &  & \\
  \hline
  & 1 & $1.1 \times 10^{9}$ & $2.9 \times 10^{8}$ & $9.8 \times 10^{7}$ & $-7.8 \times 10^{7}$ &  &  &  &  & \\ 
	& 2 & $-1.3 \times 10^{6}$ & $-7.4 \times 10^{5}$ & $-3.3 \times 10^{5}$ & $2.0 \times 10^{5}$ &  &  &  &  & \\
  & 3 & - & - & $-2.7 \times 10^{1}$ & $4.9 \times 10^{1}$ &  &  &  &  & \\
  \hline 
  &  & $c_{k33}$ & $c_{k32}$ & $c_{k31}$ & $c_{k30}$ &  &  &  &  & \\
  \hline
  & 1 & $-3.0 \times 10^{9}$ & $-7.2 \times 10^{8}$ & $-2.3 \times 10^{8}$ & $1.9 \times 10^{8}$ &  &  &  &  & \\ 
	& 2 & $3.4 \times 10^{6}$ & $1.9 \times 10^{6}$ & $8.1 \times 10^{5}$ & $-5.1 \times 10^{5}$ &  &  &  &  & \\
  \hline
  &  & $c_{k13}$ & $c_{k12}$ & $c_{k11}$ & $c_{k10}$ & $c_{k23}$ & $c_{k22}$ & $c_{k21}$ & $c_{k20}$ & \\
  \hline
  \multirow{ 3}{*}{$\omega^{\rm NQC}_{33}$} & 1 & $-2.0 \times 10^{1}$ & $-4.3 \times 10^{1}$ & $-2.8 \times 10^{1}$ & $-5.5 \times 10^{0}$ & $1.3 \times 10^{2}$ & $2.7 \times 10^{2}$ & $1.7 \times 10^{2}$ & $3.3 \times 10^{1}$ & \\
  & 2 & $-2.5 \times 10^{-2}$ & $-3.9 \times 10^{-2}$ & $-2.6 \times 10^{-2}$ & $-6.9 \times 10^{-3}$ & $1.8 \times 10^{-1}$ & $3.1 \times 10^{-1}$ & $2.1 \times 10^{-1}$ & $5.2 \times 10^{-2}$ & \\ 
	& 3 & - & - & - & - & - & - & $6.3 \times 10^{0}$ & $4.8 \times 10^{0}$ & \\
  \hline
  \multirow{ 3}{*}{$\dot{\omega}^{\rm NQC}_{33}$} & 1 & $-1.6 \times 10^{9}$ & $1.2 \times 10^{9}$ & $4.0 \times 10^{7}$ & $-1.6 \times 10^{8}$ & $9.0 \times 10^{9}$ & $-6.8 \times 10^{9}$ & $-1.9 \times 10^{8}$ & $8.9 \times 10^{8}$ & \\
  & 2 & $7.9 \times 10^{6}$ & $-6.0 \times 10^{6}$ & $9.6 \times 10^{4}$ & $6.7 \times 10^{5}$ & $-4.5 \times 10^{7}$ & $3.4 \times 10^{7}$ & $-5.3 \times 10^{5}$ & $-3.8 \times 10^{6}$ & \\ 
	& 3 & $-7.0 \times 10^{3}$ & $5.2 \times 10^{3}$ & $-4.3 \times 10^{2}$ & $-4.0 \times 10^{2}$ & $3.8 \times 10^{4}$ & $-2.7 \times 10^{4}$ & $1.5 \times 10^{3}$ & $2.4 \times 10^{3}$ & \\
  & 4 &  &  &  &  &  &  & $-1.1 \times 10^{1}$ & $9.9 \times 10^{0}$ & \\
	\hline\hline
\end{tabular}
\label{tab:nqc_coef33}
\end{table*} 

\begin{table*}[t]
	\centering    
	\caption{Fitting coefficients for \gls{NQC} extraction points from the (4,4) waveform. Here, we make a fit of a quantity $\mathcal{F}$ as $\mathcal{F}^{\rm BHNS}/\mathcal{F}^{\rm BBH}$.}
\begin{tabular}{ccccccccccc}        
	\hline\hline
	$\mathcal{F}$ & $k$ & $c_{k12}$ & $c_{k11}$ & $c_{k10}$ & $c_{k22}$ & $c_{k21}$ & $c_{k20}$ & $c_{k32}$ & $c_{k31}$ & $c_{k30}$\\
	\hline
	
	\multirow{ 3}{*}{$A^{\rm NQC}_{44}$} & 1 & $-3.8 \times 10^{8}$ & $-4.4 \times 10^{8}$ & $-1.1 \times 10^{8}$ & $2.2 \times 10^{9}$ & $2.6 \times 10^{9}$ & $6.6 \times 10^{8}$ & - & - & -\\ 
	& 2 & $2.1 \times 10^{6}$ & $2.4 \times 10^{6}$ & $6.3 \times 10^{5}$ & $-1.3 \times 10^{7}$ & $-1.4 \times 10^{7}$ & $-3.8 \times 10^{6}$ & - & - & -\\ 
	& 3 & $-2.9 \times 10^{3}$ & $-3.5 \times 10^{3}$ & $-1.0 \times 10^{3}$ & $1.7 \times 10^{4}$ & $2.1 \times 10^{4}$ & $6.4 \times 10^{3}$ & - & - & -\\
  & 4 & - & $-3.0 \times 10^{0}$ & $-2.0 \times 10^{0}$ & - & $2.2 \times 10^{1}$ & $1.5 \times 10^{1}$ & - & - & -\\
	\hline 
	
	\multirow{ 3}{*}{$\dot{A}^{\rm NQC}_{44}$} & 1 & $1.4 \times 10^{4}$ & $2.2 \times 10^{3}$ & $-3.4 \times 10^{3}$ & $-2.0 \times 10^{5}$ & $-2.3 \times 10^{4}$ & $5.7 \times 10^{4}$ & $6.5 \times 10^{5}$ & $6.0 \times 10^{4}$ & $-1.9 \times 10^{5}$\\ 
	& 2 & $1.7 \times 10^{2}$ & $3.7 \times 10^{0}$ & $-5.9 \times 10^{1}$ & $-1.5 \times 10^{3}$ & $-4.1 \times 10^{1}$ & $5.2 \times 10^{2}$ & $3.4 \times 10^{3}$ & $1.1 \times 10^{2}$ & $-1.2 \times 10^{3}$\\ 
	& 3 & $6.3 \times 10^{0}$ & $3.0 \times 10^{-1}$ & $-2.2 \times 10^{0}$ & $-3.0 \times 10^{1}$ & $-1.4 \times 10^{0}$ & $1.1 \times 10^{1}$ & - & - & -\\
  \hline

  &  & $c_{k13}$ & $c_{k12}$ & $c_{k11}$ & $c_{k10}$ & $c_{k23}$ & $c_{k22}$ & $c_{k21}$ & $c_{k20}$ & \\
  \hline
  \multirow{ 3}{*}{$\omega^{\rm NQC}_{44}$} & 1 & $-1.1 \times 10^{-1}$ & $-1.6 \times 10^{-1}$ & $1.0 \times 10^{-1}$ & $-6.8 \times 10^{-2}$ & $7.2 \times 10^{-1}$ & $9.2 \times 10^{-1}$ & $-6.2 \times 10^{-1}$ & $6.1 \times 10^{-1}$ & \\
  & 2 & $1.1 \times 10^{-3}$ & $8.7 \times 10^{-4}$ & $-1.6 \times 10^{-4}$ & $-3.2 \times 10^{-4}$ & $-6.0 \times 10^{-3}$ & $-4.9 \times 10^{-3}$ & $6.5 \times 10^{-4}$ & $1.9 \times 10^{-3}$ & \\ 
	& 3 & - & - & - & - & - & - & $-1.3 \times 10^{-1}$ & $1.8 \times 10^{0}$ & \\
  \hline

  \multirow{ 3}{*}{$\dot{\omega}^{\rm NQC}_{44}$} & 1 & - & $1.8 \times 10^{5}$ & $-1.4 \times 10^{5}$ & $2.1 \times 10^{4}$ & - & $-5.6 \times 10^{5}$ & $2.4 \times 10^{5}$ & $3.7 \times 10^{4}$ & \\
  & 2 & - & $-2.9 \times 10^{2}$ & $1.3 \times 10^{2}$ & $1.3 \times 10^{1}$ & - & $1.4 \times 10^{3}$ & $-4.8 \times 10^{2}$ & $-1.6 \times 10^{2}$ & \\ 
	& 3 & - & - & - & - & - & - & $-3.3 \times 10^{3}$ & $2.3 \times 10^{3}$ & \\
	\hline\hline
\end{tabular}
\label{tab:nqc_coef44}
\end{table*}

\clearpage

\printglossary[type=\acronymtype]
\end{document}